\documentclass[journal]{IEEEtran}
\usepackage{cite}
\usepackage{color,soul}
\usepackage{graphicx}
\usepackage{amsmath, amssymb}
\usepackage{caption, subcaption}
\usepackage{float}
\usepackage{hyperref}
\usepackage{multirow}
\usepackage{booktabs}
\usepackage{rotating}
\usepackage{silence}
\makeatletter
\newcommand\HUGES{\@setfontsize\Huge{14}{25}}
\newcommand\HUGESS{\@setfontsize\Huge{24}{45}}
\newcommand\HUGESSS{\@setfontsize\Huge{15}{45}}
\makeatother
\definecolor{indian_pines_corrected_Background!}{RGB}{0.0,0.0,127.5}
\definecolor{indian_pines_corrected_Alfalfa!}{RGB}{0.0,0.0,199.94}
\definecolor{indian_pines_corrected_Corn-notill!}{RGB}{0.0,0.0,255.0}
\definecolor{indian_pines_corrected_Corn-min!}{RGB}{0.0,63.75,255.0}
\definecolor{indian_pines_corrected_Corn!}{RGB}{0.0,127.5,255.0}
\definecolor{indian_pines_corrected_Grass/Pasture!}{RGB}{0.0,191.25,255.0}
\definecolor{indian_pines_corrected_Grass/Trees!}{RGB}{20.56,255.0,226.21}
\definecolor{indian_pines_corrected_Grass/pasture-mowed!}{RGB}{71.98,255.0,174.8}
\definecolor{indian_pines_corrected_Hay-windrowed!}{RGB}{123.39,255.0,123.39}
\definecolor{indian_pines_corrected_Oats!}{RGB}{174.8,255.0,71.98}
\definecolor{indian_pines_corrected_Soybeans-notill!}{RGB}{226.21,255.0,20.56}
\definecolor{indian_pines_corrected_Soybeans-min!}{RGB}{255.0,210.14,0.0}
\definecolor{indian_pines_corrected_Soybean-clean!}{RGB}{255.0,151.11,0.0}
\definecolor{indian_pines_corrected_Wheat!}{RGB}{255.0,92.08,0.0}
\definecolor{indian_pines_corrected_Woods!}{RGB}{255.0,33.06,0.0}
\definecolor{indian_pines_corrected_Bldg-Grass-Tree-Drives!}{RGB}{199.94,0.0,0.0}
\definecolor{indian_pines_corrected_Stone-steel towers!}{RGB}{127.5,0.0,0.0}
\definecolor{PaviaU_Background}{RGB}{0.0,0.0,127.5}
\definecolor{PaviaU_Asphalt}{RGB}{0.0,0.0,255.0}
\definecolor{PaviaU_Meadows}{RGB}{0.0,99.17,255.0}
\definecolor{PaviaU_Gravel}{RGB}{0.0,212.5,255.0}
\definecolor{PaviaU_Trees}{RGB}{77.69,255.0,169.09}
\definecolor{PaviaU_Painted metal sheets}{RGB}{169.09,255.0,77.69}
\definecolor{PaviaU_Bare Soil}{RGB}{255.0,229.81,0.0}
\definecolor{PaviaU_Bitumen}{RGB}{255.0,124.88,0.0}
\definecolor{PaviaU_Self-Blocking Bricks}{RGB}{255.0,19.94,0.0}
\definecolor{PaviaU_Shadows}{RGB}{127.5,0.0,0.0}
\definecolor{houston_data_Background!}{RGB}{0.0,0.0,127.5}
\definecolor{houston_data_Grass-healthy!}{RGB}{0.0,0.0,204.77}
\definecolor{houston_data_Grass-stressed!}{RGB}{0.0,8.5,255.0}
\definecolor{houston_data_Grass-synthetic!}{RGB}{0.0,76.5,255.0}
\definecolor{houston_data_Tree!}{RGB}{0.0,144.5,255.0}
\definecolor{houston_data_Soil!}{RGB}{0.0,212.5,255.0}
\definecolor{houston_data_Water!}{RGB}{41.13,255.0,205.65}
\definecolor{houston_data_Residential!}{RGB}{95.97,255.0,150.81}
\definecolor{houston_data_Commercial!}{RGB}{150.81,255.0,95.97}
\definecolor{houston_data_Road!}{RGB}{205.65,255.0,41.13}
\definecolor{houston_data_Highway!}{RGB}{255.0,229.81,0.0}
\definecolor{houston_data_Railway!}{RGB}{255.0,166.85,0.0}
\definecolor{houston_data_Parking-lot1!}{RGB}{255.0,103.89,0.0}
\definecolor{houston_data_Parking-lot2!}{RGB}{255.0,40.93,0.0}
\definecolor{houston_data_Tennis-court!}{RGB}{204.77,0.0,0.0}
\definecolor{houston_data_Running-track!}{RGB}{127.5,0.0,0.0}
\definecolor{Botswana_Background!}{RGB}{0.0,0.0,127.5}
\definecolor{Botswana_Water!}{RGB}{0.0,0.0,216.66}
\definecolor{Botswana_Hippo grass!}{RGB}{0.0,29.42,255.0}
\definecolor{Botswana_Floodplain grasses1!}{RGB}{0.0,107.88,255.0}
\definecolor{Botswana_Floodplain grasses2!}{RGB}{0.0,186.35,255.0}
\definecolor{Botswana_Reeds1!}{RGB}{28.47,255.0,218.3}
\definecolor{Botswana_Riparian!}{RGB}{91.75,255.0,155.02}
\definecolor{Botswana_Firescar2!}{RGB}{155.02,255.0,91.75}
\definecolor{Botswana_island interior!}{RGB}{218.3,255.0,28.47}
\definecolor{Botswana_Accacia woodlands!}{RGB}{255.0,205.6,0.0}
\definecolor{Botswana_Accacia grasslands!}{RGB}{255.0,132.95,0.0}
\definecolor{Botswana_Short mopane!}{RGB}{255.0,60.3,0.0}
\definecolor{Botswana_Mixed mopane!}{RGB}{216.66,0.0,0.0}
\definecolor{Botswana_Exposed soils!}{RGB}{127.5,0.0,0.0}
\definecolor{KSC_Background!}{RGB}{0.0,0.0,127.5}
\definecolor{KSC_Scrub!}{RGB}{0.0,0.0,216.66}
\definecolor{KSC_Willow-swamp!}{RGB}{0.0,29.42,255.0}
\definecolor{KSC_CP-hammock!}{RGB}{0.0,107.88,255.0}
\definecolor{KSC_Slash-pine!}{RGB}{0.0,186.35,255.0}
\definecolor{KSC_Oak/Broadleaf!}{RGB}{28.47,255.0,218.3}
\definecolor{KSC_Hardwood!}{RGB}{91.75,255.0,155.02}
\definecolor{KSC_Swap!}{RGB}{155.02,255.0,91.75}
\definecolor{KSC_Graminoid-marsh!}{RGB}{218.3,255.0,28.47}
\definecolor{KSC_Spartina-marsh!}{RGB}{255.0,205.6,0.0}
\definecolor{KSC_Cattail-marsh!}{RGB}{255.0,132.95,0.0}
\definecolor{KSC_Salt-marsh!}{RGB}{255.0,60.3,0.0}
\definecolor{KSC_Mud-flats!}{RGB}{216.66,0.0,0.0}
\definecolor{KSC_Water!}{RGB}{127.5,0.0,0.0}

\definecolor{trento_data_Background!}{RGB}{0.0,0.0,127.5}
\definecolor{trento_data_Apples!}{RGB}{0.0,0.0,204.77}
\definecolor{trento_data_Buildings!}{RGB}{0.0,8.5,255.0}
\definecolor{trento_data_Ground!}{RGB}{0.0,76.5,255.0}
\definecolor{trento_data_Woods!}{RGB}{0.0,144.5,255.0}
\definecolor{trento_data_Vineyard!}{RGB}{0.0,212.5,255.0}
\definecolor{trento_data_Roads!}{RGB}{41.13,255.0,205.65}

\begin{document}
\title{Hyperspectral Image Classification—Traditional
to Deep Models: A Survey for Future Prospects}
\author{
Muhammad Ahmad,
Sidrah Shabbir,
Swalpa Kumar Roy,~\IEEEmembership{Student Member,~IEEE,}
Danfeng Hong,~\IEEEmembership{Senior Member,~IEEE,}
Xin Wu,~\IEEEmembership{Member,~IEEE,}
Jing Yao,
Adil Mehmood Khan,~%
Manuel Mazzara,~%
Salvatore Distefano,~%
and Jocelyn Chanussot~\IEEEmembership{Fellow,~IEEE}

\thanks{Manuscript received October 24, 2021; revised November 19, 2021; accepted
November 30, 2021. Date of publication December 9, 2021; date of current
version January 20, 2022. This work was supported in part by the National Natural Science Foundation of China under Grant 42030111 and Grant 41722108. This work was supported by the National Natural Science Foundation of China under Grant 62101045 and the China Postdoctoral Science Foundation Funded Project No. 2021M690385. This work was supported by MIAI@Grenoble Alpes (ANR-19-P3IA-0003) and the AXA Research Fund. This research was also financially supported by The Analytical Center for the Government of the Russian Federation (Agreement No. 70-2021-00143 dd. 01.11.2021, IGK 000000D730321P5Q0002) \emph{Corresponding author: Danfeng Hong}.}
\thanks{M. Ahmad is with Department of Computer Science, National University of Computer and Emerging Sciences, Islamabad, Chiniot-Faisalabad Campus, Chiniot 35400, Pakistan, and Dipartimento di Matematica e Informatica---MIFT, University of Messina, Messina 98121, Italy; (e-mail: mahmad00@gmail.com)}
\thanks{S. Shabbir is with the Department of Computer Engineering, Khwaja Fareed University of Engineering and Information Technology (KFUEIT), Pakistan. (e-mail: sidrah.shabbir@gmail.com)}
\thanks{S. K. Roy is with the Department of Computer Science and Engineering, Jalpaiguri Government Engineering College, West Bengal 735102, India (e-mail: swalpa@cse.jgec.ac.in).}
\thanks{D. Hong and J. Yao are with the Key Laboratory of Digital Earth Science, Aerospace Information Research Institute, Chinese Academy of Sciences, Beijing 100094, China (e-mail: hongdf@aircas.ac.cn; yaojing@aircas.ac.cn).}
\thanks{X. Wu is with the School of Information and Electronics, Beijing Institute of Technology, 100081 Beijing, China, and Beijing Key Laboratory of Fractional Signals and Systems, 100081 Beijing, China. (e-mail: 040251522wuxin@163.com)}
\thanks{A. M. Khan is with the Institute of Data Science and Artificial Intelligence, Innopolis University, Innopolis, 420500, Russia. (e-mail: a.khan@innopolis.ru)}
\thanks{M. Mazzara is with Institute of Software Development and Engineering, Innopolis University, Innopolis, 420500, Russia. (e-mail: m.mazzara@innopolis.ru)}
\thanks{S. Distefano is with  Dipartimento di Matematica e Informatica---MIFT, University of Messina, Messina 98121, Italy. (e-mail: sdistefano@unime.it)}
\thanks{J. Chanussot is with the Univ. Grenoble Alpes, CNRS, Grenoble INP, GIPSA-Lab, 38000 Grenoble, France. (e-mail: jocelyn@hi.is)}
\thanks{Digital Object Identifier 10.1109/JSTARS.2021.3133021}
}
\markboth{IEEE JOURNAL OF SELECTED TOPICS IN APPLIED EARTH OBSERVATIONS AND REMOTE SENSING, VOL. 15, 2022}
{M.Ahmad \MakeLowercase{\textit{et al.}}:}
\maketitle
\begin{abstract}
Hyperspectral Imaging (HSI) has been extensively utilized in many real-life applications because it benefits from the detailed spectral information contained in each pixel. Notably, the complex characteristics i.e., the nonlinear relation among the captured spectral information and the corresponding object of HSI data make accurate classification challenging for traditional methods. In the last few years, Deep Learning (DL) has been substantiated as a powerful feature extractor that effectively addresses the nonlinear problems that appeared in a number of computer vision tasks. This prompts the deployment of DL for HSI classification (HSIC) which revealed good performance. This survey enlists a systematic overview of DL for HSIC and compared state-of-the-art strategies of the said topic. Primarily, we will encapsulate the main challenges of traditional machine learning for HSIC and then we will acquaint the superiority of DL to address these problems. This survey breakdown the state-of-the-art DL frameworks into spectral-features, spatial-features, and together spatial-spectral features to systematically analyze the achievements (future research directions as well) of these frameworks for HSIC. Moreover, we will consider the fact that DL requires a large number of labeled training examples whereas acquiring such a number for HSIC is challenging in terms of time and cost. Therefore, this survey discusses some strategies to improve the generalization performance of DL strategies which can provide some future guidelines.
\end{abstract}


\IEEEpeerreviewmaketitle
\section{Introduction}
\label{SecI}

\IEEEPARstart{H}{perspectral Imaging} (HSI) is concerned with the extraction of meaningful information based on the radiance acquired by the sensor at short or long distances without substantial contact with the object of interest \cite{Ahmad_RS}. HSI provides detailed spectral information by sampling the reflective portion of the electromagnetic spectrum covering a wide range of $0.4-2.4~m$ (i.e. visible $0.4-0.7~m$ to short wave infrared $0.7-2.4~m$) region in hundreds of narrow and contiguous spectral bands. HSI can also explore the (light) emission properties of objects in the range of mid to long infrared regions \cite{hong2021interpretable}.

\begin{figure}[!hbt]
         \centering
         \includegraphics[width=0.48\textwidth]{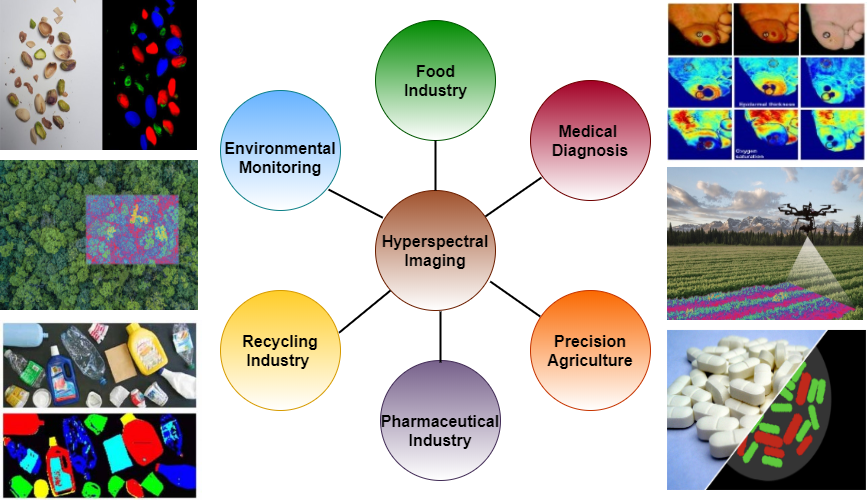}
         \caption{Various real-world applications of HSI.}
         \label{Fig.HSI_app}
\end{figure}

Despite the detailed information, it brings several challenges since traditional analysis techniques for monochromatic, RGB, and multispectral images cannot be directly exploited to extract meaningful information from Hyperspectral ones due to several reasons, e.g. HSI exhibits the unique statistical and geometrical properties of high dimensional spectral/spatial data, i.e. the volume of a hypercube and hypersphere concentrates on corners and outside shells respectively.

HSI has been adopted in several real-world applications including but not limited to the atmosphere, environmental, urban, agriculture, geological and mineral exploration, coastal zone, marine, forestry (i.e. track forest health), water quality and surface contamination, inland waters, and wetlands, snow and ice, biological, medical contexts, and food processing \cite{app10196862, app10175955, 9205804, s21093045, app10217783, HKNCAA}. There are also several military applications in camouflage, landmine detection, and littoral zone mapping. Furthermore, HSI has been used in space, air, and underwater vehicles to acquire detailed spectral information for a wide range of uses \cite{Online1, Xing2017, AHMAD2021166267, Online2}.

Infield collection and spectral library indexing of ground truth signatures for any of the said applications are critical for many reasons. For instance, the spectral information of vegetation is prejudiced by a wide range of environmental situations that make it challenging to satisfactorily represent variability without the collection of site-specific field spectra. But the real potential of HSI is mostly untapped since it allows it to go deeper than surface features considering that usually, each feature has a different spectrum band. HSI, indeed, can capture more than 200 spectral bands which help practitioners to discriminate objects that were not possible before. A few HSI application examples are shown in Fig. \ref{Fig.HSI_app}, but several other domains (e.g. smart city, Industry 4.0, Intelligent Transportation Systems) can greatly benefit from such an approach. 

\begin{figure}[!hbt]
         \centering
         \includegraphics[width=0.45\textwidth]{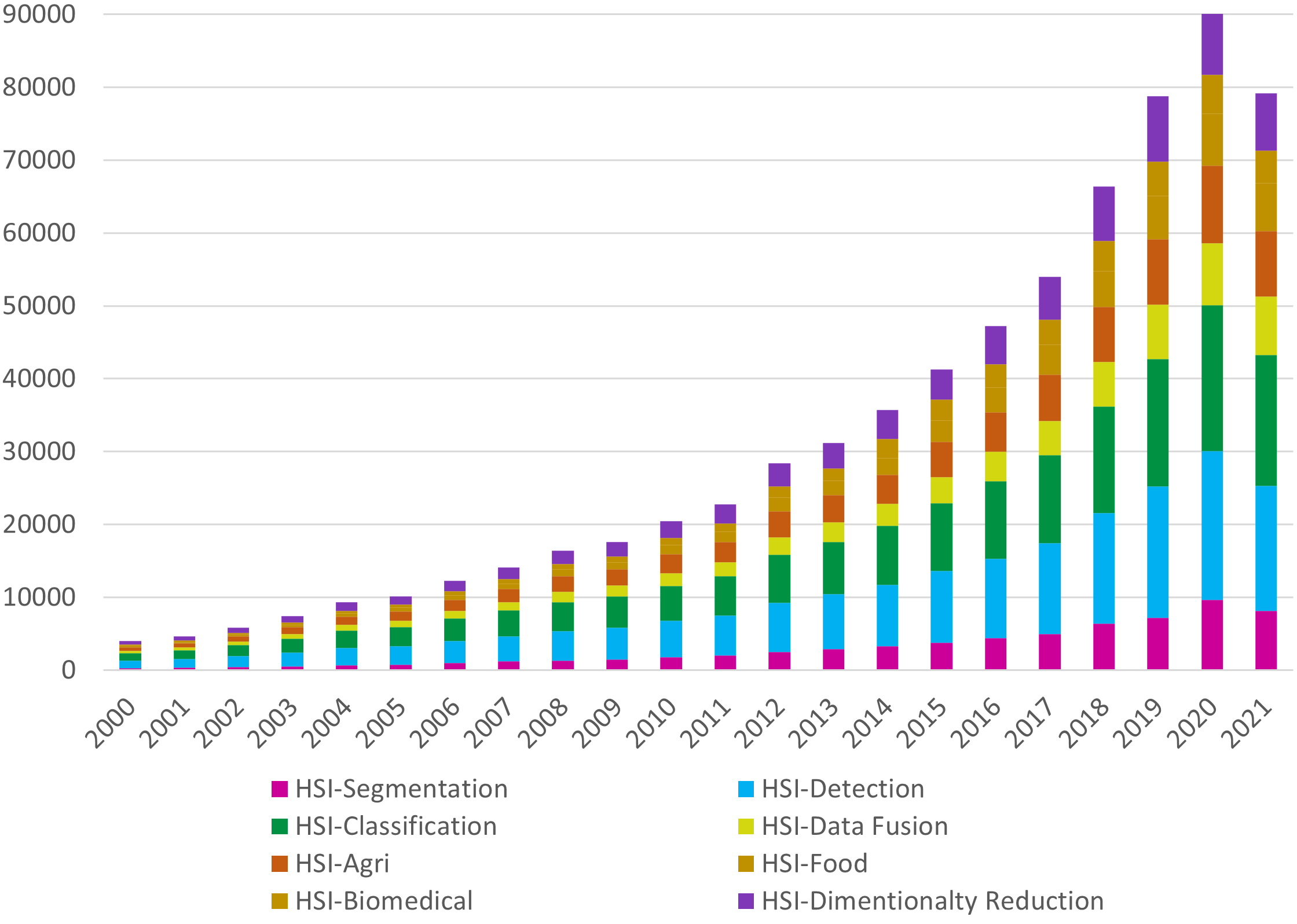}
         \caption{Various HSI related articles published per year till September 25, 2021, [Source: Google Scholar accessed on September 25, 2021 and the results (including patents and citations) were sorted by relevance].}
         \label{Fig.HSI_articles1}
\end{figure}

Considering the aforementioned limitations, HSI analysis is categorized into the following main streams: dimensionality reduction \cite{Ahmad2017, haut2018fast, hong2019learning, Ahmad122, hong2021joint}, spectral unmixing \cite{zhang2011endmember, bioucas2012hyperspectral, Ahmad2017C, zhong2016blind, Ahmad113, Ahmad11, Ahmad112, hong2019augmented, gao2021cycu}, object/change detection \cite{stein2002anomaly,liu2017multiscale, li2018hyperspectral2, wu2019orsim, liu2019review, wu2020fourier, chen2021fccdn} classification \cite{fauvel2012advances, Ahmad2018, ahmad2020}, feature learning for classification \cite{Ahmad2016, AhmadPK17, Mahmad2018,liu2020novel,hong2019cospace}, restoration and denoising \cite{gao2013comparative,wei2017structured}, resolution enhancement \cite{yi2017joint, yi2018hyperspectral}. Figure \ref{Fig.HSI_articles1} shows an exponentially growing trend in literature published per year for HSI analysis-related tasks and applications.

In this survey, we specifically focus on HSI data classification (HSIC), which has achieved a phenomenal interest of the research community due to its broad applications in the areas of land use and land cover \cite{cheng2015effective, zhu2016bag, wu2016hierarchical,cheng2017remote,hong2019learnable}, environment monitoring and natural hazards detection \cite{martha2011segment, cheng2013automatic}, vegetation mapping \cite{mishra2014mapping, li2013object} and urban planning. HSIC methodologies exploit machine learning algorithms to perform the classification task \cite{kotsiantis2006machine, kotsiantis2007supervised}. These methods are outlined in various comprehensive reviews published during/in the last decade \cite{9258970, plaza2009recent, ablin2013survey, fauvel2012advances, camps2013advances, chutia2016hyperspectral, chen2014deep, ghamisi2017advances, li2019deep2, rasti2020feature}. Nevertheless, continuous advancements in the field of Machine Learning provide improved methods from time to time. Deep learning (DL) models is one of such revolutionary advancements in machine learning that improved HSIC accuracy \cite{petersson2016hyperspectral, ahmad2020fast,hong2021spectralformer}.  

This survey aims to give an overview of the widely used DL-based techniques to perform HSIC. Specifically, we will first summarize the main challenges of HSIC which cannot be effectively overcome by traditional machine learning (TML), and later we will enlist the advantages of DL to handle the aforementioned issues. At a later stage, we will provide a framework to categorize the corresponding works among:

\begin{enumerate}
   \item Spectral and spatial feature learning, individually, and
    \item Spectral-spatial feature learning to systematically review the achievements in DL-based HSIC.
    \item Future research stems to improve the generalization performance and robustness of DL models while considering the limited availability of reliable training samples.
\end{enumerate}

The remainder of this paper is structured as follows. 
Section \ref{SecII} introduces the task of HSI Classification (HSIC) and briefly discusses the HSIC paradigm shift from Traditional (Conventional) Machine Learning to Deep Learning (DL) models, describing HSI data characteristics along with the advantages and limitations of DL that are faced while working with HSI. In section \ref{SecIII} and \ref{SecIV}, we give an overview of different forms of HSI representations and basic machine learning strategies, respectively. Section \ref{SecV} describes a few commonly used types of layers and reviews recent developments (specifically from 2017 onward) of some intensively utilized DL frameworks for HSIC. Sections \ref{CNN}, \ref{AE}, \ref{DBN}, and \ref{RNN} presents the state-of-the-art developments of Convolutional Neural Networks (CNN), Graph CNN (GCNN), Autoencoders (AEs), Deep Belief Networks (DBNs), Recurrent Neural networks (RNNs), respectively. In section \ref{SecVI}, we briefly discussed various strategies to overcome the low generalization performance of HSIC due to the limited availability of training data. Section \ref{SecVII} presents the experimental results and discussion on results obtained using different deep learning strategies. Section \ref{SecVIII} concludes the paper with a few future research directions related to joint exploitation of spectral-spatial features of HSI, limited training data, and computational complexity.

\section{Hyperspectral Image Classification (Background and Challenges)}
\label{SecII}

\subsection{\textbf{Traditional to DL Models}}

The main task of HSIC is to assign a unique label to each pixel vector of HSI cube based on its spectral or spectral-spatial properties. Mathematically, an HSI cube can be represented as \(\textbf{X} = [x_1, x_2, x_3, \dots, x_B]^T \in \mathcal{R}^{B \times (N \times M)}\), where \(B\) represent total number of spectral bands consisting of \((N \times M)\) samples per band belonging to \(\textbf{Y}\) classes where \(x_i = [x_{1,i},~x_{2,i},~x_{3,i}, \dots,x_{B,i}]^T\) is the \(i^{th}\) sample in the HSI cube with class label \(y_i \in \mathcal{R}^Y \). The classification problem can be considered as an optimization one, in which a mapping function \(f_c(.)\) takes the input data \(\textbf{X}\) and after applying some transformations over it, obtains the corresponding label \(\textbf{Y}\), to reduce the gap between obtained output and the actual one \cite{paoletti2019deep}.

\begin{equation}
Y = f_c(X,\theta)
\end{equation}
where $\theta$ is a certain adjustable parameter that may be required to apply transformations on input data $\textbf{X}$ such that $f_c: X \to Y$.

\begin{figure}[!hbt]
         \centering
         \includegraphics[width=0.48\textwidth]{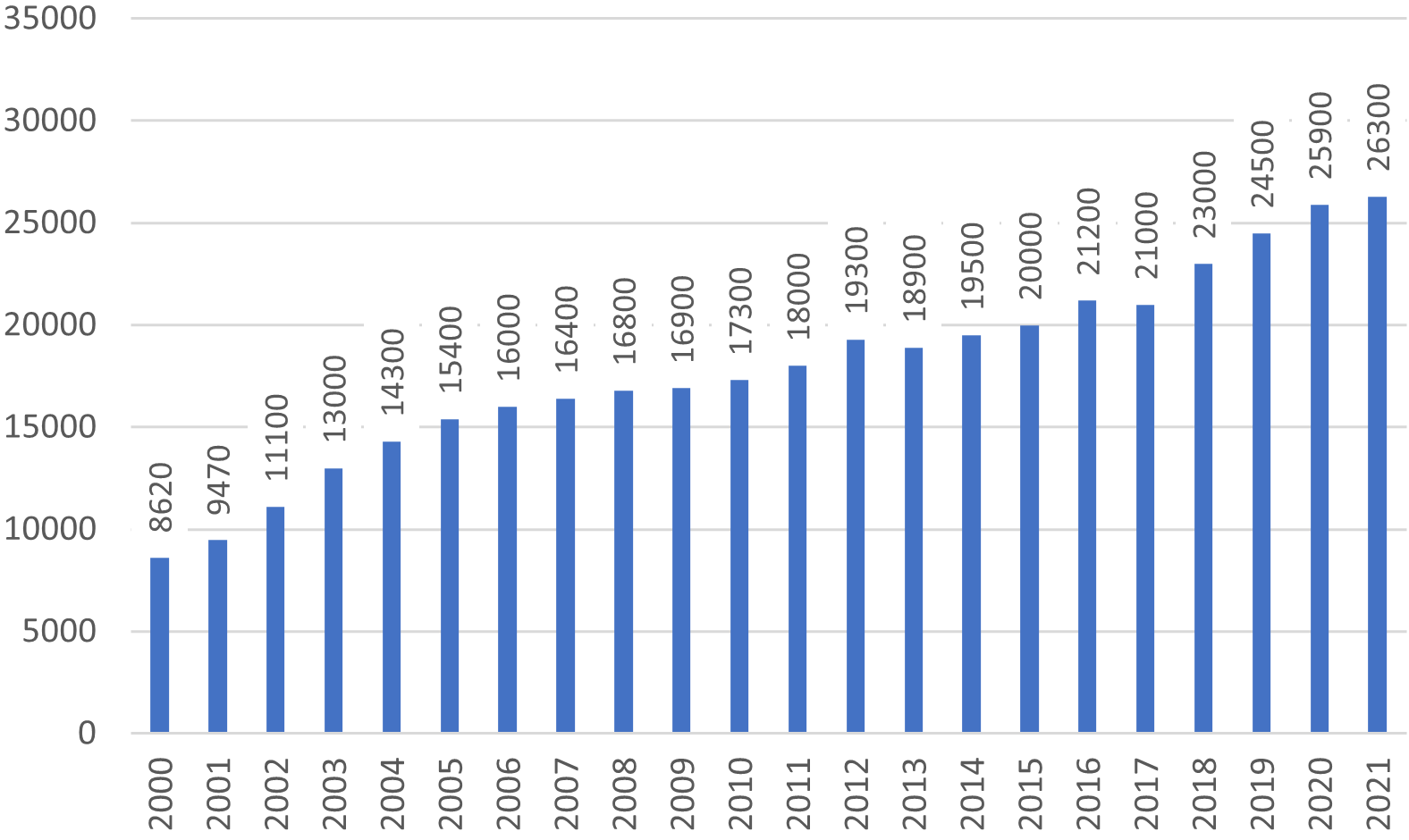}
         \caption{Remote sensing/Hyperspectral Image Classification related articles published per year till September 25, 2021, [Source: Google Scholar accessed on September 25, 2021 and the results (including patents and citations) were sorted by relevance].}
         \label{Fig.HSI_articles}
\end{figure}

In literature, substantial work has been done on HSIC and there is a growing trend in the development of such techniques as shown in Figure \ref{Fig.HSI_articles}. Most HSIC frameworks seemed to be influenced by the methodologies used in the computer vision domain \cite{hang2019cascaded}. Traditional machine learning-based HSIC approaches use hand-crafted features to train the classifier. These methods generally rely on utilizing engineering skills and domain expertise to design several human-engineered features, for instance, shape, texture, color, shape, spectral and spatial details. All these features are basic characteristics of an image and carry effective information for image classification. Commonly used hand-crafted feature extraction and classification methods include: texture descriptors such as Local Binary Patterns (LBPs) \cite{huang2016remote}, Histogram of Oriented Gradients (HOG) \cite{dalal2005histograms}, Global Image Scale-invariant Transform / Global Invariant Scalable Transform (GIST) \cite{oliva2001modeling}, Pyramid Histogram of Oriented Gradients (PHOG), Scale-invariant Feature Transform (SIFT) \cite{lowe1999object}, Random Forests \cite{ham2005investigation}, kernel-based Support Vector Machine (SVM) \cite{camps2005kernel}, K-nearest Neighbours (KNN), and Extreme Learning Machine (ELM).

Color histograms are simple and effective handcrafted features used for an image classification task. They are easy to compute and invariant to small changes in images i.e. translation and rotation. The major drawback of a color histogram is that it does not provide spatial contextual information, hence it becomes difficult to distinguish between objects of the same color but different distribution. Moreover, color histograms are sensitive to variance in illumination. HOG features represent the histogram of edge orientations of spatial sub-regions. It can effectively extract the edge and local shape details and has been utilized in various remote sensing related works \cite{article123, cheng2016object, cheng2015auto, cheng2015effective}.

Scale-invariant Feature Transform (SIFT) is a broadly used robust feature descriptor applied to image classification tasks \cite{azhar2015batik, zeglazi2016sift, xu2008classification, yang2008comparing}. The advantage of the SIFT descriptor is that it is invariant to the changes in image scale, rotation, illumination, and noise. SIFT is used to extract local features that describe a specific point in the image. The disadvantage of SIFT is that it is mathematically complex which increases its computational cost. GIST represents the global description of important aspects of an image that is the scales and orientations (gradient information) of various subregions of an image. GIST builds a spatial envelope in terms of different statistical properties like roughness, openness, and ruggedness, etc \cite{nhat2019feature}. Texture descriptors such as local binary patterns (LBPs) are used for remote sensing image analysis \cite{huang2016remote,roy2018local}. LBPs are used to describe the texture around each pixel by choosing pixels from the square neighborhood and gray level values of all neighborhood pixels are thresholded with respect to the central pixel.

The color histograms, GIST, and texture descriptors are global features that represent certain statistical characteristics of an image like color, texture \cite{haralick1973textural, ojala2002multiresolution}, and spatial structure \cite{oliva2001modeling}. While HOG and SIFT are local features that describe geometrical information. Usually they are used to construct bag-of-visual-words (BoVW) models \cite{cheng2013automatic, zhao2016feature, wu2016hierarchical, zhang2013high, yang2010bag, xu2009object, zhang2016semantic, yang2008comparing, bahmanyar2015comparative} and HOG feature-based models \cite{cheng2015effective, cheng2014multi}. Some popular feature encoding or pooling strategies to enhance the performance of BoVW are Fisher vector coding \cite{zhao2016fisher, hu2015comparative, huang2016remote}, Spatial Pyramid Matching (SPM) \cite{lazebnik2006beyond}, and Probabilistic Topic Model (PTM) \cite{zhao2015dirichlet, kusumaningrum2014integrated, zhong2015scene, bahmanyar2015comparative}. A single feature is insufficient to represent the whole image information, hence a combination of these features is used for image classification \cite{hong2020x, zhu2016bag, zhao2016feature, zhao2015dirichlet, zhong2015scene, yu2016color, risojevic2012fusion, mekhalfi2015land, sheng2012high, xie2016incorporating, hong2021more}.

Hand-crafted features can effectively represent the various attributes of an image, hence working well with the data being analyzed. However, these features may be insubstantial in the case of real data, therefore it is difficult to fine-tune between robustness and discriminability as the set of optimal features considerably vary between different data. Furthermore, human involvement in designing the features considerably affects the classification process, as it requires a high level of domain expertise to design hand-crafted features.

To mitigate the limitations of hand-crafted feature designing, a deep feature learning strategy was proposed by Hinton and Salakhutdinov in $2006$ \cite{hinton2006reducing}. Deep learning (DL) based methods can automatically learn the features from data in a hierarchical manner, to construct a model with growing semantic layers until a suitable representation is achieved. Such models have shown great potential for feature representation in remote sensing image classification \cite{zou2015deep, hu2015transferring}. 

DL architectures can learn the behavior of any data without any prior knowledge regarding the statistical distribution of the input data \cite{chen2014convolutional} and can extract both linear and non-linear features of input data without any pre-specified information. Such systems are capable of handling HSI data in both spectral and spatial domains individually, and also in a coupled fashion. DL systems possess a flexible architecture in terms of types of layers and their depth and are adaptive to various machine learning strategies like supervised, semi-supervised, and unsupervised techniques.

\subsection{\textbf{Hyperspectral Data Characteristics and DL Challenges}}

Despite the above-discussed DL potentials, there are still some challenges that need to be considered while applying DL to HSI data. Most of these challenges are related to the characteristics of HSI data i.e. hundreds of contiguous and narrow spectral channels with very high spectral resolution and low spatial resolution throughout the electromagnetic spectrum coupled with limited availability of training data. Although the pixels with rich spectral information are useful for classification purposes, however, the computation of such data takes a lot of time and resources. 

Furthermore, processing such high-dimensional data is a somewhat complex task due to an increased number of parameters. This is known as the curse of dimensionality which considerably influences the classification performance especially in the case of supervised learning \cite{bellman2015adaptive}. Since the size of training data is not adequate/insufficient and/or not reliable (i.e. the training samples may not provide any new information to the model or may have similar patterns/structures) to properly train the classifier which may lead the model to overfit. This is known as the Hughes phenomena \cite{hughes1968mean} which occurs when labeled training data is significantly smaller than the number of spectral bands present in the data. Lack of labeled HSI data is a major issue in HSIC as labeling of HSI is a time-consuming and expensive task because it usually requires human experts or investigation of real-time scenarios.

In addition to high dimensionality, HSIC suffers from various other artifacts like high intra-class variability due to unconfined variations in reflectance values caused by several environmental interferers and degradation of data caused by instrumental noise while capturing the data \cite{reichstein2019deep}. Furthermore, the addition of redundant bands due to HSI instruments affects the computational complexity of the model. Spectral mixing is another challenge related to the spatial resolution of HSI. HSI pixels with low to average spatial resolution cover vast spatial regions on the surface of earth leading to mixed spectral signatures which result in high inter-class similarity in border regions. As a result, it becomes difficult to identify the materials based on their spectral reflectance values \cite{bioucas2013hyperspectral}. Following are some main challenges that come across when DL is applied to HSIC:

\begin{itemize}
    \item \textbf{Complex Training Process}: Training of Deep Neural Network (DNN) and optimization by tuning parameters is an NP-complete problem where the convergence of the optimization process is not guaranteed \cite{nguyen2017optimization, ahmad2021artifacts, ahmad2021regularized}. Therefore, it is assumed that training of DNN is very difficult \cite{chen2014convolutional} especially in the case of HSI when a large number of parameters need to be adjusted/tuned. However, the convergence task becomes somehow easier due to the advancement of various optimization techniques for deep CNNs. Among stochastic gradient descent (SGD) \cite{bottou1991stochastic} and its momentum version (SGDM) \cite{qian1999momentum}, RMSProp \cite{hinton2012neural}, Adam \cite{kingma2014adam}, AdamW \cite{loshchilov2017decoupled}, diffGrad \cite{dubey2019diffgrad}, RAdam \cite{liu2019variance}, gradient centralization (GC) \cite{yong2020gradient}, AngularGrad \cite{roy2021angulargrad}, respectively are the successful CNN optimization techniques and widely used in any classification problems.

    \item \textbf{Limited Availability of Training Data}: As discussed above, supervised DNN requires a considerably large amount of training data otherwise their tendency to overfit increases significantly \cite{erhan2010does} leads to the Hughes phenomena. The high dimensional characteristic of HSI coupled with a small amount of labeled training data makes the DNNs ineffective for HSIC as it demands a lot of adjustments during the training phase \cite{paoletti2019deep}.

    \item  \textbf{Model's Interpretability}: The training procedure of DNNs is difficult to interpret and understand. The black box kind of nature is considered as a potential weakness of DNNs and may affect the design decisions of the optimization process. Although, a lot of work has been done to interpret the model's internal dynamics.

    \item  \textbf{High Computational Burden}: One of the main challenges of DNN is dealing with a big amount of data that involves increased memory bandwidth, high computational cost, and storage consumption \cite{alom2019state}. However, advanced processing techniques like parallel and distributed architectures \cite{plaza2008experimental, plaza2011parallel} and high-performance computing (HPC) \cite{bioucas2013hyperspectral} make it possible for DNNs to process large amounts of data.

    \item \textbf{Training Accuracy Degradation}: It is assumed that deeper networks extract more rich features from data \cite{he2016deep}, however, this is not true for all systems to achieve higher accuracy by simply adding more layers. Because by increasing the network’s depth, the problem of exploding or vanishing gradient becomes more prominent \cite{bengio1994learning} and affects the convergence of the model \cite{he2016deep}.
\end{itemize}

\section{HSI Representation}
\label{SecIII}

\begin{figure}[!hbt]
         \centering
         \includegraphics[width=0.30\textwidth]{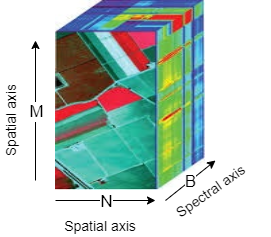}
         \caption{Hyperspectral Cube}
         \label{Fig.HSI}
\end{figure}

Hyperspectral data is represented in the form of a $3D$ hypercube, $X \in \mathcal{R}^{B \times (N \times M)}$,  which contains $1D$ spectral and $2D$ spatial details of a sample where $B$ represents the total number of spectral bands and $N$ and $M$ are spatial components i.e., width and height, respectively. The HSI cube is shown in Figure \ref{Fig.HSI}.

\subsection{\textbf{Spectral Representation}}

In such representations, each pixel vector is isolated from other pixels and processed based on spectral signatures only which means the pixel is represented only in spectral space $x_i \in \mathcal{R}^{B}$. Where $B$ can either be the actual number of spectral channels or just relevant spectral bands extracted after some dimensionality reduction (DR) method. Usually, instead of using original spectral bands, a low dimensional representation of HSI is preferred for data processing in order to avoid redundancy and achieve better class separability, without considerable loss of useful information. 

Dimensionality Reduction (DR) approaches for spectral HSI representation can either be supervised or unsupervised. Unsupervised techniques transform the high dimensional HSI into a low dimensional space without using the class label information, for example, Principal Component Analysis (PCA) and locally linear embedding \cite{fang2014dimensionality}. On the other hand, supervised DR methods utilize labeled samples to learn the data distribution i.e. to keep data points of the same classes near to each other and separate the data points of different classes. For instance, linear discriminant analysis (LDA), local Fisher discriminant analysis (LFDA) \cite{sugiyama2007dimensionality}, local discriminant embedding (LDE) \cite{chen2005local} and nonparametric weighted feature extraction (NWFE) \cite{kuo2004nonparametric}. LDA and LDFA provide better class separability by maximizing the inter-class distance of data points and minimizing the intra-class distance. However, due to the spectral mixing effect, in which the same material may appear with different spectra or different materials may have the same spectral signatures, it becomes difficult to differentiate among different classes based on the spectral reflectance values alone.

\subsection{\textbf{Spatial Representation}}

To deal with the limitations of spectral representation, another approach is to exploit the spatial information of the pixels, in which pixels in each band are represented in the form of a matrix, $x_i \in \mathcal{R}^{N \times M}$. Due to high spatial correlation, neighboring pixels have higher probabilities to belong to the same class. Therefore, in the case of spatial representation, neighboring pixels’ information is also considered and the neighborhood of a pixel can be determined using kernel or pixel centric window \cite{kumar2020feature}. Some common methods to extract spatial information from HSI cube are morphological profiles (MPs), texture features (like Gabor filters,  gray-level co-occurrence matrix (GLCM), and local binary pattern (LBP), etc.) and DNN based methods. Morphological profiles are capable of extracting geometrical characteristics. Few extensions of MPs include extended morphological profiles (EMPs) \cite{emps2005}, multiple-structure-element morphological profiles \cite{gu2016nonlinear}, invariant attribute profiles (IAPs) \cite{hong2020invariant}. 

The texture of the image provides useful spatial contextual information of HSI. For instance, a Gabor filter, a texture analysis technique, can efficiently obtain textural information at various scales and orientations. Similarly, LBP can provide rotation-invariant spatial texture representation. The GLCM can effectively determine the spatial variability of HSI by exploiting the relative positions of neighborhood pixels. The DNNs can also extract spatial information of HSI by considering the pixel as an image patch instead of representing it as a spectral vector. The spatial information contained in HSI can also be extracted by combining various of the afore discussed methods. For instance, \cite{zhang2018spatial}combined Gabor filter and differential morphological profiles \cite{pesaresi2001new} to extract local spatial sequential features for a recurrent neural network (RNN) based HSIC framework.

\subsection{\textbf{Spectral-Spatial Representation}}

This representation jointly exploits both spectral and spatial information of data. In such approaches, a pixel vector is processed based on spectral features while considering spatial-contextual information. The strategies that simultaneously use both spectral and spatial representations of HSI, either concatenate the spatial details with spectral vector \cite{chen2014deep,chen2015spectral} or process the $3D$ HSI cube to preserve the actual structure and contextual information \cite{paoletti2018deepa}.

In literature, all these HSI representations are widely exploited for HSIC. Most of the DNNs for pixel-wise classification utilized the spectral representation of HSIs \cite{jin2018classifying, wu2018discrimination}. However, to mitigate the limitations of spectral representation, many efforts have been made to incorporate the spatial information \cite{li2017hyperspectral, zhan2017hyperspectral}. Recently, joint exploitation of both spectral and spatial features has gained much popularity and led to improved classification accuracy \cite{paoletti2018deep, acquarelli2018spectral, ahmad2017graph, liu2017bidirectional, ahmad2020fast, roy2020attention}. These HSI feature exploitation approaches, for HSIC, are further discussed in the following sections.

\section{Learning Strategies}
\label{SecIV}

Deep learning models can adopt various learning strategies that can be broadly categorized into the following:

\subsection{\textbf{Supervised Learning}}

In a supervised learning approach, the model is trained based on the labeled training data which means training data is comprised of a set of inputs and their corresponding outputs or class labels. During the training phase, the model iteratively updates its parameters in order to predict the desired outputs accurately. In the testing phase, the model is tested against the new input/test data in order to validate its ability to predict the correct labels. If trained sufficiently, the model can predict the labels of new input data. However, supervised learning of DNNs requires a lot of labeled training data to fine-tune the model parameter. Therefore, they are best suited to scenarios where plentiful labeled data is available. The details of various supervised learning techniques for DNNs will be explained in the respective sections.

\subsection{\textbf{Unsupervised Learning}}

In contrast to the supervised learning approach, unsupervised learning techniques learn from the input data with no explicit labels associated with it. These approaches try to identify the underlying statistical structure of input representations or patterns in the absence of corresponding labels. As there is no ground truth available for the training data so it might be difficult to measure the accuracy of the trained model. However, such learning strategies are useful in the cases where we want to learn the inherent structure of such datasets which have a scarcity of training data. The principal component analysis (PCA) is an unsupervised learning technique that can be used to learn a low-dimensional representation of the input. Similarly, k-means clustering is another unsupervised learning method that groups the input data into homogeneous clusters.

\subsection{\textbf{Semi-supervised Learning}}

The semi-supervised learning technique is halfway between unsupervised and supervised approaches. It learns from the partially labeled datasets that are a small amount of labeled training data can be utilized to label the rest of the unlabeled data. These techniques effectively utilize all available data instead of just labeled data, therefore, these techniques have gained much popularity among the research community and are being widely used for HSIC \cite{liu2017semi, kang2019semi, wu2020semi, zhang2020semi}. The details of these methods are briefly described in section \ref{SecVI}.

\section{Development of DNNs (Types of Layers)}
\label{SecV}

In the following, we review recent developments of some widely used DNN frameworks for HSIC. We specifically surveyed the literature published from 2017 onward. DNNs exhibit a great variety of flexible and configurable models for HSIC that allow the incorporation of several types of layers. Few widely used types of layers are explained in the following.


A layer is the key building block of DNN and the type of layer has a decisive impact in terms of feature processing. A layer takes the weighted input, processes it through linear or non-linear transformation, and outputs these values to the next layer. Generally, a layer is uniform, as it has a single activation function. The first layer of the network is known as the input layer and the last layer as an output layer. All other layers in the network, in between the input and output layers, are known as hidden layers. These layers progressively find different features in the input data by performing various transformations. The choice of layer type depends on the task at hand, as some layers perform better for some tasks than others. The most commonly used layers for HSIC are explained below.

\subsection{\textbf{Fully Connected Layers}}

A fully connected (FC) layer connects every neuron in the lower layer to every neuron in the upper/next layer. Mostly, they are used as the last few layers of a model usually after convolution/pooling layers. FC takes the output of the previous layer and assigns weights to predict the probabilities for class labels. Due to a large number of connections, a large number of parameters need to be adjusted which significantly increases the computational overhead. Moreover, due to a large number of parameters, the model becomes more sensitive to overfitting \cite{cheng2017remote}. However, to mitigate the effect of overfitting, a dropout method is introduced in \cite{krizhevsky2012imagenet}.

\subsection{\textbf{Convolutional Layers}}

The convolutional (CONV) layer convolve the input data or feature maps from a lower layer with the filters (kernels). The filter contains weights whose dot product is calculated with the subset of input data by moving it across the width, height, and depth of the input region. The output of the filter is known as a feature map. CONV layer provides spatial invariance via a local connectivity approach in which the neuron in the feature map connects to a subset of input from the previous layer rather than connecting to every neuron. This reduces the number of parameters that need to train. To further reduce the number of parameters, the CONV layer uses the mechanism of parameter sharing in which the same weights are used in a particular feature map.

\subsection{\textbf{Activation Layers}}

\begin{figure}[!htb]
         \centering
         \includegraphics[width=0.35\textwidth]{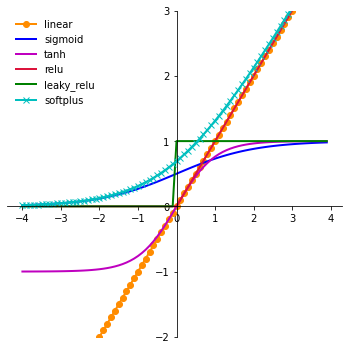}
         \caption{Graphical representation of various commonly used activation functions}
         \label{Fig.Activation}
\end{figure}

Activation layers are assumed to be a feature detector stage of DNNs \cite{goodfellow2016deep}. FC and CONV layers provide linear representations of input data or it can be said that they work similarly to linear regressors and data transformed by these layers is considered to be at the feature extraction stage \cite{paoletti2019deep}. Therefore, to learn non-linear features of data, an activation layer must be used after FC and CONV layers. In the activation layer, feature maps from previous layers go through an activation function to form an activation map. Some commonly used activation functions are sigmoid, hyperbolic tangent (tanh), rectified linear unit (ReLU), LiSHT \cite{roy2019lisht} and softmax. However, in HSI analysis, softmax and ReLU are widely employed activation functions \cite{paoletti2019deep}. Figure \ref{Fig.Activation} presents a graphical representation of a few commonly utilized activation functions.

\subsection{\textbf{Pooling or sub-sampling layers}}

\begin{figure}[!t]
         \centering
         \includegraphics[width=0.25\textwidth]{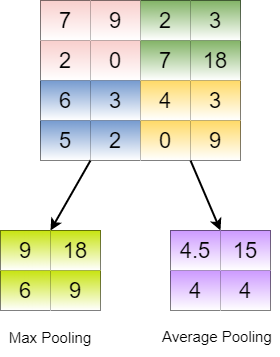}
         \caption{Max-pooling and average-pooling operations of down-sampling/pooling layer}
         \label{Fig.pooling}
\end{figure}

The pooling layer, also known as the sub-sampling or down-sampling layer, takes a certain input volume and reduces it to a single value as shown in Figure \ref{Fig.pooling}. This provides invariance to small distortions in the data. The pooling layer helps the model to control overfitting as the size of data and model parameters both are reduced which also leads to a decrease in the computational time. The commonly used down-sampling operations are max-pooling, average-pooling, and sum-pooling. Recently, a pooling technique, wavelet-pooling is introduced in \cite{williams2018wavelet} whose performance is commensurable to max-pooling and average-pooling. Alternatively, \cite{springenberg2014striving} proposed another trend in which the pooling layer is replaced by the CONV layer of increased filter stride.

\section{Convolutional Neural Network (CNN)}
\label{CNN}

The architecture of the Convolutional Neural Network (CNN) is inspired by the biological visual system presented in \cite{hubel1962receptive}. Following the natural visual recognition mechanism proposed by Hubel and Wiesel \cite{hubel1962receptive}, Neocognitron \cite{fukushima1980neocognitron} is regarded as the first hierarchical, position-invariant model for pattern recognition \cite{voulodimos2018deep} which can be considered as the predecessor of CNN \cite{gu2018recent}. The architecture of CNN can be divided into two main stages: one is Feature Extraction (FE) network and the other is a classification based on the feature maps extracted in the first stage.

The FE network consists of multiple hierarchically stacked CONV, activation, and pooling layers. The CONV layer extracts the features from input data by convolving a learned kernel with it. On each CONV layer, the kernel is spatially shared with whole input data which reduces the model’s complexity and the network becomes easier to train as the number of parameters that need to be fine-tuned is reduced. Convolved results are then passed through an activation layer which adds nonlinearities in the network to extract non-linear features of the input. This is achieved by applying a non-linear function to the convolved results. Afterward, the resolution of the feature map is reduced by applying a pooling operation to achieve shift-invariance. Generally, the pooling layer is added with every CONV layer followed by the activation function.

The classification stage consisting of FC layers and a Softmax operator gives the probability of input pattern belonging to a specific class based on the feature maps extracted at the FE stage. FC layer connects every single neuron in the previous layer to every neuron in the current layer. In \cite{lin2013network} and \cite{ gao2018joint}, the authors proposed that the FC layer can be disregarded by using a global average pooling layer. Softmax is commonly used for classification tasks \cite{zhao2017superpixel, alhichri2018multi} however, many works have also utilized SVM \cite{noor2017hyperspectral, leng2016cube} for this purpose.

\begin{figure*}[!hbt]
         \centering
         \includegraphics[width=0.90\textwidth]{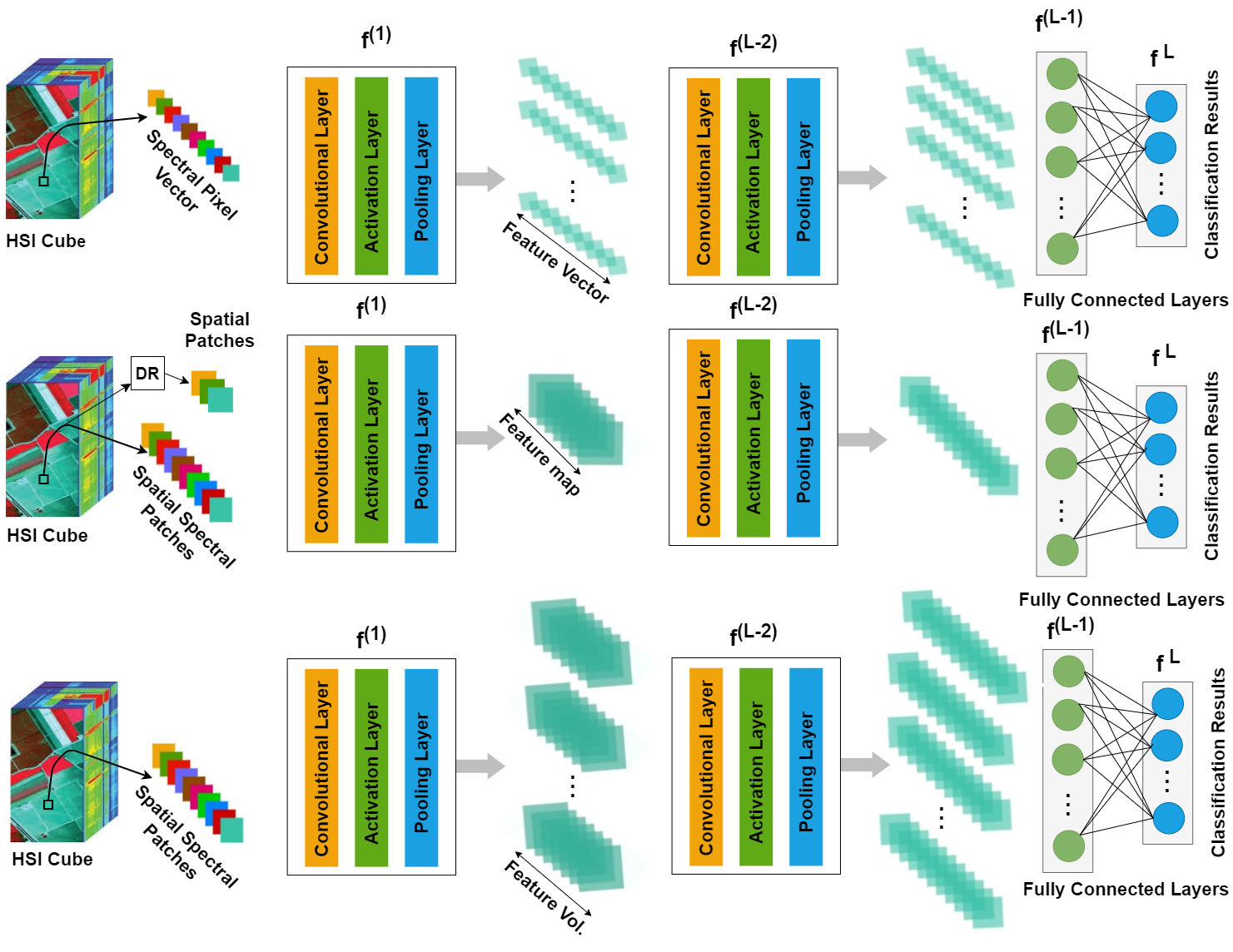}
         \caption{General architecture of Spectral CNN, Spatial CNN and Spectral-spatial CNN frameworks for HSIC.}
         \label{Fig.CNN}
\end{figure*}

In the following, we reviewed three types of CNN architectures for HSIC: i) Spectral CNN, ii) Spatial CNN and iii) Spectral-spatial CNN. Figure \ref{Fig.CNN} illustrates the general architecture of these three frameworks.

\subsection{\textbf{Spectral CNN Frameworks for HSIC}}

Spectral CNN models only consider 1D spectral information $(x_i \in \mathcal{R}^{B})$ as input, where $B$ could either be the original number of spectral bands or the appropriate number of bands extracted after some dimensionality reduction method. In \cite{yu2017convolutional}, a CNN structure was proposed to mitigate the overfitting problem and achieved a better generalization capability by utilizing $1 \times 1$ convolutional kernels and enhanced dropout rates. Moreover, a global average pooling layer is used in place of a fully connected layer in order to reduce the network parameters. To reduce high correlation among HSI bands \cite{gao2018joint} proposed a CNN architecture for HSIC which fully utilized the spectral information by transforming the 1D spectral vector to a 2D feature matrix and by cascading composite layers consisting of $1 \times 1$ and $3 \times 3$ CONV layers, the architecture achieved the feature reuse capability. Similar to \cite{yu2017convolutional}, \cite{gao2018joint} also utilized the global average pooling layer to lower the network's training parameters and to extract high dimensional features.

In \cite{wu2017convolutional} authors presented a hybrid model for HSIC in which the first few CONV layers are employed to extract position invariant middle-level features and then recurrent layers are used to extract spectral-contextual details. Similarly, \cite {jin2018classifying} used a hybrid architecture for classifying healthy and diseased Wheat heads. For the input layer, they transform spectral information into a 2D data structure. In \cite{qiu2018variety} CNN proved to be more effective as compared to SVM and KNN for the spectral-based identification of rice seed’s variety. A similar application of CNN was explored in \cite{wu2018discrimination} where various varieties of Chrysanthemum were identified using spectral data of the first five PCs of Principal component analysis (PCA). PCA is a dimensionality reduction method that is widely used in many DL applications to handle/preprocess high dimensional data. In \cite{huang2018convolutional} PCA was utilized to preprocess medical HSI and then the fusion of CNN kernels with Gabor kernels using dot product is used for classification.

The study \cite{charmisha2018dimensionally} analyzed another dimensionality reduction technique Dynamic Mode Decomposition (DMD) which converted 3D HSI data to 2D and then this data is fed to vectorized CNN (VCNN) for classification. To overcome the noise effect in pixel-wise HSIC, a method of averaged spectra is used in \cite{turra2017cnn} where an averaged spectra of a group of pixels belonging to bacterial colonies is extracted for further analysis.

\subsection{\textbf{Spatial CNN frameworks for HSIC}}

Spatial CNN models only consider spatial information and to extract the spatial information from HSI data, dimensionality reduction (DR) methods are employed on spectral-domain to lower the dimensionality of original HSI data. For instance, \cite{li2018classification} used PCA to extract the first PC with refined spatial information and fed it to a fully CNN framework for classification. Similarly, \cite{haut2019hyperspectral} trained a spatial-based 2D-CNN with one PC. In \cite{xu2018hyperspectral}, PCA whitened input data considering three PCs is fed to a random patches network as a 2D-CNN classification framework. However, the limited training samples with highly similar spectral feature make DL models prone to over-fitting. To overcome this \cite{wang2021probabilistic} proposed a probabilistic neighbourhood pooling based attention network (PNPAN) for HSI classification.

The method proposed in \cite{ding2017convolutional} cropped the patches from 2D input images (i.e. images from the different spectral bands) to train a 2D-CNN architecture that learns the data-adaptive kernels by itself. Furthermore, some authors also proposed the utilization of handcrafted features along with spectral-domain reduction. For example, \cite{chen2017hyperspectral} combined the Gabor filtering technique with 2D-CNN for HSIC to overcome the overfitting problem due to limited training samples. The Gabor filtering extracts the spatial details including edges and textures which effectively reduce the overfitting problem. The work \cite{zhu2018deformable} proposed a deformable HSIC network based on the concept of deformable sampling locations which can adaptively adjust their size and shape in accordance with HSI's spatial features. Such sampling locations are created by calculating 2D offsets for every pixel in the input image through regular convolutions by taking into account three PCs. These offsets can cover the locations of similar neighboring pixels possessing similar characteristics. Then structural information of neighboring pixels is fused to make deformable feature images. Regular convolution employed on these deformable feature images can extract more effective complex structures.

\subsection{\textbf{Spectral-Spatial CNN frameworks for HSIC}}

Spectral-spatial pixel-wise HSIC can be achieved by integrating spatial features into spectral information. For instance, \cite{ran2017hyperspectral} presented an improved pixel pair feature (PPF) approach called spatial pixel pair feature which is different from traditional PPFs with respect to two main aspects: one is the selection of pixel pair that is only the pixel from the immediate neighborhood of central pixel can be used to make a pair, second is the label of pixel pair would be as of central pixel. To extract discriminative joint representation \cite{zhong2018spectral} introduced a Supervised Spectral-Spatial Residual Network (SSRN) that uses a series of 3D convolutions in the respective spectral and spatial residual blocks. An efficient deep 3D-CNN framework was proposed in \cite{paoletti2018new} that simultaneously exploits both spectral and spatial information for HSIC. 

Similarly, to reflect the variations of spatial contexture in various hyperspectral patches, \cite{li2019adaptive} implemented an adaptive weight learning technique instead of assigning fixed weights to incorporate spatial details. Besides this, to make the convolutional kernel more flexible \cite{roy2020attention} explored a new architectural design that can adaptively find adjustable receptive filed and then an improved spectral-spatial residual network for joint feature extraction. The discriminative power of the extracted features can be further improved by combining both the max and min convolutional features before the ReLU non-linearity reported in \cite{roy2021maxmin} for the classification task. CNN's are failed to exploit rotation equivariance in a natural way \cite{paoletti2020rotation} introduced the translation equivariant representations of input features which provides extra robustness to the spatial feature locations for HSIC.

The deeper networks may suffer from the issues of overfitting and gradient vanishing problems due to the smaller number of available labeled training samples and to overcome this shortcoming the lightweight CNN's gain good attention in HSIC communities. The paper \cite{zhang2019hyperspectral} introduced an end-to-end 3D lightweight convolutional neural network to tackle the limited numbers of training samples for HSI classification. To reduce the large gap between the massive trainable parameters and the limited labeled samples \cite{jia2020lightweight} proposed to extract the spatial-spectral Schroedinger eigenmaps (SSSE) joint spatial-spectral information, and then further reduced the dimensionality using compression technique. Approximately 90\% of trainable weights of the total parameters are used immediately after the flatten operation i.e., in the fully connected layer, whereas the remaining only 10\% weights are used on the previous convolutional layers of the whole network. To overcome the paper \cite{roy2020lightweight} introduced a lightweight bag-of-feature learning paradigm into an end-to-end spectral-spatial squeeze-and-excitation residual network for HSIC.

The morphological operations i.e., erosion and dilation are powerful nonlinear feature transformations that are widely used to preserve the essential characteristics of shape and structural information of an image. Inspired by these the paper \cite{roy2021morphological} introduced a new end-to-end morphological convolutional neural network (MorphCNN) for HSIC which utilizes both the spectral and spatial features by concatenating the outputs from spectral and spatial morphological blocks extracted in a dual-path fashion.

The work \cite{li2019adaptive} proposed a two-stage framework for joint spectral-spatial HSIC which can directly extract both spectral and spatial features instead of independently concatenating them. The first stage of the proposed network is comprised of a CNN and softmax normalization that adaptively learns the weights for input patches and extracts joint shallow features. These shallow features are then fed to a network of Stacked Autoencoder (SAE) to obtain deep hierarchical features and final classification is performed with a Multinomial Logistic Regression (MLR) layer. A 3D-CNN model was introduced in \cite{li2017spectral} to jointly exploit spectral-spatial features from HSI and to validate its performance comparison is performed with spectral-based DBN, SAE, and 2D-spatial CNN for HSIC. The work \cite{roy2020fusenet} introduced a bilinear fusion mechanism over the two branches of squeeze operation based on the global and max-pooling whereas the excitation operation is performed with the fused output of squeeze operation.

The work \cite{jiao2017deep} proposed a deep multiscale spectral-spatial feature extraction approach for HSIC which can learn effective discriminant features from the images with high spatial diversity. The framework utilizes the Fully Convolutional Network (FCN) to extract deep spatial information and then, these features are fused with spectral information by using a weighted fusion strategy. Finally, pixel-wise classification is performed on these fused features.

In \cite{zhang2017spectral} a dual-channel CNN framework was implemented for spectral-spatial HSIC. In the proposed approach, 1D-CNN is used to hierarchically extract spectral features and 2D-CNN to extract hierarchical spatial features. These features are then combined together for the final classification task. Furthermore, to overcome the deficiency of training data and to achieve higher classification accuracy, the proposed framework is supported by a data augmentation technique that can increase the training samples by a factor of 6. In \cite{he2017multi}, a multiscale 3D deep CNN is introduced for end-to-end HSIC which can jointly learn both 1D spectral and 2D multiscale spatial features without any pre-processing or post-processing techniques like PCA, etc. In order to reduce the band redundancy or noise in HSI, \cite{dong2019band} explored a novel architecture for HSIC by embedding a band attention module in the traditional CNN framework. The study \cite{he2018feature} proposed an HSIC architecture in which PCA transformed images are used to obtain multi-scale cubes for handcrafted feature extraction by utilizing multi-scale covariance maps which can simultaneously exploit spectral-spatial details of HSI. These maps are then used to train the traditional CNN model for classification.

The work \cite{cheng2018exploring} combined CNN with metric learning-based HSIC framework which first utilizes CNN to extract deep spatial information using the first three PCs extracted by PCA. Then, in a metric learning-based framework, spectral and spatial features are fused for spectral-spatial feature learning by embedding a metric learning regularization factor for the classifier’s training (SVM). Similarly, \cite{gong2019cnn} combines multi-scale convolution-based CNN (MS-CNN) with diversified deep metrics based on determinantal point process (DPP) \cite{zhong2015learning} priors for (1D spectral, 2D spectral-spatial, and 3D spectral-spatial) HSIC. Multiscale filters are used in CNN to obtain multi-scale features and DPP-based diversified metric transformation is performed to increase the inter-class variance and decrease intra-class variance, and better HSI representational ability. Final classification maps are obtained by using a softmax classifier.

In recent work, \cite{liu2020multiscale} an HSIC framework is proposed to extract multi-scale spatial features by constructing a three-channel virtual RGB image from HSI instead of extracting the first three PCs through PCA. The purpose of using a three-channel RGB image is to utilize existing networks trained on natural images to extract spatial features. For multi-scale feature extraction, these images are passed to a fully convolutional network. These multi-scale spatial features are fused and further joined with PCS extracted spectral features for final classification via SVM.

A two-branch (spectral and spatial) DNN for HSIC was introduced in \cite{ma2018hyperspectral}. The spatial branch consists of a band selection layer and a convolutional and de-convolutional framework with skip architecture to extract spatial information of HSI, and in the spectral branch, a contextual DNN is used to extract spectral features. The paper \cite{sellami2019hyperspectral} introduced an adaptive band selection based semi-supervised 3D-CNN to jointly exploit spectral-spatial features whereas \cite{roy2020darecnet} explored dual-attention based autoencoder-decoder network for unsupervised hyperspectral band selection and then joint feature extraction for land cover class prediction. Similarly, in \cite{mei2019unsupervised} spectral-spatial features are simultaneously exploited in an unsupervised manner using a 3D convolution autoencoder. The pixel-wise land use and land cover (LULC) classification using traditional CNNs is often suffered by the presence of wrong / noisy labels in the training set and can easily be overfitted to the labeled noises. To overcome this problem of accurate classification \cite{roy2021lightweight} proposed a lightweight heterogeneous kernel convolution (HetConv3D) for HSI classification with noisy labels by effectively combining both the spectral and spatial kernel feature to produce discriminative and invariant feature maps for classification.

A hybrid 3D-2D-CNN architecture was presented by \cite{roy2019hybridsn} in which 3D-CNN is first used to extract joint spectral-spatial features and then 2D-CNN is further used to obtain more abstract spatial contextual features. The study \cite{zhang2011adaptive} proposed to use adaptive Markov random field for HSIC. The CNN first extracts joint spectral-spatial features and then a smooth MRF prior is placed on class labels to further refine the spatial details. Convolutional neural networks are greatly affected by overfitting and vanishing gradient problems and to overcome this a separable attention network was introduced by \cite{paoletti2021separable}. Where the input feature maps are divided into several groups and split along the channel dimension and finally an attention mask encodes global contextual information by combining them. Recently, generalized gradient centralized $3D$ convolution (G2C-Conv3D) was introduced in \cite{roy2021revisiting} to combine both the intensity level semantic information and gradient level detailed information extracted from raw HSIs during the convolutions operation. To boost the performance of accurate land-cover types classification, G2C-Conv3D can be easily plugged into the existing HSIs feature extraction networks.  

\subsection{\textbf{GCN frameworks for HSIC}}

Graph Convolutional Networks (GCNs) \cite{kipf2016semi} have been garnering increasing attention to researchers in various application fields, owing to their flexible and diversified network architecture that is capable of processing non-grid high-dimensional data. Such properties provide new insight and possibilities in processing hyperspectral data more effectively and efficiently. In detail, GCNs enable the modeling of the relations between data (or samples). Accordingly, this naturally motivates us to use the GCNs to capture the spatial relations of spectral signatures in HSIs. Due to the GCNs' limitations in the graph construction \cite{hong2021graph}, particularly for large graphs (need expensive computational cost), GCNs fail to classify or identify materials in large-scale hyperspectral scenes using normal PCs, which leads to relatively less popularity compared to CNN’s in HSIC. For this reason, there have been some tentative researches using the GCNs in the HSIC task.

For example, a second-order GCN was proposed in \cite{qin2018spectral} by modeling spatial-spectral relations on manifolds for HSIC by the attempts to reduce the computational cost on graphs. Authors of \cite{wan2019hyperspectral} first used superpixel segmentation techniques on HSIs and fed superpixels instead of pixels into GCNs. This enables the network training of GCNs on a large number of pixels in HSIs with the application to the land cover classification task. Nevertheless, these methods still fail to solve the problem of GCNs essentially. To this end, Hong \textit{et al.} \cite{hong2021graph} proposed a novel miniGCN. As the name suggests, miniGCN trains the GCNs in a mini-batch fashion, which is the same as CNN. The proposed miniGCN not only reduces the computational cost-effectively but also makes it possible to make a quantitative comparison and fusion with CNNs, further yielding a FuNet for HSIC.

\subsection{\textbf{Future directions for CNN-based HSIC}}

In the preceding section, we have reviewed the recent developments of CNNs for HSIC. Although CNN's based HSIC frameworks have achieved great success with respect to classification performance, there are still many aspects that need further investigation. For instance, there is a need to further work on such models that can jointly employ spatial and spectral information for HSIC. Many of the above-surveyed frameworks use dimensionality reduction methods to achieve better spectral-spatial representation but such approaches discard useful spectral information of HSI. Hence the development of robust HSIC approaches that can preserve spectral information is required. However, the processing of such approaches increases the computational burden, and the training process becomes slower, therefore, parallel processing of such networks using FPGAs and GPUs is desired in order to achieve the computationally fast models, that can even be suitable for mobile platforms, without the performance degradation. 

Moreover, as the CNNs are becoming deeper and deeper, more labeled training data is required for accurate classification, and as discussed before, there is a lack of labeled training data in HSI. In order to overcome this issue, more research is required to integrate the CNN with unsupervised or semi-supervised approaches. Furthermore, we should pay more attention to the generalization ability of CNNs, particularly for the input data format (not only limiting to the grid data). GCNs might be a good solution to combine with CNN's together to develop a more general CNN-based new framework. Using this, we expect to be able to further break the performance bottleneck, yielding more efficient HSIC.

\section{Autoencoders (AE)}
\label{AE}

\begin{figure}[!hbt]
         \centering
         \includegraphics[width=0.48\textwidth]{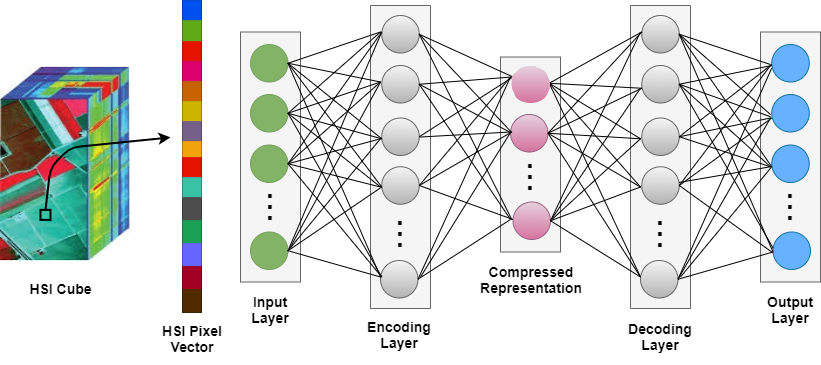}
         \caption{A general Autoencoder Architecture}
         \label{Fig.AE}
\end{figure}

Autoencoder (AE) is a popular symmetrical neural network for HSIC due to its unsupervised feature learning capability. AE itself does not perform a classification task instead it gives a compressed feature representation of high-dimensional HSI data. AE consists of an input layer, one hidden or encoding layer, one reconstruction or decoding layer, and an output layer as shown in Figure \ref{Fig.AE}. AE is trained on input data in such a manner to encode it into a latent representation that is able to reconstruct the input. To learn a compressed feature representation of input data, AE tries to reduce the reconstruction error that is minimizing the difference between the input and the output.

Whereas, the Stacked Autoencoder (SAE) is built by stacking multiple layers of AEs in such a way that the output of one layer is served as an input of the subsequent layer. Denoising autoencoder (DAE) is a variant of AE that has a similar structure as AE except for the input data. In DAE, the input is corrupted by adding noise to it, however, the output is the original input signal without noise. Therefore, DAE, different from AE, can recover original input from a noisy input signal.

To learn high-level representation from data, the work \cite{zhu2017auto} proposed a combination of multi-layer AEs with maximum noise fraction which reduces the spectral dimensionality of HSI, while a softmax logistic regression classifier is employed for HSIC. The study reported in \cite{hassanzadeh2017unsupervised} combined multi-manifold learning framework proposed by \cite{wang2010multi} with Counteractive Autoencoder \cite{rifai2011contractive} for improved unsupervised HSIC. The work \cite{zhang2017recursive} jointly exploited spectral-spatial features of HSI through an unsupervised feature extracting framework composed of recursive autoencoders (RAE) network. It extracts the features from the neighborhood of the target pixel and weights are assigned based on the spectral similarity between target and neighboring pixels. A two-stream DNN with a class-specific fusion scheme was introduced in \cite{hao2017two} which learns the fusion weights adaptively. One stream composed of stacked denoising auto-encoder is used to extract spectral features and the second stream is implemented to extract spatial information using Convolutional Neural Network (CNN), while final classification is performed by fusing the class prediction scores obtained from the classification results of both streams.

Another work proposed a hybrid architecture for multi-feature based spectral-spatial HSIC which utilizes PCA for dimensionality reduction, guided filters \cite{he2012guided} to obtain spatial information, and sparse AE for high-level feature extraction. The framework proposed in \cite{sun2017encoding} exploited both spectral and spatial information for HSIC by adopting batch-based training of AEs and features are generated by fusing spectral and spatial information via a mean pooling scheme. Another work \cite{zhao2017spectral} developed a spectral-spatial HSIC framework by extracting appropriate spatial resolution of HSI and utilization of stacked sparse AE for high-level feature extraction followed by Random Forest (RF) for the final classification task.

Similarly, \cite{wan2017stacked} also used stacked sparse AE for various types of representation that is spectral-spatial and multi-fractal features along with other higher-order statistical representations. A combination of SAE and extreme learning machine was proposed in \cite{lv2017remote} for HSIC, which segments the features of the training set and transform them via SAE, after transformation, feature subsets are rearranged according to the original order of the training set and fed to extreme learning machine-based classifiers, while Q-statistics is used for final classification result. This processing of feature subsets helps to improve variance among base classifiers \cite{lv2017remote}. Similarly, in a recent work \cite{ahmad2019multi} implemented a computationally efficient multi-layer extreme learning machine-based AE which learns the features in three folds, as proposed in \cite{Mahmad2018} for HSIC.

To overcome the issue of high intra-class variability and high inter-class similarity in HSI, \cite{zhou2019learning} developed an SAE-based HSIC which can learn compact and discriminative features by imposing a local fisher discriminant regularization. Similarly, in the latest work \cite{lan2019hyperspectral} a k-sparse denoising AE is spliced with and spectral–restricted spatial features that overcome the high intra-class variability of spatial features for HSIC. The study \cite{paul2018spectral} proposed an HSIC architecture that first makes the spectral segments of HSI based on mutual information measure to reduce the computation time during feature extraction via SAE, while spatial information is incorporated by using extended morphological profiles (EMPs) and SVM/RF is used for final classification. Recently, \cite{liu2019spatial} used SAE for the classification of an oil slick on the sea surface by jointly exploiting spectral-spatial features of HSI.

\subsection{\textbf{Future Directions for AE-based HSIC}}

In the above section, we have surveyed the recent developments of AEs based techniques for HSIC. Although such frameworks provide powerful predictive performance and show good generalization capabilities, more sophisticated work is still desired. Many of the discussed approaches do not fully exploit abundant spatial information so further techniques need to be developed that can fully employ joint spatial and spectral information for HSIC. Moreover, the issue of high intra-class variability and high inter-class similarity in HSI also hinders the classification performance. Many of the above-reviewed works have addressed this issue but further research to overcome this aforesaid issue is required. One direction could be further exploring approaches like pre-training, co-training, and adaptive neural networks, etc for AE-based HSIC frameworks.

\section{Deep Belief Network (DBN)}
\label{DBN}

\begin{figure}[!hbt]
         \centering
         \includegraphics[width=0.20\textwidth]{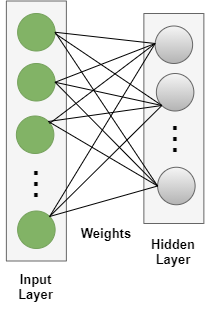}
         \caption{Basic architecture of RBM}
         \label{Fig.RBM}
\end{figure}

Deep Belief Network (DBN) \cite{hinton2006fast} is a hierarchical deep DNN that learns the features from input in an unsupervised, layer-by-layer approach. The layers in DBN are built using Restricted Boltzmann Machine (RBM) comprised of a two-layer architecture in which visible units are connected to hidden units \cite{zhang2018overview} as shown in Figure \ref{Fig.RBM}.

\begin{figure}[!hbt]
         \centering
         \includegraphics[width=0.48\textwidth]{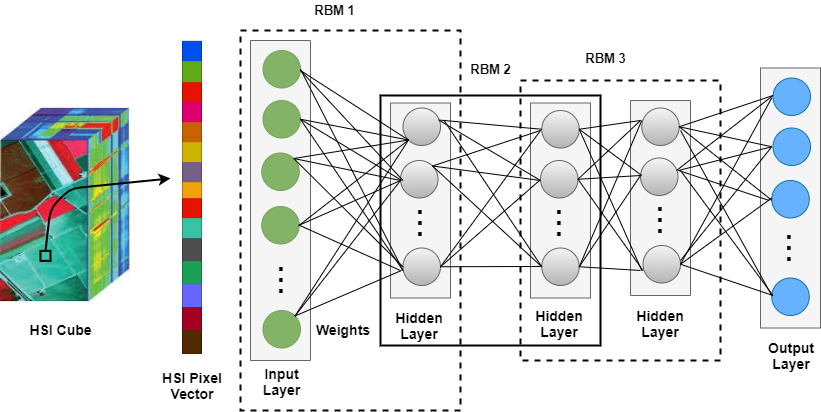} 
         \caption{A three layer DBN architecture}
         \label{Fig.DBN}
\end{figure}

A detailed overview of RBM can be found at \cite{zhang2018overview}. To extract more comprehensive features from input data, the hidden unit of one RBM can be fed to the visible units of other RBM. This type of layer-by-layer architecture builds a DBN, which is trained greedily and can capture deep features from HSI. The architecture of three-layer DBN is shown in Figure \ref{Fig.DBN}.

In literature, several works implemented DBN for HSIC. For instance, \cite{ayhan2017application} used DBN for land cover classification by combining spectral-spatial information and making a comparison with some other classification approaches. The usual learning process of DBN involves two steps: one is unsupervised pre-training with unlabeled samples and the second is supervised fine-tuning with the help of labeled samples. However, this training process may result in two problems: first, multiple hidden units may tend to respond similarly \cite{shaham2016deep} due to co-adaptation \cite{hinton2012improving} and second is linked with the sparsity and selectivity of activations neurons that are some neurons may always be dead or always responding \cite{xiong2015diversity}. To mitigate these two problems, \cite{zhong2017learning} introduced a diversified DBN model through regularizing the pre-training and fine-tuning process by imposing a diversity prior to enhancing the DBN's classification accuracy for HSI.

To extract efficient texture features for the HSIC, the work \cite{li2018hyperspectral} proposed a DBN based texture feature enhancement framework that combines band grouping and sample band selection approach with a guided filter to enhance the texture features, which are then learned by a DBN model and final classification results are obtained by a softmax classifier. The work \cite{tan2019parallel} implemented a parallel layers framework consisting of Gaussian-Bernoulli RBM which extracts high-level, local invariant, and nonlinear features from HSI and a logistic regression layer is used for classification.

To improve the classification accuracy, some works are considered to jointly exploit the spectral and spatial information contained in HSI. For instance, \cite{li2019deep} introduced a DBN framework with the logistics regression layer and verified that the joint exploitation of spectral-spatial features leads to improved classification accuracy. Similarly, \cite{sellami2019spectra} proposed a spectral-spatial graph-based RBM method for HSIC which constructs the spectral-spatial graph through joint similarity measurement based on spectral and spatial details, then an RBM is trained to extract useful joint spectral-spatial features from HSI, and finally, these features are passed to a DBN and logistic regression layer for classification.

\subsection{\textbf{Future directions for DBN-based HSIC}}

In the preceding section, we have reviewed the latest developments of DBN-based HSIC frameworks. We have observed that relative to other DNNs, very few works have utilized the DBNs for HSIC. Therefore, there is a need to further explore the DBN-based robust techniques that can jointly employ spatial and spectral features for HSIC. In addition, another research direction can be the regularization of the pretraining and fine-tuning processes of DBN to efficiently overcome the issue of dead or potentially over-tolerant (always responding) neurons.

\section{Recurrent Neural Network (RNN)}
\label{RNN}

\begin{figure}[!hbt]
         \centering
         \includegraphics[width=0.48\textwidth]{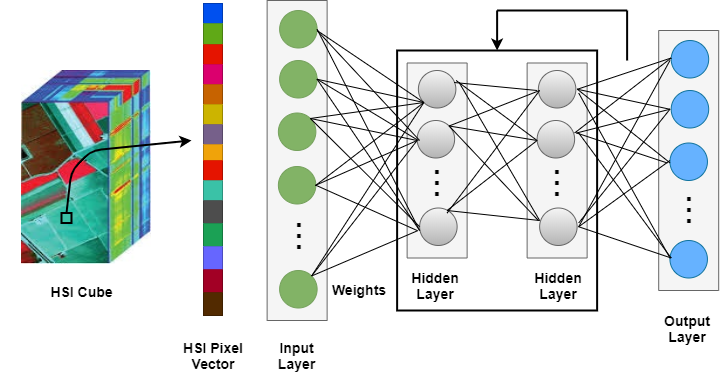}
         \caption{RNN architecture}
         \label{Fig.RNN}
\end{figure}

The architecture of the Recurrent Neural Network (RNN), shown in Figure \ref{Fig.RNN}, comprises loop connections, where the node activation of the next step depends on the previous step \cite{williams1989learning}. Therefore, RNNs are capable of learning temporal sequences. RNN models process the spectral information of HSI data as time sequence considering the spectral bands as time steps \cite{paoletti2020scalable}. There are three basic models of RNN a) Vanilla, b) Long-Short-Term Memory (LSTM) and c) Gated Recurrent Unit (GRU). 

\begin{figure}[!hbt]
         \centering
         \includegraphics[width=0.48\textwidth]{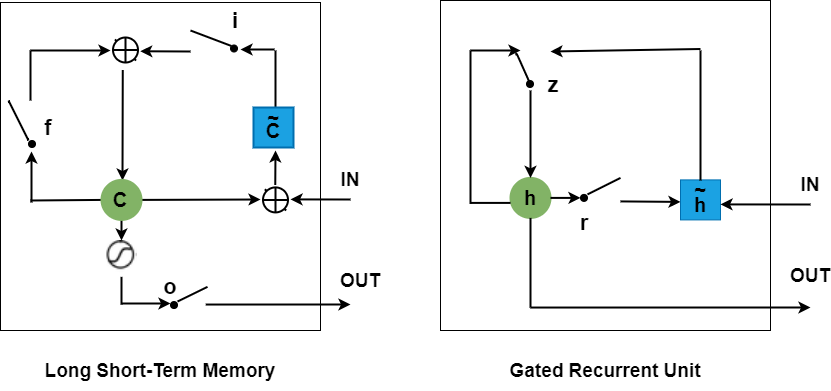} 
         \caption{Internal architecture of LSTM and GRU}
         \label{Fig.gru_lstm}
\end{figure}

Vanilla is the simplest RNN model and leads to information degradation while processing high-dimensional data. LSTM models composed of two states overcome this issue by controlling the information flow through three gates: input, forget, and output gates. It learns the relevant information over time by discarding the extraneous information. However, the gate controlling strategy makes the LSTM a considerably complex approach. GRU variant of LSTM enjoys the simplicity of the Vanilla model and provides high performance similar to LSTM. GRU is a simpler version of LSTM which modifies the input and forget gate as an update ($z_t$) and reset ($r_t$) gate and removes the output gate. A comparison of LSTM and GRU's internal architecture is presented in Figure \ref{Fig.gru_lstm}.

The work \cite{hang2019cascaded} proposed an RNN based HSIC framework with a novel activation function (parametric rectified tanh) and GRU, which utilizes the sequential property of HSI to determine the class labels. In \cite{zhang2018spatial} a local spatial sequential (LSS) method based RNN framework was introduced which first extracts low-level features from HSI by using Gabor filter and differential morphological profiles \cite{pesaresi2001new}  and then fuse these features to obtain LSS features from the proposed method, these LSS features are further passed to an RNN model to extract high-level features, while a softmax layer is used for final classification.

Keeping in view the usefulness of spatial information to achieve improved classification accuracies, the work \cite{zhou2019hyperspectral} proposed a spectral-spatial LSTM based network that learns spectral and spatial features of HSI by utilizing two separate LSTM followed softmax layer for classification, while a decision fusion strategy is implemented to get joint spectral-spatial classification results. Similarly, \cite{sharma2018land} proposed a patch-based RNN with LSTM cells that incorporate multi-temporal and multi-spectral information along with spatial characteristics for land cover classification.

In literature, several works proposed CNN-based hybrid RNN architectures (CRNN) for HSIC. For instance, \cite{wu2017convolutional} implemented a convolutional RNN in which the first few CONV layers are employed to extract position invariant middle-level features, and then recurrent layers are used to extract spectral-contextual details for HSIC. Similarly, \cite{wu2017semi} utilized such a model for semi-supervised HSIC by using pseudo labels. The study \cite{zhou2018integrating} suggested an HSIC framework in which CNN is used to extract spatial features from HSI, then these features are passed to a GRU-based fusion network that performs feature level and decision level fusion.  

Similarly, Luo, \textit{et al.}, \cite{luo2018shorten} exploited both spectral and spatial information contained in HSI by combining CNN with parallel GRU-based RNN which simplifies the training of GRU and improves performance. Bidirectional Convolutional LSTM (CLSTM) was proposed in \cite{liu2017bidirectional} to jointly exploit spectral-spatial feature of HSI for classification. In, \cite{shi2018multi} combined multiscale local spectral-spatial features extracted by 3D-CNN with a hierarchical RNN which learns the spatial dependencies of local spectral-spatial features at multiple scales. Recurrent 2D-CNN and recurrent 3D-CNN for HSIC were proposed in \cite{yang2018hyperspectral} and along with an interesting comparison of these frameworks with their corresponding 2D and 3D-CNN models, which validates the superiority of recurrent CNN. The work \cite{seydgar20193} integrated CNN with CLSTM in which a 3D-CNN model is used to capture low-level spectral-spatial features and CLSTM recurrently analyzes this low-level spectral-spatial information. Recently, \cite{hang2019cascaded}, introduced a cascade RNN for HSIC which consist of two layers of GRU-based RNN, the first layer is used to reduce the redundant spectral bands and the second layer is used to learn the features from HSI, furthermore, a few convolutional layers are employed to incorporate the rich spatial information contained in HSI.

\subsection{\textbf{Future directions for RNN-based HSIC}}

In the above section, we have surveyed the recent developments of AEs based techniques for HSIC. Although RNN-based HSIC frameworks have attracted considerable attention to the remote sensing community and achieved great success for classification performance, there are still many aspects that need further investigation. For instance, the construction of sequential input data for RNN. Most of the surveyed methods considered HSI pixel as a sequential point that is the pixel from each spectral band that forms a data sequence. However, This increases the length of RNN’s input sequence considerably large which can lead to an overfitting issue. 

Moreover, processing such large data sequences increases the computational time and the learning process becomes slower. Therefore, the use of parallel processing tools needs to be further investigated to achieve good generalization performance of RNN-based HSIC. In addition, approaches like a grouping of spectral bands to decrease the data sequence length and utilization of the entire spectral signature to better discriminate between various classes can further be explored to construct the sequential input of the RNN model. Another interesting future direction may involve the implementation of RNN-based HSIC frameworks in a real multi-temporal HSI context.

\section{Strategies for Limited Labeled Samples}
\label{SecVI}

Although DNNs have been successfully exploited for the task of HSIC however, they require a considerably large amount of labeled training data. However, as discussed earlier, the collection of labeled HSI is very critical and expensive due to numerous factors that either demand human experts or exploration of real-time scenarios. The limited availability of labeled training data hinders classification performance. To overcome the aforesaid issue, many effective strategies have been proposed in the literature. In this section, we will briefly discuss some of these strategies while focusing on active learning algorithms.

\subsection{\textbf{Data Augmentation}}

To combat the issue of limited training samples, data augmentation is proven to be an effective tool for HSIC. It generates new samples from the original training samples without introducing additional labeling costs. Data augmentation approaches can be categorized into two main strategies as i) data wrapping; ii) oversampling \cite{shorten2019survey}. Data wrapping usually encodes several invariances (translational, size, viewpoint, and/or illumination) by conducting geometric and color-based transformations while preserving the labels, and oversampling-based augmentation methods inflate the training data by generating synthetic samples based on original data distributions. Oversampling techniques include mixture-based instance generation, feature space augmentations \cite{shorten2019survey}, and Generative Adversarial Networks (GANs) \cite{JIA2021179}.

Referring to HSIC literature, several data augmentation-based frameworks have been employed to improve the classification performance by avoiding potential overfitting, which is generally caused by the limited availability of training data. For instance, \cite{yu2017deep} enhanced the training data by using three data augmentation operations (flip, rotate, and translation), and then this enhanced data is exploited to train CNN for HSIC.  The work \cite{li2018data} presented a comprehensive comparison of various extensively utilized HSI data augmentation techniques and proposed a pixel-block pair-based data augmentation that utilized both spectral and spatial information of HSI to synthesis new instances, to train a CNN model for HSIC. The work \cite{cao2020hyperspectral} compared the classification performance of a combination of CNN and AL with and without data augmentation techniques and demonstrated that the data augmentation leads to higher classification accuracies. Similarly, in another comparison \cite{rochac2019data}, data augmentation-based CNN exhibited a 10\% increase in HSIC accuracy when compared to a PCA-based CNN model.

The above-discussed methods utilize offline data augmentation techniques that increase the training data by creating new instances during/before the training process of a model. Recently, a novel data augmentation framework for HSI is proposed in \cite{nalepa2019training} which, rather than inflating the training data, generates the samples at test time, and a DNN trained over original training data along with a voting scheme is used for the final class label. To improve the generalization capability of DNN models, the work \cite{nalepa2019training} also proposed two fast data augmentation techniques for high-quality data syncretization. A similar PCA-based online data augmentation strategy is proposed in \cite{nalepa2019hyperspectral} which also synthesis new instances during the inference, instead of training.

\subsection{\textbf{Semi-Supervised/Unsupervised Learning}}

Semi-Supervised Learning (SSL) approaches learn data distribution by jointly exploiting both labeled and unlabeled data. These techniques expand the training data by utilizing unlabeled samples along with labeled ones in order to construct a relationship between feature space and class labels. Several SSL-based HSIC frameworks have been proposed in the literature that can mainly be categorized as follows: i) Co-training, ii) Self-training, iii) GANs, iv) Graph-based SSL models and v) Semi-supervised SVM. A recent comprehensive survey on these SSL techniques can be found in \cite{van2020survey}. Moreover, another in-depth survey of SSL approaches is also presented in \cite{pise2008survey}.

The SSL-based HSIC techniques are briefly summarized in \cite{sawant2017semi}, where authors also made a detailed comparison of these methods. The method presented in \cite{wu2017semi} used pseudo or cluster-labeled samples to pre-train a CRNN for HSIC and small-sized labeled data is used to fine-tune the network. Similarly, \cite{kang2019semi} proposed a semi-supervised HSIC framework that exploits PCA and extended morphological attribute profiles to extract pseudo-labeled samples which are fed to a CNN-based deep feature fusion network. 

The work \cite{fang2018semi} proposed a dual strategy co-training approach based on spectral and spatial features of HSI. Similarly, \cite{zhou2019semisupervised} separately pre-trained two SAEs, one using spectral and the other using spatial features of HSI, and fine-tuning is achieved via a co-training approach. \cite{li2017st} proposed a region information-based self-training approach to enhance the training data. A graph-based self-training framework was developed in \cite{aydemir2017semisupervised} where initial sampling is achieved through subtractive clustering. Recently, \cite{wu2020semi} improved the HSIC performance by pseudo-labeling the unlabeled samples through a clustering-based self-training mechanism and regulating the self-training by employing spatial constraints.

\subsection{\textbf{Generative Adversarial Network (GAN)}}

\begin{figure}[!hbt]
         \centering
         \includegraphics[width=0.40\textwidth]{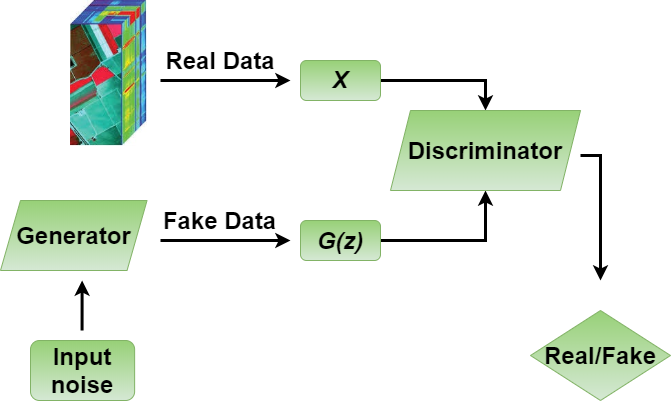} 
         \caption{A general architecture of Generative Adversarial Network (GAN)}
         \label{Fig.GAN}
\end{figure}

GAN proposed by \cite{goodfellow2014generative}, is comprised of two neural networks, one is known as a generator and the other is known as discriminator (Figure \ref{Fig.GAN}). GANs can learn to replicate the samples by exploiting the data distribution details. The work \cite{zhan2017semisupervised} proposed a spectral feature-based GAN for SSL-based HSIC. 

Similarly, \cite{he2017generative} proposed a GAN-based spectral-spatial HSIC framework. Similarly, \cite{zhu2018generative} developed CNN-based 1D-GAN and 3D-GAN architectures to enhance the classification performance. A 1D customized GAN is used to generate the spectral features \cite{zhan2018semi}, which is further used by CNN for feature extraction, and then majority voting is performed HSIC. Very recently, \cite{feng2019classification} introduced a spatial-spectral multi-class GAN (MSGAN) which utilizes two generators to produce spatial and spectral information with the help of multiple adversarial objectives. To address the data imbalance problem for HSI classification \cite{zhong2019generative} proposed a new semi-supervised model which combines GAN with conditional random fields (CRFs). 

Similarly,  \cite{wang2019caps} investigated a Caps-TripleGAN model which effectively generates new samples using a 1D structure Triple Generative Adversarial Network (TripleGAN) and classifying the generated HSI samples using the capsule network (CapsNet). The work \cite{xue2019semi} proposed to utilize a 3D CNN-based generator network and a 3D deep residual network-based discriminator network for HSIC. To learn high-level contextual features combination of both capsule network and convolutional long short-term memory (ConvLSTM) based discriminator model has been proposed in \cite{wang2020generative} for HSIC. 

The work \cite{alipour2020structure} proposed to address the scarcity of training examples by utilizing a GAN model where the performance of the discriminator is further improved by an auxiliary classifier to produce more structurally coherent virtual training samples. Besides this, to enhance the model performance \cite{roy2021generative} proposed a generative adversarial minority oversampling-based technique for addressing the long-standing problem of class-wise data imbalanced imposed by HSIC.

\subsection{\textbf{Transfer Learning}}

Transfer learning enhances the performance of a model by using prior knowledge of a relevant primary task to perform a secondary task. In other words, information extracted from the relevant source domain is transferred to the target domain to learn unseen/unlabeled data. Therefore, transfer learning can be effectively employed in domains with insufficient or no training data. Based on the availability of labeled training instances, transfer learning frameworks can further be categorized as supervised or unsupervised transfer learning. Generally, both source and target domains are assumed to be related but not exactly similar. However, they may follow different distributions as in the case of HSIC where categories of interest are the same but data in two domains may vary due to different acquisition circumstances.

In DNN based HSIC, the model learns features in a hierarchical manner, where lower layers usually extract generic features, when trained on various images. Therefore, the features learned by these layers can be transferred to learn a new classifier for the target dataset. For instance, \cite{yang2017learning} pertained to a two-branch spectral-spatial CNN model with an ample amount of training data from other HSIs and then applied the lower layers of the pre-trained model to the target network for the robust classification of target HSI. To learn the target-specific features, higher layers of the target network are randomly initialized and the whole network is fine-tuned by utilizing limited labeled instances of target HSI. Similarly, \cite{windrim2018pretraining} proposed a suitable method to pre-train and fine-tune a CNN network to utilize it for the classification of new HSIs. The study \cite{liu2018hyperspectral} combined data augmentation and transfer learning approaches to combat the shortage of training data in order to improve HSIC performance.

As discussed before, data in source and target domain may vary in many aspects, for instance, in the case of HSIs, the dimensions of two HSIs may vary due to the acquisition from different sensors. Handling such cross-domain variations and transferring the knowledge between them is known as heterogeneous transfer learning (a detailed survey of such methods can be found in \cite{day2017survey}). In HSIC literature, several works have been proposed to bridge the gap for transferring the knowledge between two HSIs, with varying dimensions and/or distributions. 

For example, \cite{lin2018deep} proposed an effective heterogeneous transfer learning-based HSIC framework that works well with both homogeneous and heterogeneous HSIs, and \cite{li2017iterative} used an iterative re-weighting mechanism-based heterogeneous transfer learning for HSIC. Similarly, a recent work \cite{liu2020transfer} proposed a band selection-based transfer learning approach to pre-train a CNN, which retains the same number of dimensions for various HSIs. Furthermore, \cite{lin2017structure} proposed an unsupervised transfer learning technique to classify completely unknown target HSI and \cite{pires2020convolutional} demonstrate that the networks trained on natural images can enhance the performance of transfer learning for remote sensing data classification as compared to the networks trained from scratch using smaller HSI data.

\subsection{\textbf{Active Learning}}

\begin{figure}[!hbt]
         \centering
         \includegraphics[width=0.45\textwidth]{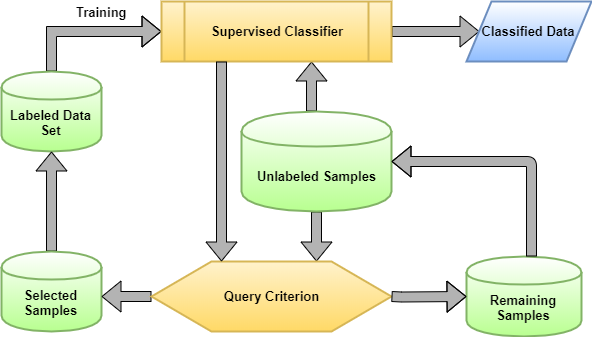} 
         \caption{A general overview of Active Learning}
         \label{Fig.AL}
\end{figure}

Active Learning (AL) iteratively enhances the predictive performance of a classifier by actively increasing the size of training data, for each training iteration, by utilizing an unlabeled pool of samples. In each iteration, AL enhances the training dataset by actively selecting the most valuable instances from the pool of unlabeled data and an oracle (Human or machine-based) assigns the true class labels to these instances. Finally, these useful instances are added to the existing training dataset and the classifier is retrained on this new training dataset. The process continues until a stopping criterion, that maybe the size of the training dataset, the number of iterations, or the desired accuracy score, is achieved. A general framework of AL is illustrated in Figure \ref{Fig.AL}.

The selection of the most useful/effective samples is made in such a way that the samples should be informative and representative of the overall input distribution in order to improve accuracy. Based on the criteria of adding new instances to the training set, AL frameworks can be designated as either stream-based or pool-based. In stream-based selection, one instance at a time is drawn from an actual set of unlabeled samples and the model decides whether to label it or not based on its usefulness. While in pool-based strategy, samples are queried from a pool/subset of unlabeled data based on ranking scores computed from various measures to evaluate the sample's usefulness. 

The work \cite{ganti2012upal} found that streamed-based selection gives poorer learning rates as compared to pool-based selection as the former tends to query extra instances. In pool-based selection, it is important to incorporate diversity in the pool of samples, in order to avoid redundancy within the pool of samples. Generally, the following three aspects are focused on while selecting/querying the most valuable samples: heterogeneity behavior, model’s performance, and representativeness of samples. A brief introduction of these sampling approaches is given below:

\subsubsection{\textbf{Heterogeneity-based selection}}

These approaches select the samples that are more heterogeneous to the already seen instances with respect to model diversity, classification uncertainty, and contention between a committee of various classifiers. Uncertainty sampling, expected model change, and query-by-committee are examples of heterogeneity-based models.

\begin{itemize}
    \item \textbf{Uncertainty Sampling}: In this approach, the classifier iteratively tries to query the label of those samples for which it is most uncertain while predicting the label. The selection of new instances is based on ranking scores against a specified threshold and the instances with scores closest to that threshold are queried for labels.  One simple example of such a scheme could be implementing the probabilistic classifier on a sample in a scenario of binary classification and querying its label if the predicted class probability is close to $0.5$. 

    \item \textbf{Query-by-Committee}: Such heterogeneity-based approaches perform the sampling process based on the dissimilarities in the predictions of various classifiers trained on the same set of labeled samples. A committee of various classifiers trained on the same set of training data is used to predict the class labels of unlabeled samples and the samples for which classifiers differ more are selected for querying labels. The committee of different classifiers can either be built by using ensemble learning algorithms like Bagging and Boosting \cite{melville2004diverse} or by changing the model parameters \cite{aggarwal2014active}. Generally, a less number of diverse classifiers is adequate for constructing a committee \cite{seung1992query, melville2004diverse}.

    \item \textbf{Expected Model Change}: Such a heterogeneity-based approach chooses the instances which result in a significant change from the current model in terms of the gradient of the objective function. Such techniques attempt to query the label for those instances that are considerably different from the current model. These sampling techniques only fit the models which follow gradient-based training procedures/optimization.
\end{itemize}

\subsubsection{\textbf{Performance-based Selection}}

Such methods consider the effect of adding queried samples to the model performance. They try to optimize the performance of the model by reducing variance and error. There are two types of performance-based sampling:

\begin{itemize}
    \item \textbf{Expected Error Reduction}: This approach is interrelated to uncertainty sampling in such a way that uncertainty measures maximize the label uncertainty of the sample to be queried for the label while expected error reduction reduces the label uncertainty of the queried sample. Referring to the already discussed example of the binary classification problem, the expected error reduction approach would choose the samples with a probability far away from $0.5$ in order to reduce the error rate. Such techniques are also known as the greatest certainty models \cite{aggarwal2014active}.
    
    \item \textbf{Expected Variance Reduction}: Reducing the variance of the model is guaranteed to reduce future generalization error \cite{settles2009active}. Therefore, expected variance reduction techniques attempt to indirectly reduce the generalization error by minimizing the model variance. Such approaches query the instances that result in the lowest model variance. The Fisher information ratio is a well-known variance minimization framework.
\end{itemize}

\subsubsection{\textbf{Representativeness-based selection}}

Heterogeneity-based models are prone to include outlier and controversial samples but performance-based approaches implicitly avoid such samples by estimating future errors. Representative sampling tends to query such instances that are representative of the overall input distribution, hence, avoid outliers and unrepresentative samples. These approaches weigh the dense input region to a higher degree while the querying process. Density-weighted techniques like information density are examples of representativeness sampling approaches that consider the representativeness of samples along with heterogeneity behavior, and are also known as hybrid models \cite{aggarwal2014active}.

Recently, AL has been intensively utilized in HSIC. \cite{liu2017feature} proposed a feature-driven AL framework to define a well-constructed feature space for HSIC. \cite{zhang2019active} proposed a Random Forest-based semi-supervised AL method that exploits spectral-spatial features to define a query function to select the most informative samples as target candidates for the training set.

Spatial information has been intensively exploited in many AL-based HSIC. For instance, \cite{guo2016superpixel} presented an AL framework that splice together the spectral and spatial features of superpixels. Similarly, \cite{xue2018active} considered the neighborhood and superpixel information to enhance the uncertainty of queried samples. In recent work, \cite{bhardwaj2020spectral} exploited the attribute profiles to incorporate spatial information in an AL-based HSIC framework.

Batch-mode AL frameworks have been widely employed to accelerate the learning process. Such approaches select a batch of samples, in each iteration, to be queried for a label. Therefore, the diversity of the samples is extremely critical in batch mode AL techniques in order to avoid redundancy. A multi-criteria batch-mode AL method proposed by \cite{patra2017spectral} defines a novel query function based on diversity, uncertainty, and cluster assumption measures. These criteria are defined by exploiting the properties of KNN, SVM, and K-means clustering respectively, and finally, genetic algorithms are used to choose the batch of most effective samples. Similarly, \cite{zhang2017batch} proposed a regularized multi-metric batch-mode AL framework for HSIC that exploits various features of HSI.

A multiview AL (MVAL) framework was proposed in \cite{xu2017multiview} that analyzes the object from various views and measure the informativeness of the sample through multiview Intensity-based query criteria. Similarly, \cite{pradhan2018fisher} also exploited the concept of multiview learning using the Fisher Discriminant Ratio to generate multiple views. In another work, \cite{zhang2019adaptive} proposed a novel adaptive MVAL framework for HSIC which jointly exploits the spatial and spectral features in each view. Recently, \cite{li2020subpixel} proposed an MVAL technique that utilizes pixel-level, subpixel-level, and superpixel-level details to generate multiple views for HSIC. Moreover, the proposed method exploits joint posterior probability estimation and dissimilarities among multiple views to query the representative samples.

In the HSIC literature, several works have combined the AL and DNN. For instance, \cite{sun2016active} joined autoencoder with AL technique and \cite{liu2016active} proposed a DBN-based AL framework for HSIC. Similarly, \cite{haut2018active} coupled Bayesian CNN with AL paradigm for spectral-spatial HSIC. Recently, \cite{cao2020hyperspectral} proposed a CNN-based AL framework to better exploit the unlabeled samples for HSIC.  

Many works integrated AL with transfer learning for HSIC. For example, \cite{lin2018active} proposed an AL-based transfer learning framework that extracts the salient samples and exploits high-level features to correlate the source and target domain data. Another work, \cite{deng2018AT} proposed a Stacked Sparse AE-based Active Transfer Learning technique that jointly utilizes both spectral and spatial features for HSIC. Another work \cite{deng2018active} combined domain adaptation and AL methods based on multiple kernels for HSIC.

AL-based HSIC offers some sophisticated frameworks to enhance the generalization capabilities of models. For instance, \cite{Ahmad2018} proposed a fuzziness-based AL method to improve the generalization performance of discriminative and generative classifiers. The method computes the fuzziness-based distance of each instance and estimated class boundary, and the instances having greater fuzziness values and smaller distances from class boundaries are selected to be the candidates for the training set. Recently, \cite{article} proposed a non-randomized spectral-spatial AL framework for multiclass HSIC that combines the spatial prior Fuzziness approach with Multinomial Logistic Regression via a Splitting and Augmented Lagrangian classifier. The authors also made a comprehensive comparison of the proposed framework with state-of-the-art sample selection methods along with diverse classifiers.

\section{Experimental Evaluation}
\label{SecVII}

The most research-oriented works published in the literature present a comprehensive experimental evaluation to highlight the pros and cons of the work/s proposed. However, to some extent, these works may have chosen different experimental settings, for instance, training, validation, and test samples may have the same number or percentage but the samples may be different as these samples are normally chosen randomly. Therefore, to make a fair comparison among different works proposed in the literature, one must need to have the same experimental settings.

These experimental settings include the same samples (geographical locations should remain the same for all chosen models no the different ones) and the number of samples should have been selected for each round of training in the cross-validation process. Normally, these samples have been chosen randomly, thus high likely, they may be different for different models if the models are executed at different times. 

The other issue with most of the literature proposed in recent years is overlapping between training/test samples, i.e., training/validation samples have been randomly selected (including or excluding the above point) for training and validation however, the entire dataset has been passed at a testing phase which leads to a highly biased model (as the training samples have already been seen by the model) and produces high accuracy. Thus, in this work, the training/test samples are though chosen randomly (because all the models have been executed at the same time) however, the above point has been taken seriously and the intersection among these samples remain empty.

\subsection{Experimental Datasets}

\begin{figure}[!htb]
        \centering
            \begin{subfigure}{0.30\columnwidth}
                \centering
		        \includegraphics[width=0.70\columnwidth]{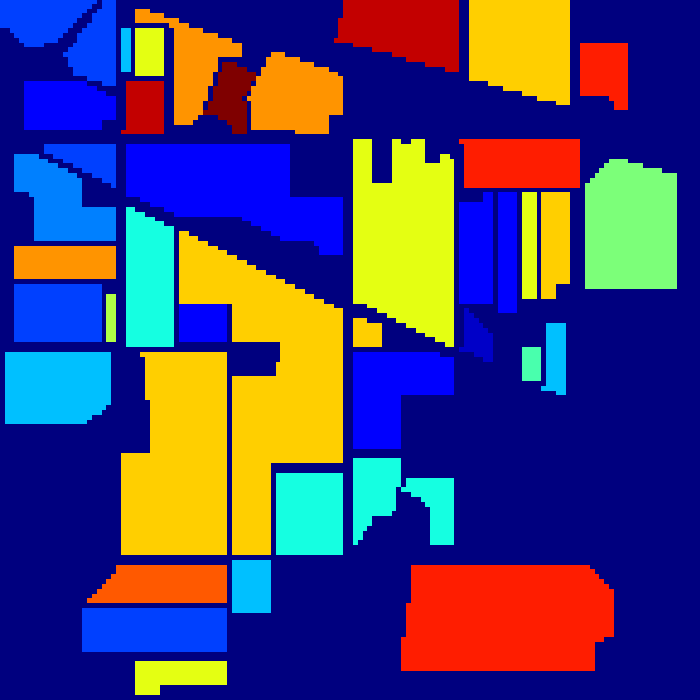}
    		    \caption{GT Maps}
		        \label{Fig.3A}
	        \end{subfigure}
	    \begin{subfigure}{0.30\columnwidth}
	        \centering
		    \includegraphics[width=0.70\columnwidth]{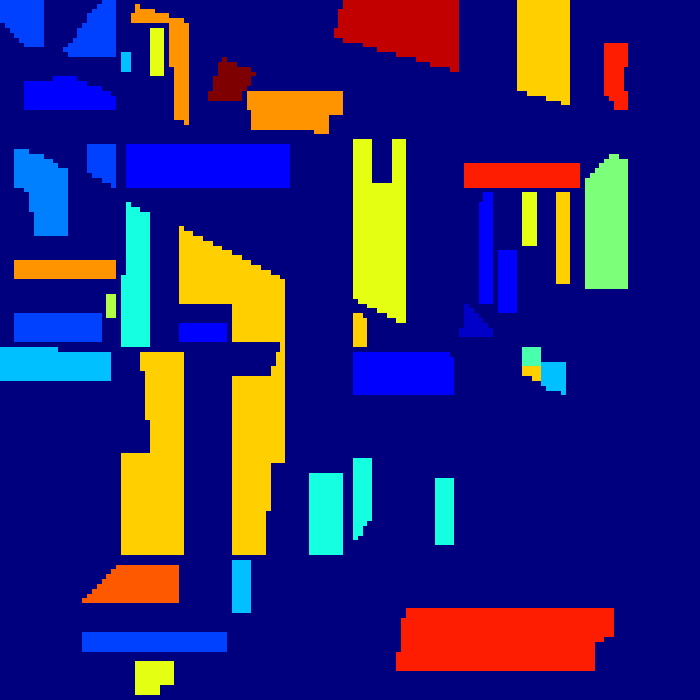}
    		\caption{Disjoint Training} 
		    \label{Fig.3B}
	    \end{subfigure}
	    \begin{subfigure}{0.30\columnwidth}
	        \centering
		    \includegraphics[width=0.70\columnwidth]{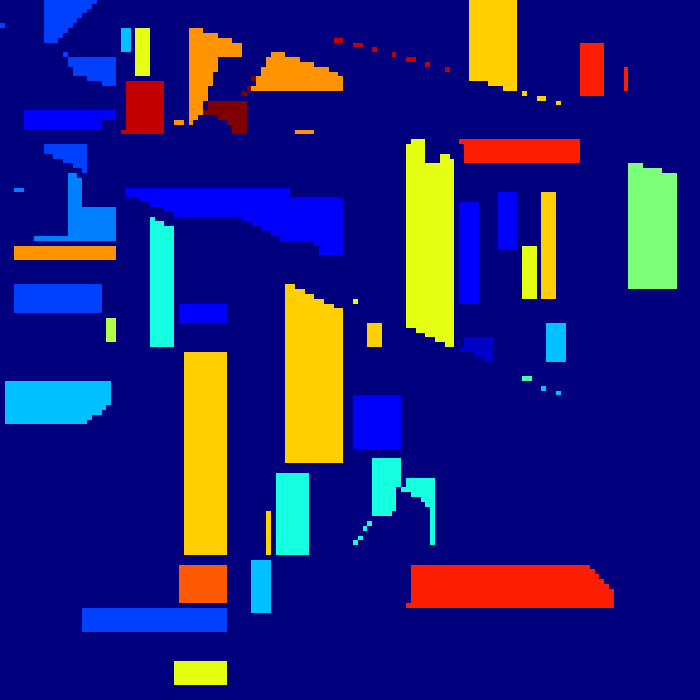}
    		\caption{Disjoint Test}
		    \label{Fig.3C}
	    \end{subfigure}
        \\ \vspace{0.2cm}
        
        \begin{minipage}[adjusting]{0.98\linewidth}
            \resizebox{\linewidth}{!}{\begin{tabular}{ccc|ccc|ccc} 
                \toprule
                \fboxsep=1mm \fboxrule=1mm
                \fcolorbox{indian_pines_corrected_Background!}{indian_pines_corrected_Background!}{\null} & Background & 10776 &  \fboxsep=1mm \fboxrule=1mm \fcolorbox{indian_pines_corrected_Alfalfa!}{indian_pines_corrected_Alfalfa!}{\null} & Alfalfa & 46 & \fboxsep=1mm \fboxrule=1mm \fcolorbox{indian_pines_corrected_Corn-notill!}{indian_pines_corrected_Corn-notill!}{\null} & Corn notill & 1428 \\ \fboxsep=1mm \fboxrule=1mm
                
                \fcolorbox{indian_pines_corrected_Corn-min!}{indian_pines_corrected_Corn-min!}{\null} & Corn min & 830 & \fboxsep=1mm \fboxrule=1mm \fcolorbox{indian_pines_corrected_Corn!}{indian_pines_corrected_Corn!}{\null} & Corn & 237 & \fboxsep=1mm \fboxrule=1mm \fcolorbox{indian_pines_corrected_Grass/Pasture!}{indian_pines_corrected_Grass/Pasture!}{\null} & Grass/Pasture & 483 \\ \fboxsep=1mm \fboxrule=1mm

                \fcolorbox{indian_pines_corrected_Grass/Trees!}{indian_pines_corrected_Grass/Trees!}{\null} & Grass/Trees & 730 & \fboxsep=1mm \fboxrule=1mm \fcolorbox{indian_pines_corrected_Grass/pasture-mowed!}{indian_pines_corrected_Grass/pasture-mowed!}{\null} & Grass/pasture-mowed & 28 & \fboxsep=1mm \fboxrule=1mm \fcolorbox{indian_pines_corrected_Hay-windrowed!}{indian_pines_corrected_Hay-windrowed!}{\null} & Hay windrowed & 478 \\ \fboxsep=1mm \fboxrule=1mm
                
                \fcolorbox{indian_pines_corrected_Oats!}{indian_pines_corrected_Oats!}{\null} & Oats & 20 & \fboxsep=1mm \fboxrule=1mm \fcolorbox{indian_pines_corrected_Soybeans-notill!}{indian_pines_corrected_Soybeans-notill!}{\null} & Soybeans notill & 972 & \fboxsep=1mm \fboxrule=1mm \fcolorbox{indian_pines_corrected_Soybeans-min!}{indian_pines_corrected_Soybeans-min!}{\null} & Soybeans min & 2455 \\ \fboxsep=1mm \fboxrule=1mm
                
                \fcolorbox{indian_pines_corrected_Soybean-clean!}{indian_pines_corrected_Soybean-clean!}{\null} &Soybean clean & 593 & \fboxsep=1mm \fboxrule=1mm \fcolorbox{indian_pines_corrected_Wheat!}{indian_pines_corrected_Wheat!}{\null} & Wheat & 205 & \fboxsep=1mm \fboxrule=1mm \fcolorbox{indian_pines_corrected_Woods!}{indian_pines_corrected_Woods!}{\null} & Woods & 1265 \\ \fboxsep=1mm \fboxrule=1mm
                
                \fcolorbox{indian_pines_corrected_Bldg-Grass-Tree-Drives!}{indian_pines_corrected_Bldg-Grass-Tree-Drives!}{\null} & Bldg Grass Tree Drives & 386 & \fboxsep=1mm \fboxrule=1mm \fcolorbox{indian_pines_corrected_Stone-steel towers!}{indian_pines_corrected_Stone-steel towers!}{\null} & Stone steel towers & 93 & & \fboxsep=1mm \fboxrule=1mm Total samples & 21025 \\ \bottomrule
            \end{tabular}}
        \end{minipage}
        \caption{The type associated with the land-cover classes and number of available samples in the Indian Pines (IP) dataset. Moreover, Spatially disjoint training and test samples for the IP dataset are also presented.}
        \label{fig:GT_IP}
    \end{figure}

The \textbf{Indian Pines} (IP) dataset was gathered by the Airborne Visible/Infrared Imaging Spectrometer (AVIRIS) \cite{green1998imaging} over the Indian Pines test site in North-western Indiana. It contains $224$ spectral bands within a wavelength range of $400$ to $2500$ $nm$. The $24$ null and corrupted bands have been removed. The spatial size of the image is $145\times{145}$ pixels, and it comprises of $16$ mutually exclusive vegetation classes. The spatial resolution is 20 meters per pixel (MPP). The detailed class description and ground truth maps are presented in Figure \ref{Fig.3A}. Moreover, the disjoint Training/Test sample maps are presented in Figures \ref{Fig.3B} and \ref{Fig.3C}.

\begin{figure}[!htb]
        \centering
            \begin{subfigure}{0.30\columnwidth}
                \centering
		        \includegraphics[width=0.70\columnwidth]{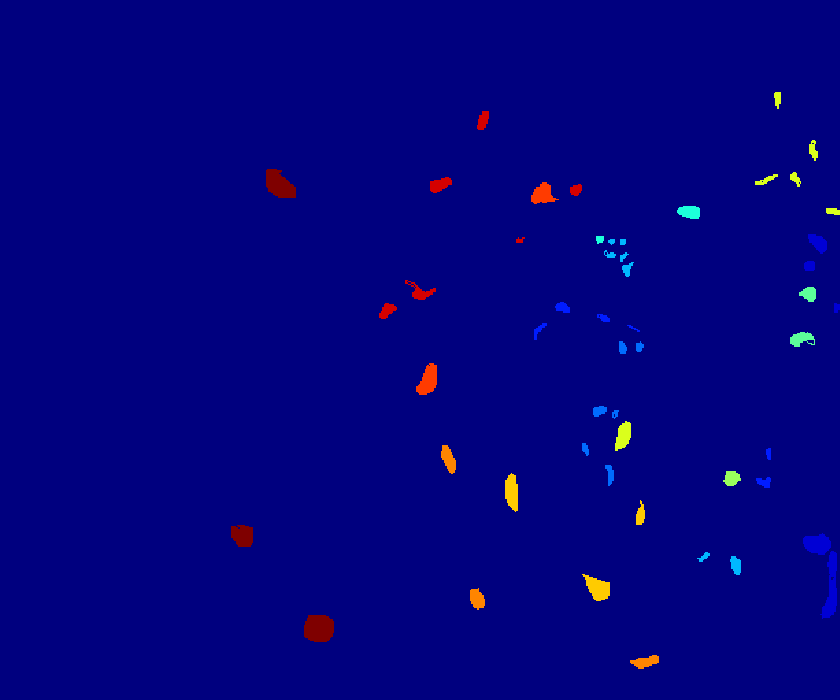}
    		    \caption{GT Maps}
		        \label{Fig.5A}
	        \end{subfigure}
	    \begin{subfigure}{0.30\columnwidth}
	        \centering
		    \includegraphics[width=0.70\columnwidth]{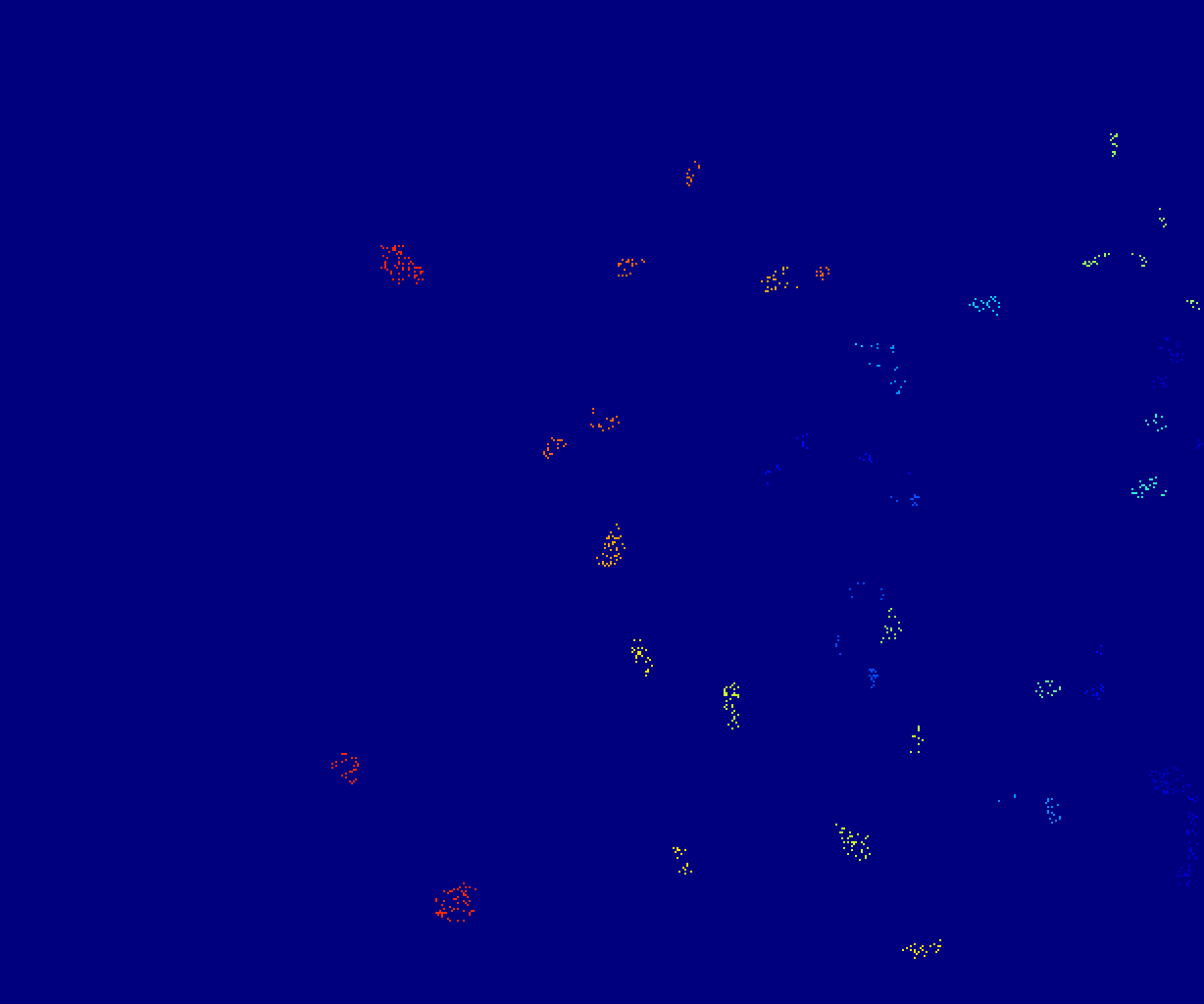}
    		\caption{Disjoint Training} 
		    \label{Fig.5B}
	    \end{subfigure}
	    \begin{subfigure}{0.30\columnwidth}
	        \centering
		    \includegraphics[width=0.70\columnwidth]{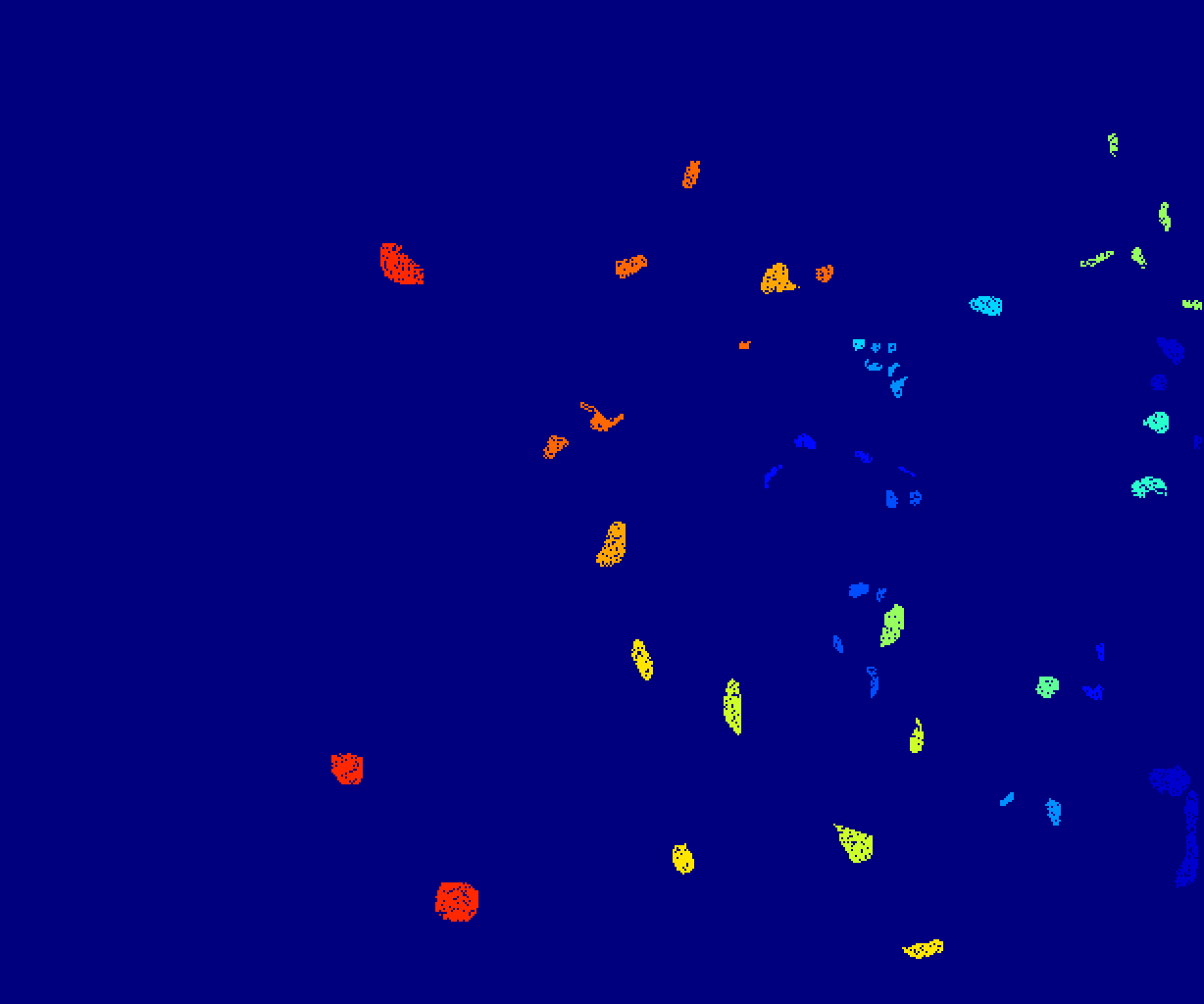}
    		\caption{Disjoint Test}
		    \label{Fig.5C}
	    \end{subfigure}
        \\ \vspace{0.2cm}
        
        \begin{minipage}[adjusting]{0.98\linewidth}
            \resizebox{\linewidth}{!}{\begin{tabular}{ccc|ccc|ccc}
                \toprule
                \fboxsep=1mm \fboxrule=1mm \fcolorbox{KSC_Background!}{KSC_Background!}{\null} & Background & 309157 & \fboxsep=1mm \fboxrule=1mm  \fcolorbox{KSC_Scrub!}{KSC_Scrub!}{\null} & Scrub & 761 &  \fboxsep=1mm \fboxrule=1mm \fcolorbox{KSC_Willow-swamp!}{KSC_Willow-swamp!}{\null} & Willow swamp & 243 \\ \fboxsep=1mm \fboxrule=1mm

                \fcolorbox{KSC_CP-hammock!}{KSC_CP-hammock!}{\null} & CP hammock & 256 & \fboxsep=1mm \fboxrule=1mm \fcolorbox{KSC_Slash-pine!}{KSC_Slash-pine!}{\null} & Slash pine & 252 & \fboxsep=1mm \fboxrule=1mm \fcolorbox{KSC_Oak/Broadleaf!}{KSC_Oak/Broadleaf!}{\null} & Oak/Broadleaf & 161 \\ \fboxsep=1mm \fboxrule=1mm
                
                \fcolorbox{KSC_Hardwood!}{KSC_Hardwood!}{\null} & Hardwood & 229 & \fboxsep=1mm \fboxrule=1mm \fcolorbox{KSC_Swap!}{KSC_Swap!}{\null} & Swap & 105 & \fboxsep=1mm \fboxrule=1mm \fcolorbox{KSC_Graminoid-marsh!}{KSC_Graminoid-marsh!}{\null} & Graminoid marsh & 431 \\ \fboxsep=1mm \fboxrule=1mm

	            \fcolorbox{KSC_Spartina-marsh!}{KSC_Spartina-marsh!}{\null} & Spartina marsh & 520 & \fboxsep=1mm \fboxrule=1mm \fcolorbox{KSC_Cattail-marsh!}{KSC_Cattail-marsh!}{\null} & Cattail marsh & 404 & \fboxsep=1mm \fboxrule=1mm \fcolorbox{KSC_Salt-marsh!}{KSC_Salt-marsh!}{\null} & Salt marsh & 419 \\ \fboxsep=1mm \fboxrule=1mm

                \fcolorbox{KSC_Mud-flats!}{KSC_Mud-flats!}{\null} & Mud flats & 503 & \fboxsep=1mm \fboxrule=1mm \fcolorbox{KSC_Water!}{KSC_Water!}{\null} & Water & 927 & & \fboxsep=1mm \fboxrule=1mm Total samples & 207400            
                \\ \bottomrule
            \end{tabular}}
        \end{minipage}

        \caption{The type associated with the land-cover classes and number of available samples in the Kennedy Space Center (KSC) dataset. Moreover, Spatially disjoint training and test samples for the KSC dataset are also presented.}
        \label{fig:GT_KSC}
    \end{figure}

The \textbf{Kennedy Space Center} (KSC) dataset was gathered in 1996 by AVIRIS~\cite{green1998imaging}, with wavelengths ranging from $400$ to $2500$ $nm$. The image has $512\times{614}$ pixels and $176$ spectral bands after removal of some low signal-to-noise ratio~(SNR) bands. The KSC dataset comprises $5202$ labeled samples, with a total of $13$ upland and wetland classes. The detailed class description and ground truth maps are presented in Figure \ref{Fig.5A}. Moreover, the disjoint Training/Test sample maps are presented in Figures \ref{Fig.5B} and \ref{Fig.5C}.  

\begin{figure}[!htb]
        \centering
            \begin{subfigure}{0.30\columnwidth}
                \centering
		        \includegraphics[width=0.60\columnwidth]{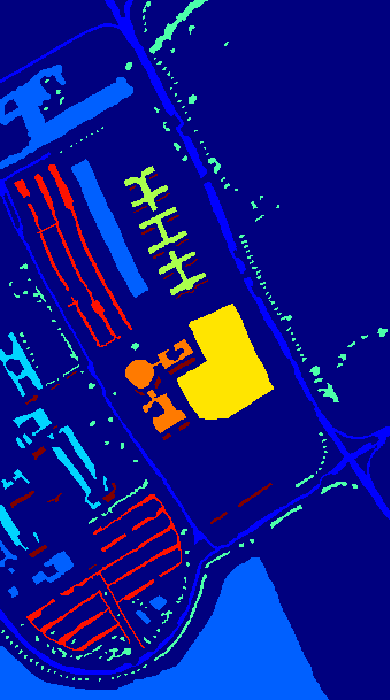}
    		    \caption{GT Maps}
		        \label{Fig.4A}
	        \end{subfigure}
	    \begin{subfigure}{0.30\columnwidth}
	        \centering
		    \includegraphics[width=0.60\columnwidth]{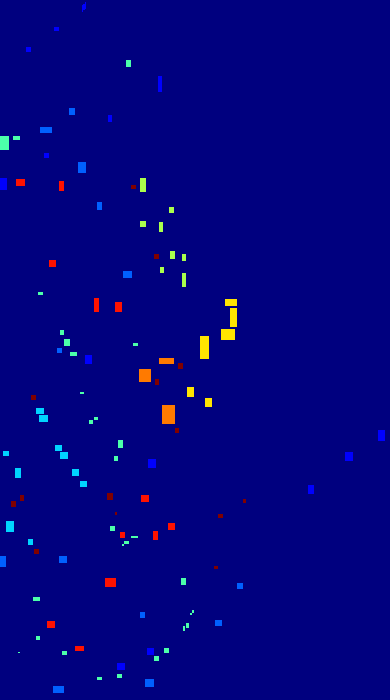}
    		\caption{Disjoint Training} 
		    \label{Fig.4B}
	    \end{subfigure}
	    \begin{subfigure}{0.30\columnwidth}
	        \centering
		    \includegraphics[width=0.60\columnwidth]{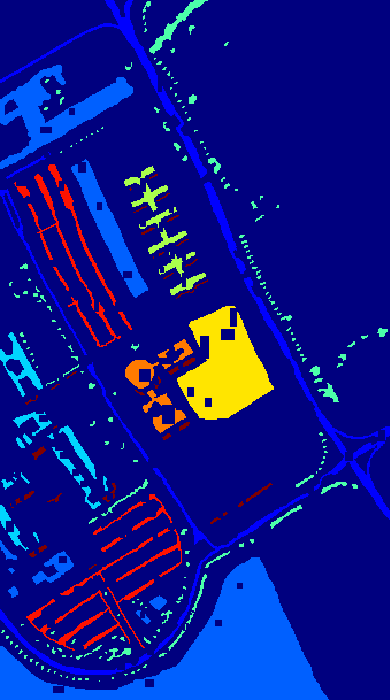}
    		\caption{Disjoint Test}
		    \label{Fig.4C}
	    \end{subfigure}
        \\ \vspace{0.2cm}

        \begin{minipage}[adjusting]{0.98\linewidth}
            \resizebox{\linewidth}{!}{\begin{tabular}{ccc|ccc|ccc}  
                \toprule
                \fboxsep=1mm \fboxrule=1mm
                \fcolorbox{PaviaU_Background}{PaviaU_Background}{\null} & Background & 164624 &  \fboxsep=1mm \fboxrule=1mm \fcolorbox{PaviaU_Asphalt}{PaviaU_Asphalt}{\null} & Asphalt & 6631 & \fboxsep=1mm \fboxrule=1mm \fcolorbox{PaviaU_Meadows}{PaviaU_Meadows}{\null} & Meadows & 18649\\ \fboxsep=1mm \fboxrule=1mm
                
                \fcolorbox{PaviaU_Gravel}{PaviaU_Gravel}{\null} & Gravel & 2099 & \fboxsep=1mm \fboxrule=1mm \fcolorbox{PaviaU_Trees}{PaviaU_Trees}{\null} & Trees & 3064 & \fboxsep=1mm \fboxrule=1mm \fcolorbox{PaviaU_Painted metal sheets}{PaviaU_Painted metal sheets}{\null} & Painted metal sheets & 1345 \\ \fboxsep=1mm \fboxrule=1mm
                
                \fcolorbox{PaviaU_Bare Soil}{PaviaU_Bare Soil}{\null} & Bare Soil & 5029 & \fboxsep=1mm \fboxrule=1mm \fcolorbox{PaviaU_Bitumen}{PaviaU_Bitumen}{\null} & Bitumen & 1330 & \fboxsep=1mm \fboxrule=1mm \fcolorbox{PaviaU_Self-Blocking Bricks}{PaviaU_Self-Blocking Bricks}{\null} & Self Blocking Bricks & 3682 \\ \fboxsep=1mm \fboxrule=1mm
               \fcolorbox{PaviaU_Shadows}{PaviaU_Shadows}{\null} & Shadows & 947 & & \fboxsep=1mm \fboxrule=1mm Total samples & 207400 \fboxsep=1mm \fboxrule=1mm \\ \bottomrule                
            \end{tabular}}
        \end{minipage}

        \caption{The type associated with the land-cover classes and number of available samples in the Pavia University (PU) dataset. Moreover, Spatially disjoint training and test samples for the PU dataset are also presented.}
        \label{fig:GT_PU}
    \end{figure}

The \textbf{University of Pavia} (UP) dataset was acquired by the Reflective Optics System Imaging Spectrometer (ROSIS) sensor during a flight campaign over the university campus at Pavia, Northern Italy \cite{huang2009comparative}. It consists of $610\times{340}$ pixels with $103$ spectral bands in the wavelength range from $430$ to $860~{nm}$ and 2.5 MPP. It comprises 9 urban land-cover classes. The detailed class description and ground truth maps are presented in Figure \ref{Fig.4A}. Moreover, the disjoint Training/Test sample maps are presented in Figures \ref{Fig.4B} and \ref{Fig.4C}.

The IEEE Geoscience and Remote Sensing Society published the \textbf{University of Houston} (UH) dataset--collected by the Compact Airborne Spectrographic Imager (CASI)-- in 2013 \cite{xu2016fusion}, as part of its Data Fusion Contest. It is composed of $340\times{1905}$ pixels with 144 spectral bands. The spatial resolution of this dataset is $2.5$ MPP with a wavelength ranging from $0.38$ to $1.05$ $\mu$m. Finally, the ground truth comprises 15 different land-cover classes. The detailed class description and ground truth maps are presented in Figure \ref{Fig.6A} and disjoint Training/Test sample maps are presented in Figures \ref{Fig.6B} and \ref{Fig.6C}.

  \begin{figure}[!htb]
        \centering
        \begin{minipage}[adjusting]{0.98\linewidth}
           \begin{subfigure}{0.98\textwidth}
                \centering          \includegraphics[width=0.70\columnwidth]{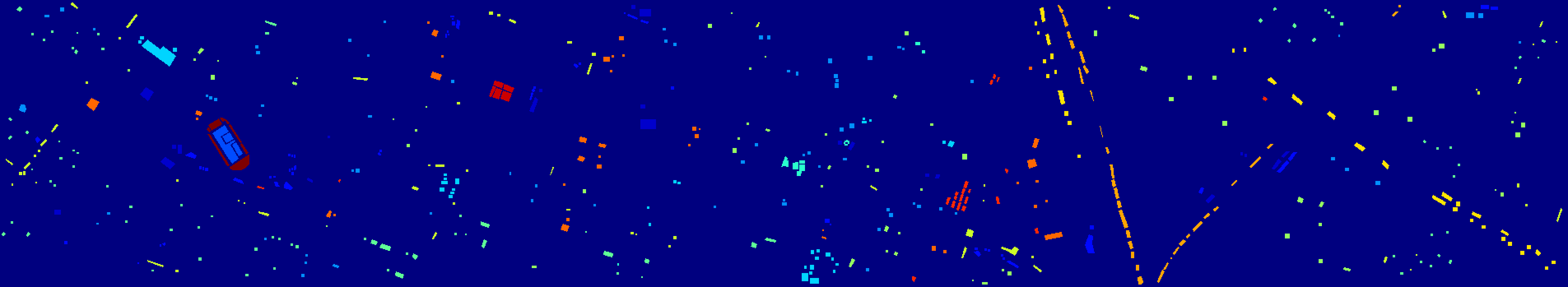}
    		    \caption{GT Maps}
		        \label{Fig.6A}
	        \end{subfigure} \\
	        \begin{subfigure}{0.98\textwidth}
                \centering	        \includegraphics[width=0.70\columnwidth]{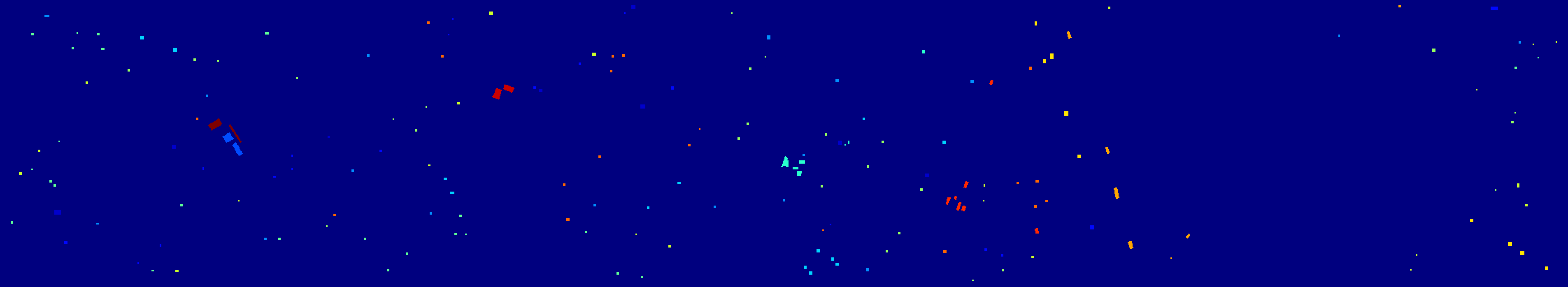}
    		    \caption{Disjoint Training}
		        \label{Fig.6B}
	        \end{subfigure} \\ 
	    \begin{subfigure}{0.98\textwidth}
	        \centering		    \includegraphics[width=0.70\columnwidth]{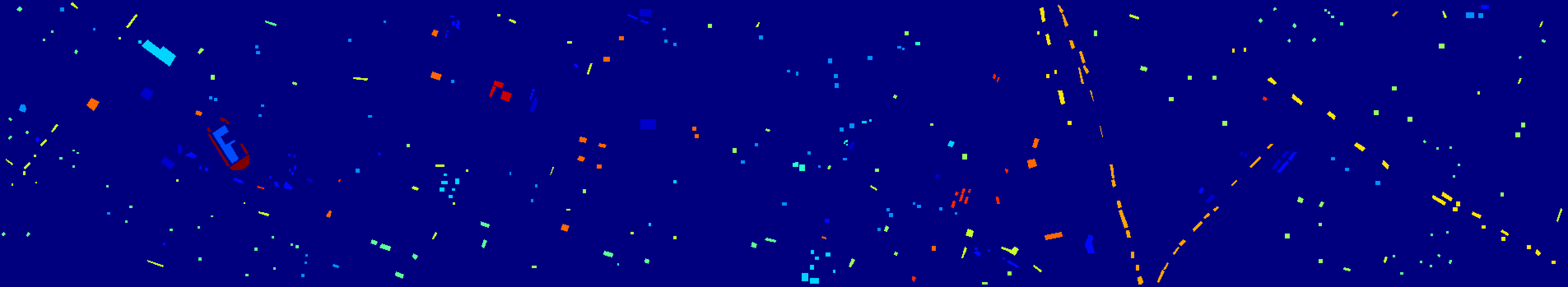}
    		\caption{Disjoint Testing} 
		    \label{Fig.6C}
	    \end{subfigure}
        \end{minipage}
        \\ \vspace{0.2cm}

        \begin{minipage}[adjusting]{0.98\linewidth}
            \centering            \resizebox{0.98\linewidth}{!}{\begin{tabular}{ccc|ccc|ccc}
            \toprule
            \fboxsep=1mm \fboxrule=1mm \fcolorbox{houston_data_Background!}{houston_data_Background!}{\null} & Background & 1314661 &  \fboxsep=1mm \fboxrule=1mm \fcolorbox{houston_data_Grass-healthy!}{houston_data_Grass-healthy!}{\null} & Grass-healthy & 1251 & \fboxsep=1mm \fboxrule=1mm \fcolorbox{houston_data_Grass-stressed!}{houston_data_Grass-stressed!}{\null} & Grass-stressed & 1254 \\ \fboxsep=1mm \fboxrule=1mm 
            
            \fcolorbox{houston_data_Grass-synthetic!}{houston_data_Grass-synthetic!}{\null} & Grass-synthetic & 697 &  \fboxsep=1mm \fboxrule=1mm \fcolorbox{houston_data_Tree!}{houston_data_Tree!}{\null} & Tree & 1244 & \fboxsep=1mm \fboxrule=1mm \fcolorbox{houston_data_Soil!}{houston_data_Soil!}{\null} & Soil & 1242 \\ \fboxsep=1mm \fboxrule=1mm 

            \fcolorbox{houston_data_Water!}{houston_data_Water!}{\null} & Water & 325  &  \fboxsep=1mm \fboxrule=1mm \fcolorbox{houston_data_Residential!}{houston_data_Residential!}{\null} & Residential & 1286 & \fboxsep=1mm \fboxrule=1mm \fcolorbox{houston_data_Commercial!}{houston_data_Commercial!}{\null} & Commercial & 1244 \\ \fboxsep=1mm \fboxrule=1mm

            \fcolorbox{houston_data_Road!}{houston_data_Road!}{\null} & Road & 1252 &  \fboxsep=1mm \fboxrule=1mm \fcolorbox{houston_data_Highway!}{houston_data_Highway!}{\null} & Highway & 1227 & \fboxsep=1mm \fboxrule=1mm \fcolorbox{houston_data_Railway!}{houston_data_Railway!}{\null} & Railway & 1235 \\ \fboxsep=1mm \fboxrule=1mm

            \fcolorbox{houston_data_Parking-lot1!}{houston_data_Parking-lot1!}{\null} & Parking-lot1 & 1233 &  \fboxsep=1mm \fboxrule=1mm \fcolorbox{houston_data_Parking-lot2!}{houston_data_Parking-lot2!}{\null} & Parking-lot2 & 469 & \fboxsep=1mm \fboxrule=1mm \fcolorbox{houston_data_Tennis-court!}{houston_data_Tennis-court!}{\null} & Tennis-court & 428 \\ \fboxsep=1mm \fboxrule=1mm

            \fcolorbox{houston_data_Running-track!}{houston_data_Running-track!}{\null} & Running-track & 660  & & Total samples & 1329690 \\ \bottomrule

            \end{tabular}}
        \end{minipage}
        \caption{The type associated with the land-cover classes and number of available samples in the Houston (UH) dataset. 
        }
        \label{fig:houstonGT}
    \end{figure}

\begin{figure}[!hbt]
    \centering
    \begin{minipage}[adjusting]{0.98\linewidth}
        \begin{subfigure}{0.98\textwidth}
            \centering          \includegraphics[width=0.60\columnwidth]{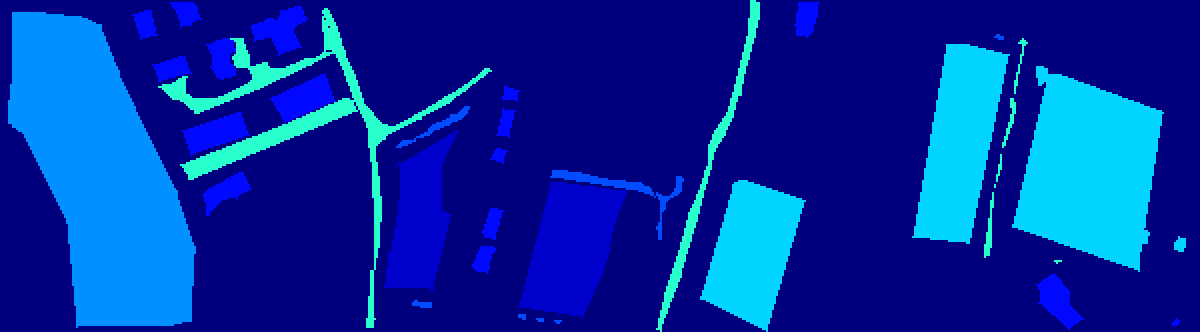}
    		\caption{GT Maps}
		    \label{Fig.6A}
	   \end{subfigure} \\
	   \begin{subfigure}{0.98\textwidth}
            \centering	        \includegraphics[width=0.60\columnwidth]{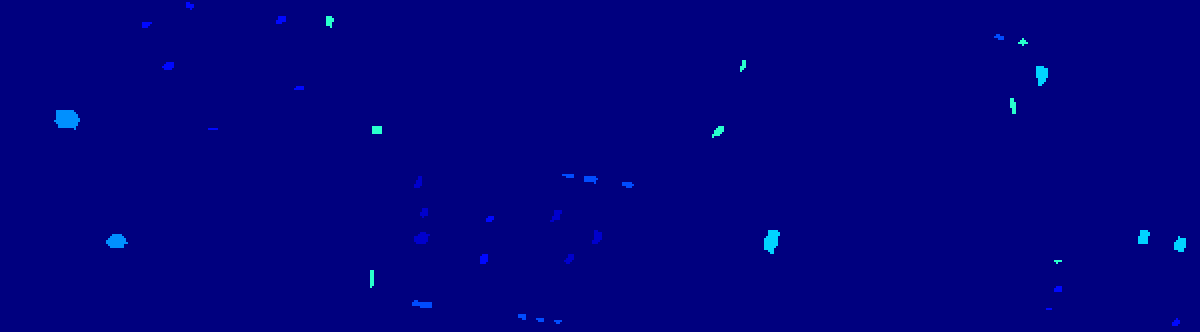}
    		\caption{Disjoint Training}
		    \label{Fig.6B}
	   \end{subfigure} \\ 
	   \begin{subfigure}{0.98\textwidth}
	        \centering		    \includegraphics[width=0.60\columnwidth]{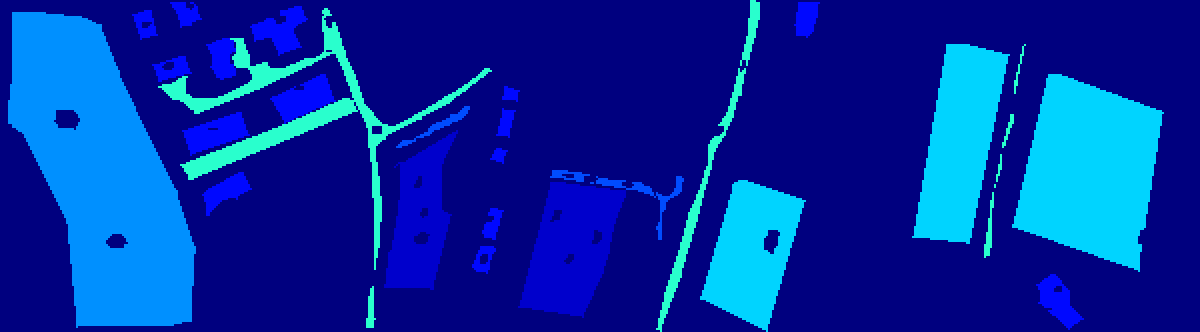}
    		\caption{Disjoint Testing} 
		    \label{Fig.6C}
	   \end{subfigure}
       \end{minipage}
       \\ \vspace{0.2cm}
       
    \resizebox{0.90\linewidth}{!}{\begin{tabular}{ccc|ccc} \hline
        \fboxsep=1mm \fboxrule=1mm
        \fcolorbox{trento_data_Background!}{trento_data_Background!}{\null} & Background & 168986 & \fboxsep=1mm \fboxrule=1mm
        \fcolorbox{trento_data_Apples!}{trento_data_Apples!}{\null} & Apples & 4034\\
        \fboxsep=1mm \fboxrule=1mm
    
        \fcolorbox{trento_data_Buildings!}{trento_data_Buildings!}{\null} & Buildings & 2903 & \fboxsep=1mm \fboxrule=1mm
        \fcolorbox{trento_data_Ground!}{trento_data_Ground!}{\null} & Ground & 479\\
        \fboxsep=1mm \fboxrule=1mm
    
        \fcolorbox{trento_data_Woods!}{trento_data_Woods!}{\null} & Woods & 9123 & \fboxsep=1mm \fboxrule=1mm
        \fcolorbox{trento_data_Vineyard!}{trento_data_Vineyard!}{\null} & Vineyard & 10501\\
        \fboxsep=1mm \fboxrule=1mm
    
        \fcolorbox{trento_data_Roads!}{trento_data_Roads!}{\null} & Roads & 3174 & &
        \fboxsep=1mm \fboxrule=1mm Total samples & 199200  \\ \bottomrule
    \end{tabular}}
    \caption{The type associated with the land-cover classes and number of available samples in the University of Trento (UT) dataset.}
    \label{fig:Trento}
\end{figure}

The \textbf{University of Trento} (UT) dataset was gathered by the using AISA eagle sensor over the rural regions in the south of Trento, Italy. The HSI contains 63 spectral bands within a wavelength of range $0.42$ to $0.99$ $\mu{m}$ \cite{xu2017multisource}. The scene has $600 \times 166$ pixels, which comprises of $6$ mutually exclusive vegetation land-cover classes where the spectral resolution is 9.2 $nm$, and the spatial resolution is 1 meter per pixel (MPP). In addition, the available samples are divided into disjoint training and test samples of 6 classes and Fig. \ref{fig:Trento} lists the information about the per class number of samples for six different land-covers.

Table \ref{Tab.1} provides a summary description of each dataset used in the following experiments whereas, Table \ref{Tab.2} enlists the numbers of disjoint samples (i.e., Train/Test samples selected from each class) used for all the experimental results. Please note that the number of train/test (i.e. percentage) samples and geographical locations of train/test samples remain the same for all experimental methods (competing methods).

\begin{table}[!hbt]
    \centering
    \caption{Summary of the HSI datasets used for experimental evaluation.}
    \resizebox{\columnwidth}{!}{\begin{tabular}{l||c||c||c||c|c} \hline 
        --- & \textbf{IP} & \textbf{PU} & \textbf{KSC} & \textbf{UH} & \textbf{UT}\\  \hline 
        \textbf{Year} & 1992 & 2001 & 1996 & 2013 & \\
        \textbf{Source} & AVIRIS & ROSIS-03 & AVIRIS & CASI & AISA \\
        \textbf{Spatial} & $145\times 145$ & $610 \times 610$ & $512\times 614$ & $340\times 1905$ & $600 \times 166$ \\
        \textbf{Spectral} & 220 & 115 & 176 & 144 & 63 \\
        \textbf{Wavelength} & $400-2500$ & $430-860$ & $400-2500$ & $0.35-1.05$ & $0.42-0.99$ \\
        \textbf{Samples} & 21025 & 207400 & 314368 & 1329690 & 199200 \\
        \textbf{Classes} & 16 & 9 & 13 & 15 & 6 \\
        \textbf{Sensor} & Aerial & Aerial & Aerial & Aerial & Aerial \\
        \textbf{Resolution} & $20~m$ & $1.3~m$ & $10~nm$ & $2.5~mpp$ & $1~mpp$\\ \hline 
    \end{tabular}}
    \label{Tab.1}
\end{table}

\begin{table*}[!hbt]
    \centering
    \caption{Number of Disjoint Train/Test Samples used for the experimental results. Where \textsc{TrS} and \textsc{TeS} stands for disjoint Train and Test samples, respectively.}
    \resizebox{0.95\textwidth}{!}{\color{black}
\begin{tabular}{cccc|cccc|cccc|cccc|cccc} \hline 
        \multicolumn{4}{c|}{\textbf{IP Data}} & \multicolumn{4}{c|}{\textbf{KSC Data}} & \multicolumn{4}{c|}{\textbf{PU Data}} & \multicolumn{4}{c|}{\textbf{UH Data}}
        & \multicolumn{4}{c}{\textbf{UT Data}} \\ \hline 
        Class & Land Cover & \textsc{TrS} & \textsc{TeS} & Class & Land Cover & \textsc{TrS} & \textsc{TeS} & Class & Land Cover & \textsc{TrS} & \textsc{TeS} & Class & Land Cover & \textsc{TrS} & \textsc{TeS} & Class & Land Cover & \textsc{TrS} & \textsc{TeS} \\ \hline 
         1 & \fcolorbox{indian_pines_corrected_Alfalfa!}{indian_pines_corrected_Alfalfa!}{\null} & 29 & 25 &  1  & \fcolorbox{KSC_Scrub!}{KSC_Scrub!}{\null} & 114  &  602 &  1 & \fcolorbox{PaviaU_Asphalt}{PaviaU_Asphalt}{\null} & 548 & 6083 & 1 & \fcolorbox{houston_data_Grass-healthy!}{houston_data_Grass-healthy!}{\null} & 198 & 1053 & \fcolorbox{trento_data_Apples!}{trento_data_Apples!}{\null} & Apples & 129 & 3905 \\
         2 & \fcolorbox{indian_pines_corrected_Corn-notill!}{indian_pines_corrected_Corn-notill!}{\null} & 762 & 675   &  2  & \fcolorbox{KSC_Willow-swamp!}{KSC_Willow-swamp!}{\null} & 36 & 207 & 2 & \fcolorbox{PaviaU_Meadows}{PaviaU_Meadows}{\null} & 540 &  18109 & 2 & \fcolorbox{houston_data_Grass-stressed!}{houston_data_Grass-stressed!}{\null}  & 190 & 1064 &  \fcolorbox{trento_data_Buildings!}{trento_data_Buildings!}{\null} & Buildings & 125  & 2778 \\
         3 & \fcolorbox{indian_pines_corrected_Corn-min!}{indian_pines_corrected_Corn-min!}{\null} & 435 & 404   &  3  & \fcolorbox{KSC_CP-hammock!}{KSC_CP-hammock!}{\null} & 38  & 218 & 3 & \fcolorbox{PaviaU_Gravel}{PaviaU_Gravel}{\null} & 392 & 1707 & 3 & \fcolorbox{houston_data_Grass-synthetic!}{houston_data_Grass-synthetic!}{\null} & 192 & 505 &  \fcolorbox{trento_data_Ground!}{trento_data_Ground!}{\null} & Ground & 105   & 374 \\
         4 & \fcolorbox{indian_pines_corrected_Corn!}{indian_pines_corrected_Corn!}{\null} & 146 & 99  &  4  & \fcolorbox{KSC_Slash-pine!}{KSC_Slash-pine!}{\null} & 38  & 214 & 4 & \fcolorbox{PaviaU_Trees}{PaviaU_Trees}{\null} & 524 & 2540 & 4 & \fcolorbox{houston_data_Tree!}{houston_data_Tree!}{\null} & 188 & 1056 &  \fcolorbox{trento_data_Woods!}{trento_data_Woods!}{\null} & Woods & 154   & 8969 \\
         5 & \fcolorbox{indian_pines_corrected_Grass/Pasture!}{indian_pines_corrected_Grass/Pasture!}{\null} & 232 & 274  &  5  & \fcolorbox{KSC_Oak/Broadleaf!}{KSC_Oak/Broadleaf!}{\null} & 24 & 137 & 5 & \fcolorbox{PaviaU_Painted metal sheets}{PaviaU_Painted metal sheets}{\null} & 265 & 1080 & 5 & \fcolorbox{houston_data_Soil!}{houston_data_Soil!}{\null} & 186 & 1056 &  \fcolorbox{trento_data_Vineyard!}{trento_data_Vineyard!}{\null} & Vineyard & 184   & 10317 \\
         6 & \fcolorbox{indian_pines_corrected_Grass/Trees!}{indian_pines_corrected_Grass/Trees!}{\null} & 394 & 354  &  6  & \fcolorbox{KSC_Hardwood!}{KSC_Hardwood!}{\null} & 34  & 195 & 6 & \fcolorbox{PaviaU_Bare Soil}{PaviaU_Bare Soil}{\null} & 532 & 4497 & 6 & \fcolorbox{houston_data_Water!}{houston_data_Water!}{\null} & 182 & 143 & \fcolorbox{trento_data_Roads!}{trento_data_Roads!}{\null} & Roads & 122   & 3052 \\
         7 & \fcolorbox{indian_pines_corrected_Grass/pasture-mowed!}{indian_pines_corrected_Grass/pasture-mowed!}{\null} & 16 & 2 &  7  & \fcolorbox{KSC_Swap!}{KSC_Swap!}{\null} & 16  & 89 &  7 & \fcolorbox{PaviaU_Bitumen}{PaviaU_Bitumen}{\null} & 375 & 955 & 7 & \fcolorbox{houston_data_Residential!}{houston_data_Residential!}{\null} & 196 & 1072 &  &   &  & \\
         8 & \fcolorbox{indian_pines_corrected_Hay-windrowed!}{indian_pines_corrected_Hay-windrowed!}{\null} & 235 & 250  &  8  & \fcolorbox{KSC_Graminoid-marsh!}{KSC_Graminoid-marsh!}{\null} & 65 & 366 &  8 & \fcolorbox{PaviaU_Self-Blocking Bricks}{PaviaU_Self-Blocking Bricks}{\null} & 514 & 3168 & 8 & \fcolorbox{houston_data_Commercial!}{houston_data_Commercial!}{\null} & 191 & 1053 &  &   &  & \\
         9 & \fcolorbox{indian_pines_corrected_Oats!}{indian_pines_corrected_Oats!}{\null} & 10 & 10  &  9  & \fcolorbox{KSC_Spartina-marsh!}{KSC_Spartina-marsh!}{\null} & 78 & 442 &  9 & \fcolorbox{PaviaU_Shadows}{PaviaU_Shadows}{\null} & 231 & 716 & 9 & \fcolorbox{houston_data_Road!}{houston_data_Road!}{\null} & 193 & 1059 &  &   &  & \\
         10 & \fcolorbox{indian_pines_corrected_Soybeans-notill!}{indian_pines_corrected_Soybeans-notill!}{\null} & 470 & 503  &  10  & \fcolorbox{KSC_Cattail-marsh!}{KSC_Cattail-marsh!}{\null} & 61 & 343 &  &  &  &  & 10 & \fcolorbox{houston_data_Highway!}{houston_data_Highway!}{\null} & 191 & 1036 &  &   &  & \\
         11 & \fcolorbox{indian_pines_corrected_Soybeans-min!}{indian_pines_corrected_Soybeans-min!}{\null} & 1424 & 1065  &  11  & \fcolorbox{KSC_Salt-marsh!}{KSC_Salt-marsh!}{\null} & 63 & 356 &  &  &  &  & 11 & \fcolorbox{houston_data_Railway!}{houston_data_Railway!}{\null} & 181 & 1054 &  &   &  & \\
         12 & \fcolorbox{indian_pines_corrected_Soybean-clean!}{indian_pines_corrected_Soybean-clean!}{\null} & 328 & 282  &  12  & \fcolorbox{KSC_Mud-flats!}{KSC_Mud-flats!}{\null} & 75 & 428 &  &  &  &  & 12 & \fcolorbox{houston_data_Parking-lot1!}{houston_data_Parking-lot1!}{\null} & 192 & 1041 &  &   &  & \\
         13 & \fcolorbox{indian_pines_corrected_Wheat!}{indian_pines_corrected_Wheat!}{\null} & 132 & 80  &  13  & \fcolorbox{KSC_Water!}{KSC_Water!}{\null} & 139 & 788 &   &  &  &  & 13 & \fcolorbox{houston_data_Parking-lot2!}{houston_data_Parking-lot2!}{\null} & 184 & 285 &  &   &  & \\
         14 & \fcolorbox{indian_pines_corrected_Woods!}{indian_pines_corrected_Woods!}{\null} & 728 & 545 & & & & & & & & & 14 & \fcolorbox{houston_data_Tennis-court!}{houston_data_Tennis-court!}{\null} & 181 & 247 &  &   &  & \\
         15 & \fcolorbox{indian_pines_corrected_Bldg-Grass-Tree-Drives!}{indian_pines_corrected_Bldg-Grass-Tree-Drives!}{\null} & 291 & 99 & & & & & & & & & 15 & \fcolorbox{houston_data_Running-track!}{houston_data_Running-track!}{\null} & 187 & 473 &  &   &  & \\
         16 &  \fcolorbox{indian_pines_corrected_Stone-steel towers!}{indian_pines_corrected_Stone-steel towers!}{\null} & 57 & 44 & & & & & & & & & & & & &  &   &  & \\ \hline
\end{tabular}}
    \label{Tab.2}
\end{table*}

\begin{table*}[!hbt]
\let\center\empty
\let\endcenter\relax
\centering
\caption{Classification results obtained by RF \cite{ham2005investigation}, MLR \cite{li2010semisupervised}, SVM \cite{melgani2004classification}, MLP \cite{paoletti2019deep}, RNN \cite{hang2019cascaded}, LSTM \cite{hochreiter1997long}, GRU \cite{cho2014properties}, CNN-1D \cite{hong2021graph}, CNN-2D \cite{makantasis2015deep}, CNN-3D \cite{hamida2018deep}, HybridSN \cite{roy2019hybridsn}, and MorphCNN \cite{roy2021morphological} on the disjoint train-test dataset for the PU scene.}
\resizebox{\linewidth}{!}{\color{black}
\begin{tabular}{c|ccccccccccccccccccc} \hline
\textbf{Class} & RF \cite{ham2005investigation} & MLR \cite{li2010semisupervised} & SVM \cite{melgani2004classification}& MLP \cite{paoletti2019deep}& RNN \cite{hang2019cascaded}& LSTM \cite{hochreiter1997long}& GRU \cite{cho2014properties}& CNN-1D \cite{hong2021graph}& CNN-2D \cite{makantasis2015deep}& CNN-3D \cite{hamida2018deep}& HybridSN \cite{roy2019hybridsn} & MorphCNN \cite{roy2021morphological} \\ \hline 

1 & 89.98$\pm$0.15 & 77.68$\pm$0.0 & 82.23$\pm$0.0 & 84.53$\pm$1.89 & 83.08$\pm$3.3 & 82.63$\pm$2.39 & 77.25$\pm$6.92 & 87.18$\pm$2.11 & 93.4$\pm$1.89 & 85.66$\pm$4.0 & 89.74$\pm$5.19 & \textbf{94.52}$\pm$1.9 
\\

2 & 74.39$\pm$0.01 & 58.79$\pm$0.01 & 65.81$\pm$0.0 & 75.13$\pm$2.4 & 67.9$\pm$2.92 & 78.74$\pm$1.99 & 80.1$\pm$5.12 & 89.64$\pm$2.53 & 96.84$\pm$1.93 & 95.88$\pm$1.71 & 81.78$\pm$3.15 & \textbf{97.12}$\pm$8.71 
\\

3 & 38.42$\pm$0.13 & 67.21$\pm$0.02 & 66.72$\pm$0.0 & 68.37$\pm$5.17 & 65.17$\pm$7.99 & 60.73$\pm$11.0 & 54.79$\pm$14.82 & 71.1$\pm$5.98 & 65.48$\pm$13.94 & 68.11$\pm$6.47 & 82.88$\pm$1.54 & \textbf{85.08}$\pm$4.53 
\\

4 & 98.24$\pm$0.05 & 74.27$\pm$0.05 & \textbf{97.77}$\pm$0.0 & 93.5$\pm$2.32 & 90.72$\pm$2.56 & 97.1$\pm$1.22 & 92.05$\pm$2.31 & 95.32$\pm$1.49 & 95.55$\pm$2.14 & 97.02$\pm$0.83 & 83.66$\pm$4.58 & 97.0$\pm$1.03 
\\

5 & 95.98$\pm$0.04 & 98.88$\pm$0.04 & 99.37$\pm$0.0 & 99.37$\pm$0.08 & 99.23$\pm$0.09 & 99.28$\pm$0.08 & 99.51$\pm$0.12 & 99.48$\pm$0.26 & 98.03$\pm$0.92 & 98.9$\pm$0.56 & \textbf{99.94}$\pm$0.04 & 99.25$\pm$0.22 
\\

6 & 51.43$\pm$0.19 & 93.53$\pm$0.02 & 91.62$\pm$0.0 & 89.94$\pm$4.14 & 85.07$\pm$3.14 & 65.94$\pm$5.92 & 74.86$\pm$11.38 & 88.28$\pm$2.33 & 80.52$\pm$9.39 & 68.85$\pm$11.29 & 72.43$\pm$13.23 & \textbf{93.92}$\pm$3.88 
\\

7 & 80.63$\pm$0.36 & 85.08$\pm$0.05 & 87.36$\pm$0.0 & 87.2$\pm$3.05 & 82.94$\pm$3.79 & 84.95$\pm$4.02 & 90.17$\pm$3.9 & 86.77$\pm$3.38 & 89.29$\pm$9.48 & 73.09$\pm$9.53 & \textbf{96.16}$\pm$1.88 & 84.98$\pm$10.74 
\\

8 & 97.64$\pm$0.14 & 87.58$\pm$0.01 & 90.46$\pm$0.0 & 90.37$\pm$1.24 & 85.85$\pm$4.97 & 88.89$\pm$7.83 & 90.42$\pm$4.39 & 90.43$\pm$3.34 & 94.5$\pm$5.44 & 95.21$\pm$1.69 & 92.80$\pm$0.90 & \textbf{96.62}$\pm$2.21 
\\

9 & 94.92$\pm$0.05 & \textbf{99.22}$\pm$0.05 & 93.71$\pm$0.0 & 98.44$\pm$1.17 & 94.52$\pm$4.79 & 98.29$\pm$1.47 & 93.51$\pm$7.93 & 97.33$\pm$3.31 & 95.8$\pm$0.76 & 93.54$\pm$1.76 & 94.04$\pm$3.99 & 97.05$\pm$0.46 
\\ \hline\hline

OA & 77.44$\pm$0.06 & 72.23$\pm$0.0 & 77.8$\pm$0.0 & 82.05$\pm$0.88 & 77.07$\pm$0.95 & 80.38$\pm$0.52 & 80.7$\pm$0.56 & 89.09$\pm$0.97 & 92.55$\pm$1.02 & 89.43$\pm$1.37 & 84.18$\pm$1.40 & \textbf{95.51}$\pm$0.66 
\\

AA & 80.18$\pm$0.06 & 82.47$\pm$0.01 & 86.12$\pm$0.0 & 87.43$\pm$1.03 & 83.83$\pm$0.72 & 84.06$\pm$0.74 & 83.63$\pm$2.03 & 89.5$\pm$1.03 & 89.94$\pm$1.37 & 86.25$\pm$1.98 & 88.16$\pm$1.94 & \textbf{93.95$\pm$0.96} 
\\ 

k(x100) & 70.44$\pm$0.07 & 65.44$\pm$0.0 & 72.06$\pm$0.0 & 76.89$\pm$1.07 & 70.84$\pm$1.04 & 74.32$\pm$0.68 & 74.76$\pm$1.02 & 85.5$\pm$1.22 & 89.9$\pm$1.42 & 85.61$\pm$1.94 & 79.13$\pm$1.42 & \textbf{93.95}$\pm$0.88 \\ \hline

\end{tabular}}
\label{table:compdisjointUP}
\end{table*}

\begin{figure*}[!t]
\resizebox{\textwidth}{!}{
\begin{tabular}{cccccc}
\includegraphics[]{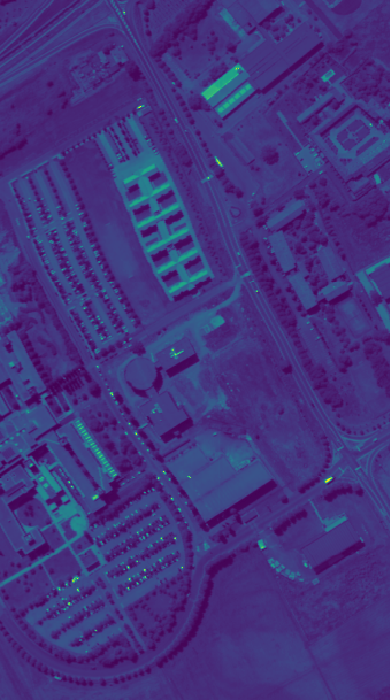} &
\includegraphics[]{gt/UP.png} &
\includegraphics[]{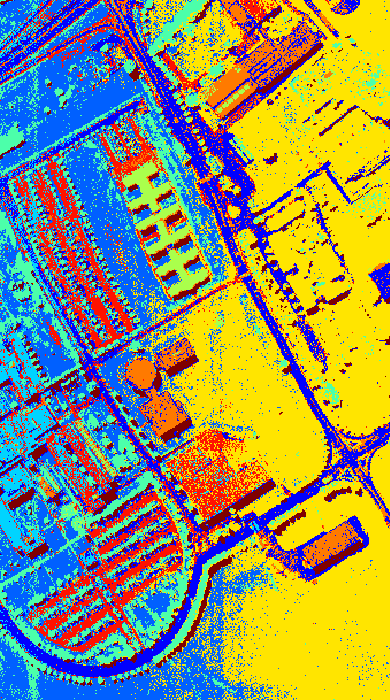} &
\includegraphics[]{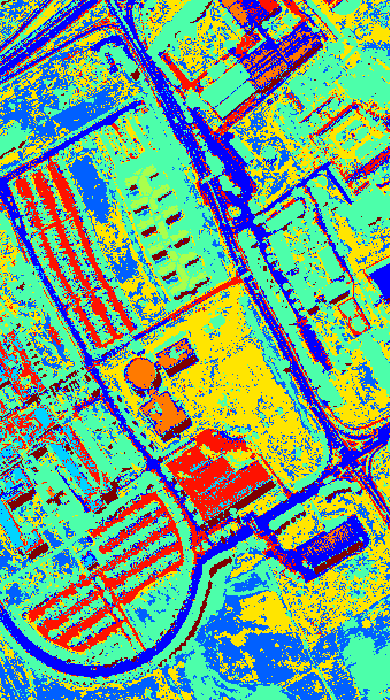} &
\includegraphics[]{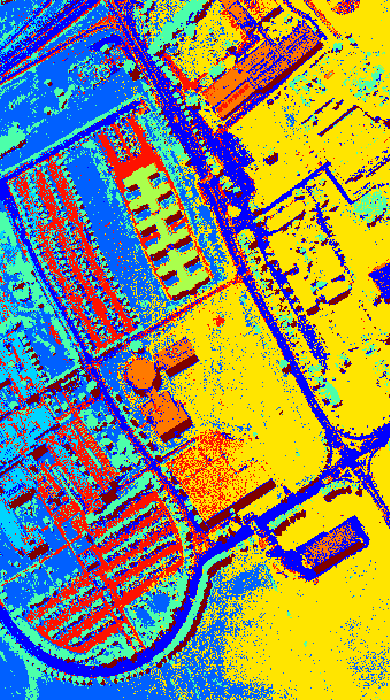} &
\includegraphics[]{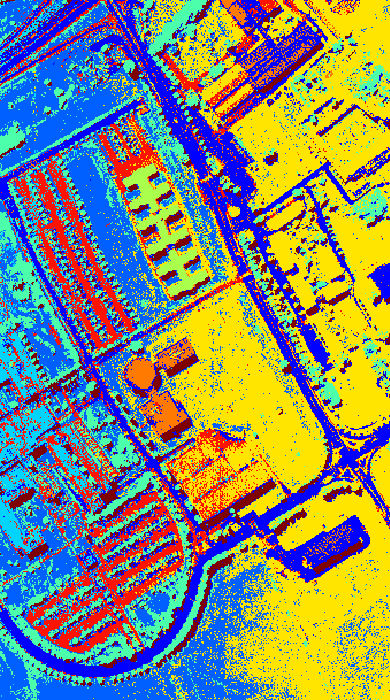} \\
[8pt]
\HUGES\textit{a}) 1PC &
\HUGES\textit{b}) GT &
\HUGES\textit{c}) MLR  & 
\HUGES\textit{d}) SVM  & 
\HUGES\textit{e}) MLP  &
\HUGES\textit{f}) RNN  \\
[8pt]
\includegraphics[]{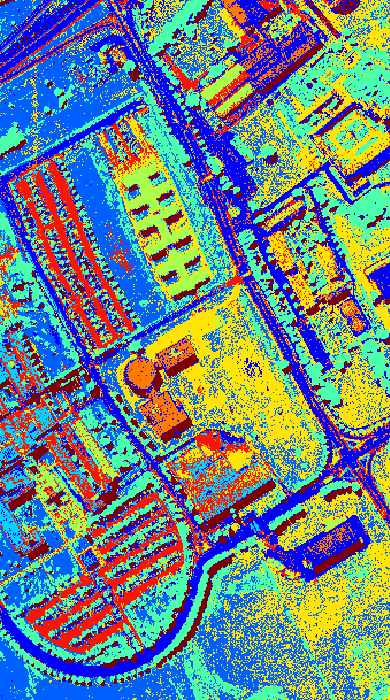} &
\includegraphics[]{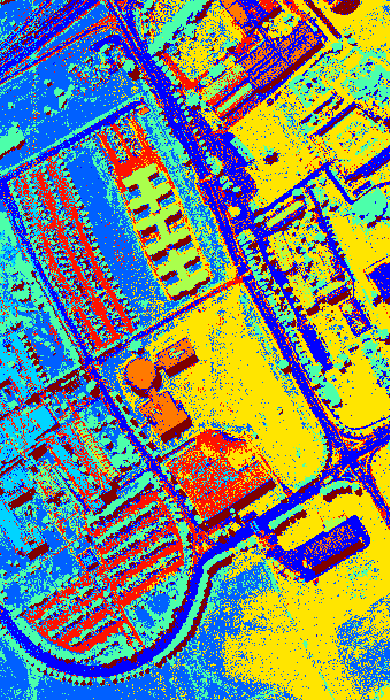} &
\includegraphics[]{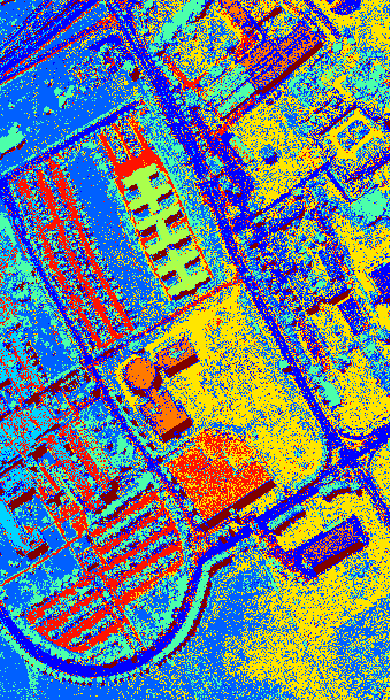} &
\includegraphics[]{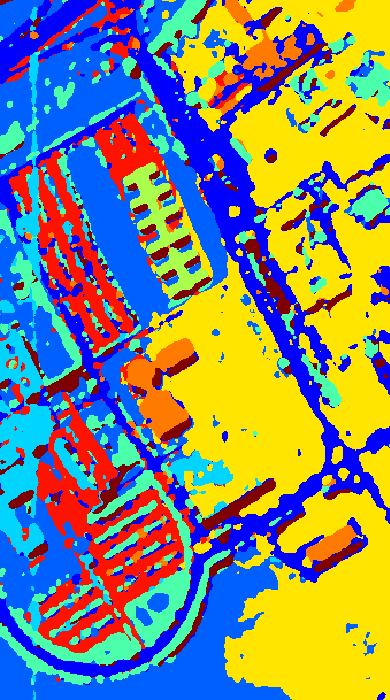} &
\includegraphics[]{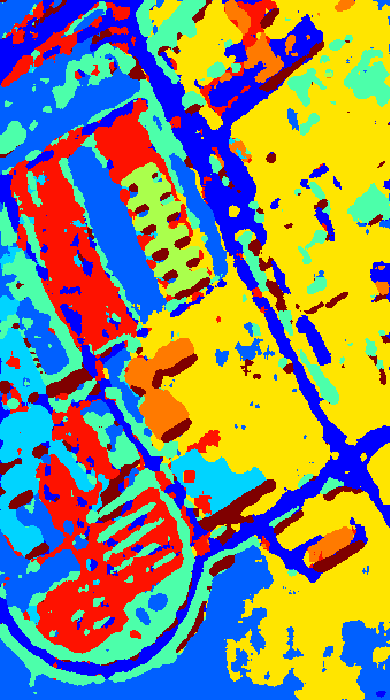} &
\includegraphics[]{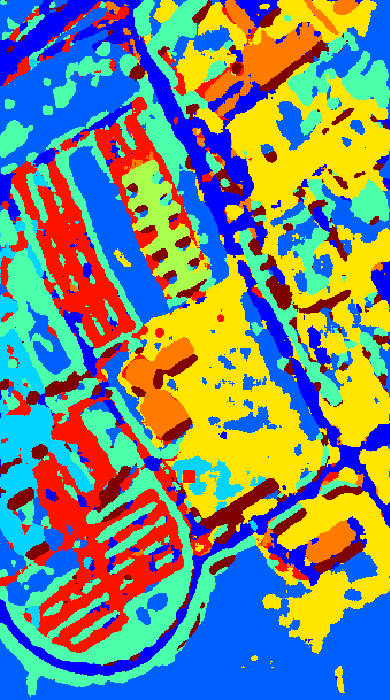} \\
[8pt]
\HUGES\textit{g}) LSTM  &
\HUGES\textit{h}) GRU   &
\HUGES\textit{i}) CNN1D &
\HUGES\textit{j}) CNN2D  &
\HUGES\textit{k}) CNN3D  & 
\HUGES\textit{l}) MorphCNN (90.55\%) \\
\end{tabular}}
\caption{Classification Maps obtained by MLR \cite{li2010semisupervised}, SVM \cite{melgani2004classification}, MLP \cite{paoletti2019deep}, RNN \cite{hang2019cascaded}, LSTM \cite{hochreiter1997long}, GRU \cite{cho2014properties}, CNN-1D \cite{hong2021graph}, CNN-2D \cite{makantasis2015deep}, CNN-3D \cite{hamida2018deep} and MorphCNN \cite{roy2021morphological} on the disjoint train-test dataset for the UP scene.}
\label{fig:comparativeDUP}
\end{figure*}

\subsection{Experimental Results on Disjoint Train/Test Samples}

To strengthen the ideas highlighted in this survey and to make the claims valid, the main contributions made in recent years include MLR, SVM, MLP, RNN, LSTM, GRU, CNN-1D, CNN-2D, CNN-3D, and MorphCNN have been considered to compare the experimental results. Some of the representative works for each above are as follows; Cloud Implementation of Logistic Regression for HSIC \cite{li2010semisupervised, li2011hyperspectral, haut2017cloud} (MLR), Classification of Hyperspectral Remote Sensing Images with SVM \cite{melgani2004classification}, (SVM), Deep Recurrent Neural Networks for HSIC \cite{hang2019cascaded} (RNN), Long Short-Term Memory \cite{hochreiter1997long} (LSTM), On the properties of Neural Machine Translation: Encoder-Decoder Approaches \cite{cho2014properties} (GRU), Deep Convolutional Neural Networks for HSIC \cite{hong2021graph} (CNN1D), Deep Supervised Learning for Hyperspectral Data Classification through Convolutional Neural Networks \cite{makantasis2015deep} (CNN2D), 3-D Deep Learning Approach for Remote Sensing Image Classification \cite{hamida2018deep} (CNN3D), Morphological Convolutional Neural Networks for HSIC \cite{roy2021morphological} (MorphCNN), and MLP \cite{paoletti2019deep}. 

To some extent, all the aforesaid works are based on Convolutional and Recurrent Networks and are evaluated on four benchmark HSI datasets namely IP, PU, KSC, Houston Scene, and the University of Toronto. This survey only pays attention to the robustness of all these models while considering the small sample size of training data to classify HSI for joint spatial-spectral classification.

\begin{table*}[!t]
\let\center\empty
\let\endcenter\relax
\centering
\caption{Classification results obtained by RF \cite{ham2005investigation}, MLR \cite{li2010semisupervised}, SVM \cite{melgani2004classification}, MLP \cite{paoletti2019deep}, RNN \cite{hang2019cascaded}, LSTM \cite{hochreiter1997long}, GRU \cite{cho2014properties}, CNN-1D \cite{hong2021graph}, CNN-2D \cite{makantasis2015deep}, CNN-3D \cite{hamida2018deep}, HybridSN \cite{roy2019hybridsn}, and MorphCNN \cite{roy2021morphological} on the disjoint train-test dataset for the IP scene.}
\resizebox{\linewidth}{!}{\color{black}
\begin{tabular}{c|cccccccccccc} \hline
\textbf{Class} & RF \cite{ham2005investigation} & MLR \cite{li2010semisupervised} & SVM \cite{melgani2004classification}& MLP \cite{paoletti2019deep}& RNN \cite{hang2019cascaded}& LSTM \cite{hochreiter1997long}& GRU \cite{cho2014properties}& CNN-1D \cite{hong2021graph}& CNN-2D \cite{makantasis2015deep}& CNN-3D \cite{hamida2018deep}& HybridSN \cite{roy2019hybridsn} & MorphCNN \cite{roy2021morphological} \\ \hline

1 & 85.33$\pm$1.88 & 80.0$\pm$0.0 & 88.0$\pm$0.0 & 73.6$\pm$7.42 & 58.4$\pm$4.8 & 89.6$\pm$1.96 & 77.6$\pm$9.33 & 80.8$\pm$12.75 & 73.64$\pm$14.77 & 48.18$\pm$22.84 & 82.66$\pm$13.59 & \textbf{92.27}$\pm$3.55 \\

2 & 55.11$\pm$0.32 & 81.48$\pm$0.0 & 80.0$\pm$0.0 & 81.45$\pm$1.07 & 75.5$\pm$1.48 & 82.22$\pm$1.26 & 81.1$\pm$2.77 & 79.38$\pm$4.16 & 83.12$\pm$6.12 & \textbf{85.12}$\pm$7.88 & 82.17$\pm$2.64 & 84.05$\pm$8.03 \\

3 & 22.77$\pm$0.20 & 54.11$\pm$0.12 & 69.55$\pm$0.0 & 64.55$\pm$2.85 & 63.37$\pm$1.93 & 64.16$\pm$5.44 & 70.35$\pm$1.36 & 74.26$\pm$6.12 & \textbf{81.98}$\pm$3.89 & 77.22$\pm$13.04 & 76.73$\pm$4.02 & 79.34$\pm$3.45 \\

4 & 13.13$\pm$1.64 & 38.38$\pm$0.0 & 48.48$\pm$0.0 & 47.07$\pm$10.41 & 29.49$\pm$5.4 & \textbf{55.35}$\pm$10.62 & 53.33$\pm$11.15 & 31.92$\pm$11.55 & 45.39$\pm$6.36 & 50.11$\pm$10.04 & 33.33$\pm$3.59 & 52.14$\pm$6.24 \\

5 & 41.60$\pm$0.78 & \textbf{91.97}$\pm$0.0 & 87.23$\pm$0.0 & 86.94$\pm$1.07 & 87.59$\pm$1.5 & 89.27$\pm$0.68 & 88.4$\pm$0.85 & 90.73$\pm$1.07 & 89.11$\pm$5.55 & 80.28$\pm$6.52 & 81.14$\pm$8.89 & 91.66$\pm$1.69 \\

6 & 94.06$\pm$0.23 & 94.63$\pm$0.0 & 96.33$\pm$0.0 & 95.93$\pm$0.97 & 95.31$\pm$0.83 & 96.39$\pm$0.87 & 96.38$\pm$1.14 & \textbf{96.39}$\pm$0.9 & 95.02$\pm$5.68 & 89.81$\pm$4.03 & 97.36$\pm$2.54 & 95.74$\pm$2.28 \\

7 & 0.0$\pm$0.0 & 0.0$\pm$0.0 & \textbf{50.0}$\pm$0.0 & 10.0$\pm$20.0 & 0.0$\pm$0.0 & 0.0$\pm$0.0 & 0.0$\pm$0.0 & 0.0$\pm$0.0 & 0.0$\pm$0.0 & 0.0$\pm$0.0 & 0.0$\pm$0.0 & 0.0$\pm$0.0 \\

8 & 91.33$\pm$0.18 & \textbf{100.0}$\pm$0.0 & \textbf{100.0}$\pm$0.0 & 99.84$\pm$0.2 & 99.52$\pm$0.3 & 99.2$\pm$0.91 & 99.12$\pm$0.64 & 99.84$\pm$0.32 & 99.96$\pm$0.13 & 95.96$\pm$6.78 & 96.53$\pm$3.78 & \textbf{100.0}$\pm$0.0 \\

9 & 40.0$\pm$0.0 & 0.0$\pm$0.0 & 50.0$\pm$0.0 & \textbf{80.0}$\pm$15.49 & 56.0$\pm$10.2 & 76.0$\pm$8.0 & 66.0$\pm$4.9 & 50.0$\pm$8.94 & 26.66$\pm$15.87 & 77.78$\pm$21.66 & 66.66$\pm$24.94 & 44.44$\pm$19.88 \\

10 & 26.83$\pm$1.26 & 66.76$\pm$0.08 & 76.54$\pm$0.0 & 75.35$\pm$5.02 & 71.13$\pm$5.93 & 81.51$\pm$2.76 & 78.53$\pm$4.64 & \textbf{81.83}$\pm$4.69 & 77.44$\pm$8.99 & 77.9$\pm$6.2 & 74.35$\pm$9.46 & 80.77$\pm$3.77 \\

11 & 81.06$\pm$0.49 & 84.13$\pm$0.0 & 87.7$\pm$0.0 & 83.19$\pm$1.51 & 78.86$\pm$1.45 & 80.4$\pm$2.43 & 82.29$\pm$1.82 & 80.39$\pm$3.65 & \textbf{89.4}$\pm$5.47 & 82.73$\pm$3.81 & 79.18$\pm$4.92 & 88.54$\pm$5.03 \\

12 & 28.95$\pm$0.44 & 66.31$\pm$0.0 & 77.3$\pm$0.0 & 78.58$\pm$2.95 & 71.91$\pm$5.07 & 76.31$\pm$1.39 & 83.19$\pm$1.16 & 84.75$\pm$7.5 & 87.72$\pm$3.06 & 82.64$\pm$14.49 & 71.04$\pm$4.11 & \textbf{88.46}$\pm$4.26 \\

13 & 86.25$\pm$1.02 & 95.0$\pm$0.0 & 97.5$\pm$0.0 & \textbf{98.0}$\pm$0.61 & 97.0$\pm$1.7 & 97.25$\pm$0.94 & 97.75$\pm$0.94 & 97.75$\pm$0.5 & 95.28$\pm$4.74 & 89.72$\pm$6.89 & 96.25$\pm$4.44 & 87.64$\pm$3.43 \\

14 & 91.07$\pm$0.87 & 90.64$\pm$0.0 & 91.38$\pm$0.0 & 92.92$\pm$1.48 & 90.28$\pm$1.09 & 94.13$\pm$1.18 & 92.88$\pm$1.79 & 93.32$\pm$2.34 & \textbf{98.94}$\pm$0.55 & 98.31$\pm$1.41 & 91.68$\pm$4.33 & 98.82$\pm$1.01 \\

15 & 10.10$\pm$0.0 & 89.9$\pm$0.0 & 80.81$\pm$0.0 & 87.88$\pm$3.78 & 75.56$\pm$6.43 & 90.71$\pm$2.34 & \textbf{93.54}$\pm$1.64 & 89.9$\pm$4.78 & 82.02$\pm$14.83 & 55.17$\pm$27.57 & 45.45$\pm$21.39 & 69.44$\pm$15.86 \\

16 & 71.96$\pm$3.86 & \textbf{97.73}$\pm$0.0 & 97.73$\pm$0.0 & 87.27$\pm$4.45 & 88.64$\pm$4.31 & 94.09$\pm$2.32 & 95.45$\pm$2.49 & 96.82$\pm$2.32 & 82.0$\pm$6.69 & 82.5$\pm$12.5 & 84.09$\pm$4.90 & 84.0$\pm$4.21 \\ \hline\hline

OA & 60.80$\pm$0.14 & 80.33$\pm$0.02 & 84.12$\pm$0.0 & 82.95$\pm$0.23 & 79.07$\pm$0.33 & 83.55$\pm$0.39 & 84.2$\pm$0.21 & 84.0$\pm$0.28 & 87.25$\pm$1.03 & 83.6$\pm$1.41 & 80.86$\pm$1.74 & \textbf{87.45}$\pm$1.01 \\

AA & 52.47$\pm$0.21 & 70.69$\pm$0.01 & \textbf{79.91}$\pm$0.0 & 77.66$\pm$1.98 & 71.16$\pm$0.87 & 79.16$\pm$0.75 & 78.49$\pm$0.36 & 76.76$\pm$0.75 & 75.48$\pm$2.12 & 73.34$\pm$3.46 & 72.41$\pm$2.32 & 77.33$\pm$1.56 \\

k(x100) & 54.41$\pm$0.18 & 77.47$\pm$0.02 & 81.87$\pm$0.0 & 80.56$\pm$0.26 & 76.12$\pm$0.4 & 81.27$\pm$0.44 & 82.01$\pm$0.26 & 81.81$\pm$0.35 & 85.48$\pm$1.15 & 81.36$\pm$1.62 & 78.24$\pm$1.98 & \textbf{85.75}$\pm$1.14 \\ \hline

\end{tabular}}
\label{table:compdisjointIP}
\end{table*}

\begin{figure*}[!t]
\resizebox{\textwidth}{!}{
\begin{tabular}{cccccc}
\includegraphics[]{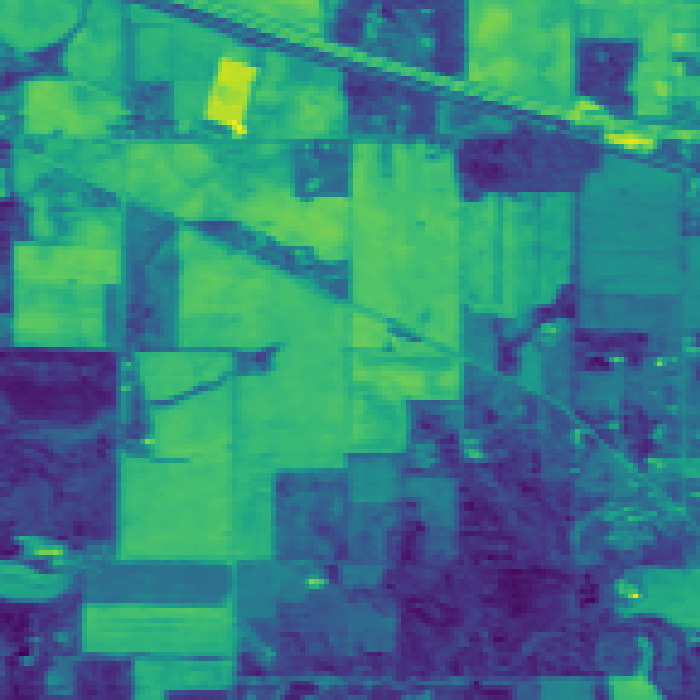} &
\includegraphics[]{gt/IP.png} &
\includegraphics[]{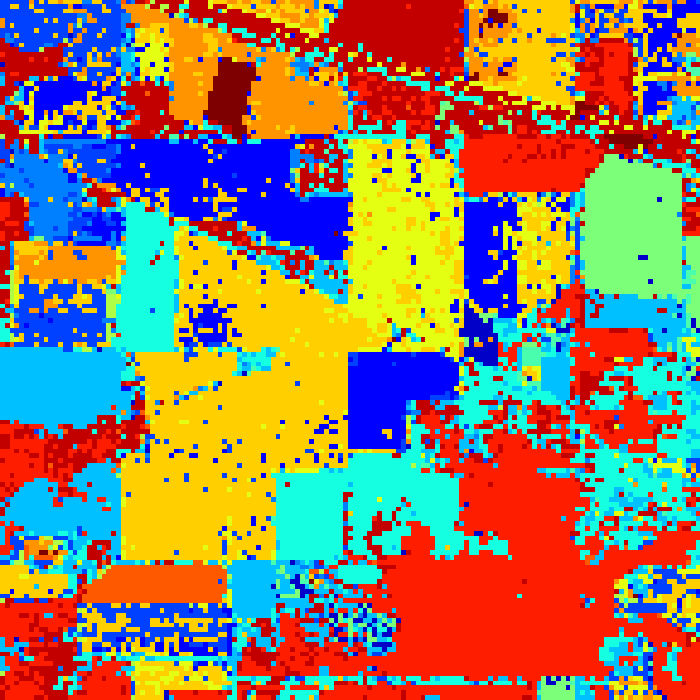} &
\includegraphics[]{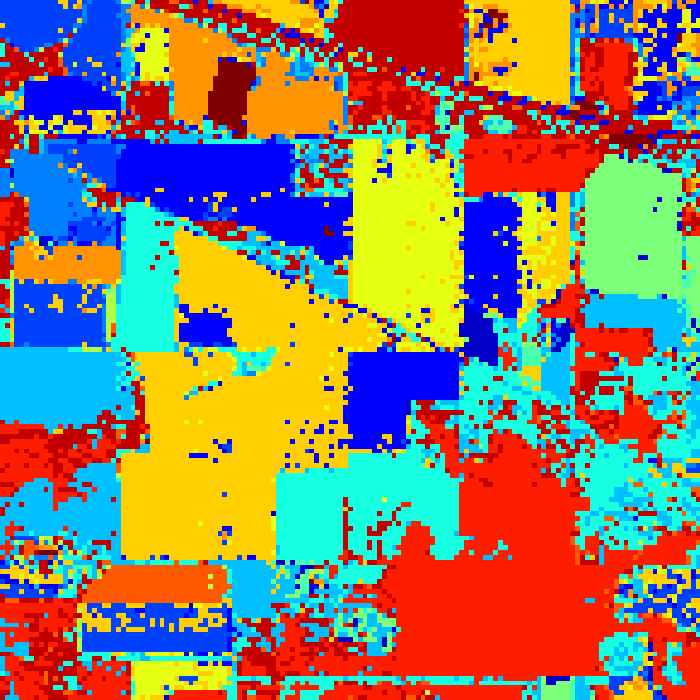} &
\includegraphics[]{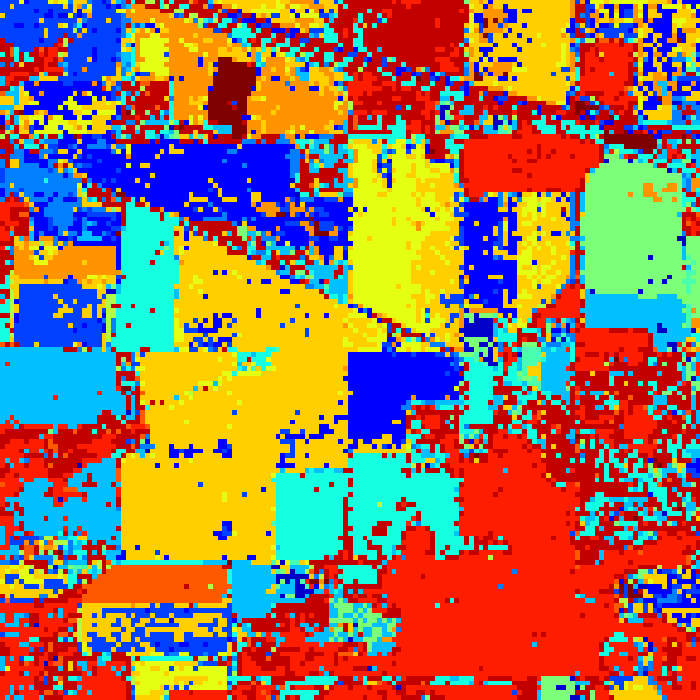} &
\includegraphics[]{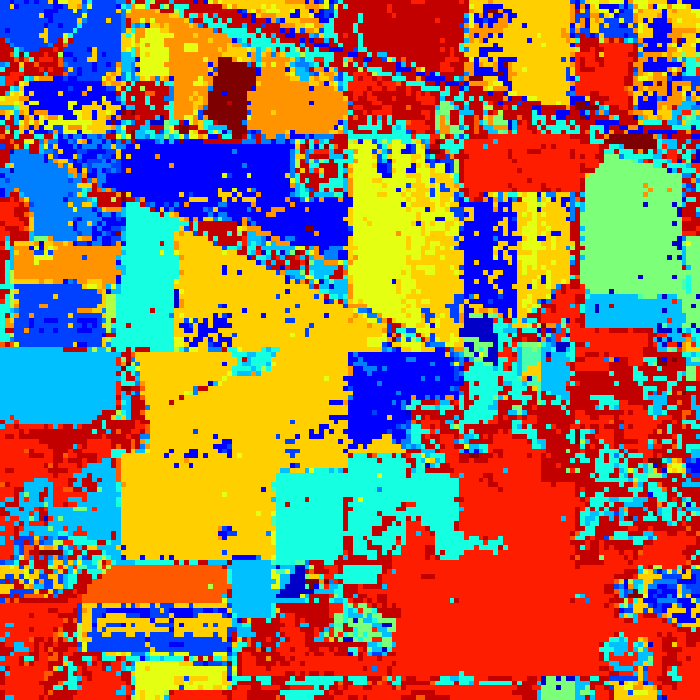} \\
[8pt]
\HUGESS\textit{a}) 1PC &
\HUGESS\textit{b}) GT &
\HUGESS\textit{c}) MLR  & 
\HUGESS\textit{d}) SVM  & 
\HUGESS\textit{e}) MLP  &
\HUGESS\textit{f}) RNN  \\
[8pt]
\includegraphics[]{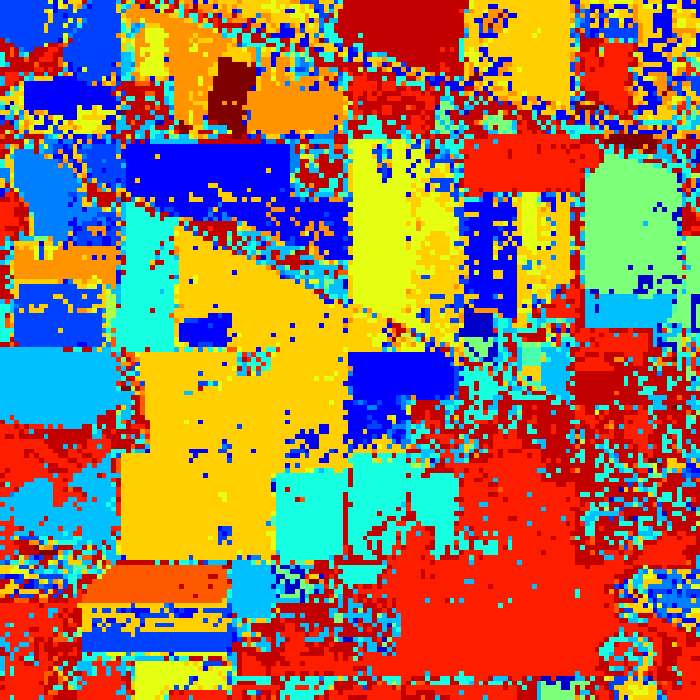} &
\includegraphics[]{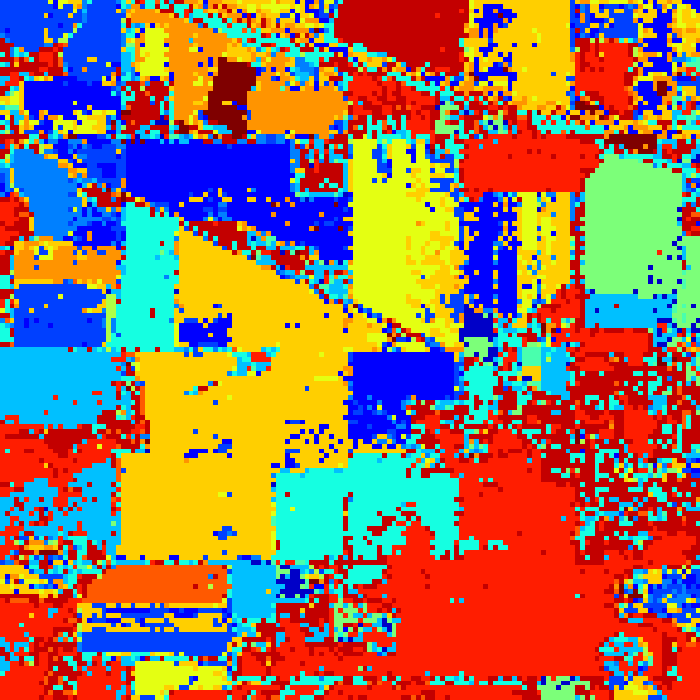} &
\includegraphics[]{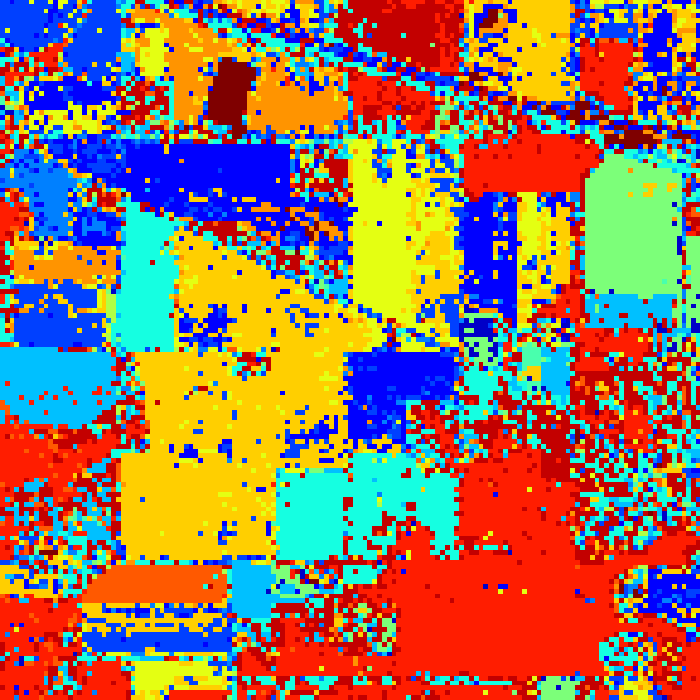} &
\includegraphics[]{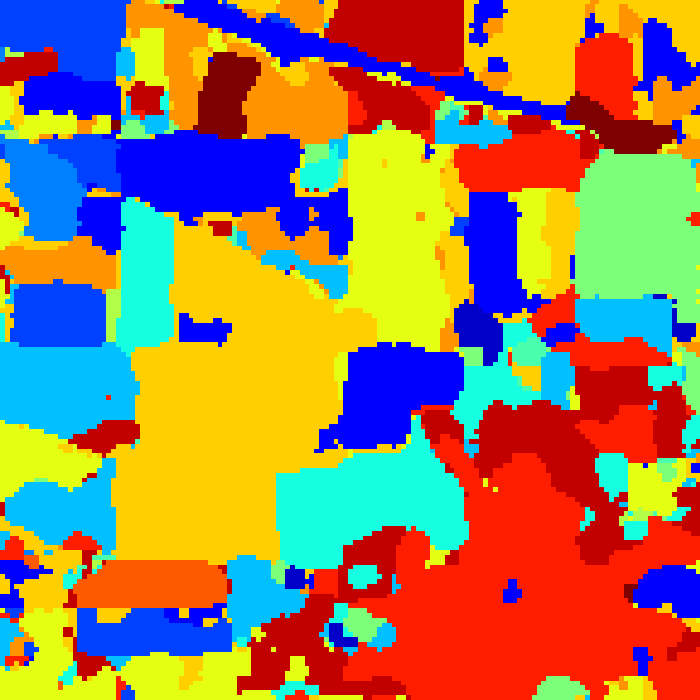} &
\includegraphics[]{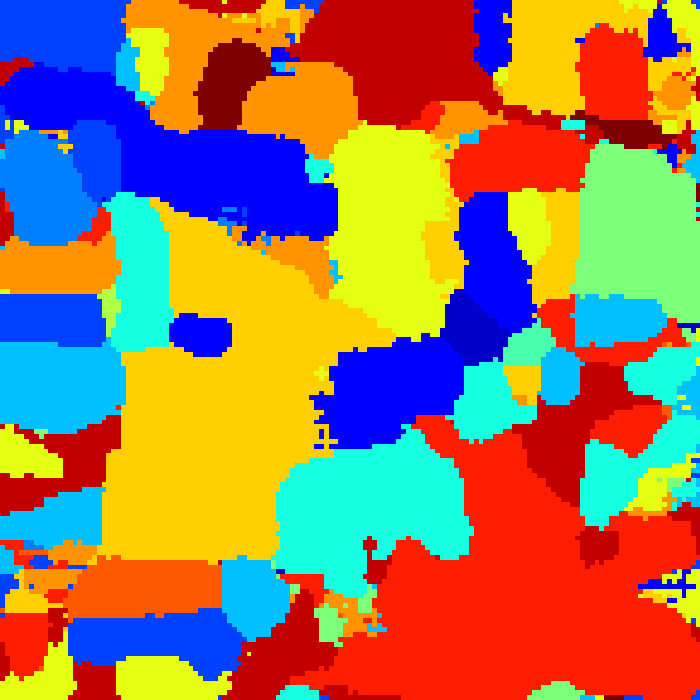} &
\includegraphics[]{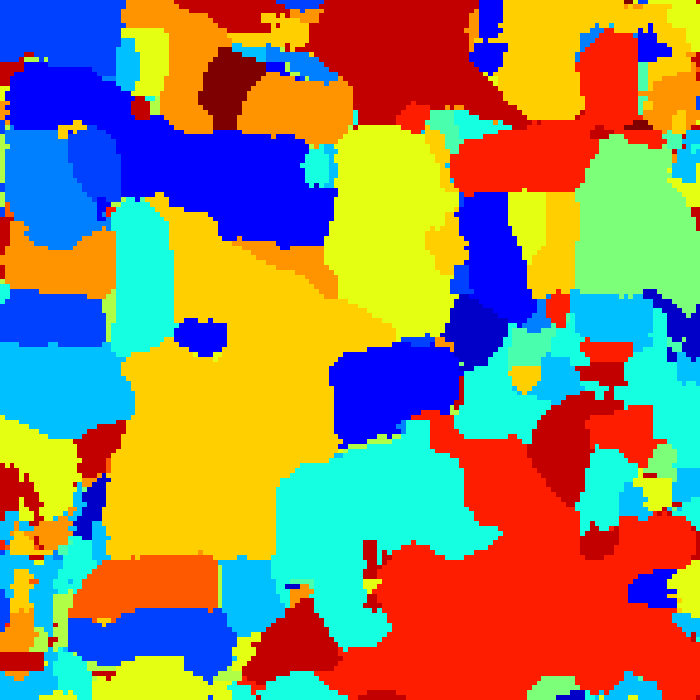} \\
[8pt]
\HUGESS\textit{g}) LSTM  &
\HUGESS\textit{h}) GRU  &
\HUGESS\textit{i}) CNN1D  &
\HUGESS\textit{j}) CNN2D  &
\HUGESS\textit{k}) CNN3D  & 
\HUGESS\textit{l}) MorphCNN \\
\end{tabular}}
\caption{Classification Maps obtained by MLR \cite{li2010semisupervised}, SVM \cite{melgani2004classification}, MLP \cite{paoletti2019deep}, RNN \cite{hang2019cascaded}, LSTM \cite{hochreiter1997long}, GRU \cite{cho2014properties}, CNN-1D \cite{hong2021graph}, CNN-2D \cite{makantasis2015deep}, CNN-3D \cite{hamida2018deep} and MorphCNN \cite{roy2021morphological} on the disjoint train-test dataset for the IP scene.}
\label{fig:comparativeDIP}
\end{figure*}

\begin{table*}[!t]
\let\center\empty
\let\endcenter\relax
\centering
\caption{Classification results obtained by RF \cite{ham2005investigation}, MLR \cite{li2010semisupervised}, SVM \cite{melgani2004classification}, MLP \cite{paoletti2019deep}, RNN \cite{hang2019cascaded}, LSTM \cite{hochreiter1997long}, GRU \cite{cho2014properties}, CNN-1D \cite{hong2021graph}, CNN-2D \cite{makantasis2015deep}, CNN-3D \cite{hamida2018deep}, HybridSN \cite{roy2019hybridsn}, and MorphCNN \cite{roy2021morphological} on the disjoint train-test dataset for the UH scene.}
\resizebox{\linewidth}{!}{\color{black}
\begin{tabular}{c|cccccccccccc}
\hline
\textbf{Class} & RF \cite{ham2005investigation} & MLR \cite{li2010semisupervised} & SVM \cite{melgani2004classification}& MLP \cite{paoletti2019deep}& RNN \cite{hang2019cascaded}& LSTM \cite{hochreiter1997long}& GRU \cite{cho2014properties}& CNN-1D \cite{hong2021graph}& CNN-2D \cite{makantasis2015deep}& CNN-3D \cite{hamida2018deep}& HybridSN \cite{roy2019hybridsn} & MorphCNN \cite{roy2021morphological} \\ \hline

1 & 82.87$\pm$0.04 & 82.24$\pm$0.06 & 82.34$\pm$0.0 & 81.23$\pm$0.28 & 82.22$\pm$0.28 & \textbf{82.76}$\pm$0.36 & 82.58$\pm$0.35 & 82.28$\pm$0.98 & 82.25$\pm$0.65 & 82.1$\pm$0.39 & 82.74$\pm$0.29 & 82.43$\pm$0.33 \\

2 & 82.51$\pm$0.38 & 82.5$\pm$0.07 & 83.36$\pm$0.0 & 82.29$\pm$0.55 & 82.87$\pm$0.33 & 80.19$\pm$1.36 & 81.64$\pm$0.71 & \textbf{91.78}$\pm$6.46 & 84.15$\pm$0.28 & 84.14$\pm$0.45 & 90.91$\pm$6.22 & 84.42$\pm$0.19 \\

3 & 64.09$\pm$0.49 & 99.8$\pm$0.0 & 99.8$\pm$0.0 & 99.72$\pm$0.1 & 99.72$\pm$0.2 & 99.68$\pm$0.16 & 99.88$\pm$0.1 & \textbf{99.92}$\pm$0.16 & 90.31$\pm$4.41 & 77.85$\pm$4.8 & 98.81$\pm$0.74 & 97.21$\pm$1.23 \\

4 & 92.04$\pm$0.0 & 98.3$\pm$0.0 & \textbf{98.96}$\pm$0.0 & 87.58$\pm$1.28 & 93.5$\pm$2.02 & 91.23$\pm$1.29 & 93.22$\pm$2.84 & 94.36$\pm$3.12 & 87.24$\pm$3.21 & 89.24$\pm$1.1 & 83.96$\pm$1.24 & 92.37$\pm$0.33 \\

5 & 99.81$\pm$0.07 & 97.44$\pm$0.0 & 98.77$\pm$0.0 & 97.35$\pm$0.49 & 97.76$\pm$0.29 & 97.65$\pm$0.31 & 97.37$\pm$0.16 & 98.77$\pm$0.13 & 99.51$\pm$0.48 & 98.97$\pm$0.59 & 99.46$\pm$0.75 & \textbf{99.77}$\pm$0.53 \\

6 & 96.27$\pm$0.32 & 94.41$\pm$0.0 & 97.9$\pm$0.0 & 94.55$\pm$0.28 & 95.1$\pm$0.0 & 97.06$\pm$1.79 & 98.32$\pm$1.63 & 95.8$\pm$1.88 & 96.43$\pm$2.14 & 98.91$\pm$1.44 & 98.60$\pm$1.97 & \textbf{99.46}$\pm$1.15 \\

7 & 86.19$\pm$0.34 & 73.37$\pm$0.07 & 77.43$\pm$0.0 & 75.24$\pm$2.27 & 81.4$\pm$0.43 & 78.88$\pm$1.0 & 77.03$\pm$2.18 & 82.78$\pm$2.23 & 86.44$\pm$2.18 & 85.48$\pm$1.98 & 75.62$\pm$3.89 & \textbf{88.07}$\pm$1.78 \\

8 & 41.69$\pm$0.23 & 63.82$\pm$0.0 & 60.3$\pm$0.0 & 57.0$\pm$6.97 & 40.06$\pm$1.07 & 40.11$\pm$1.92 & 53.62$\pm$2.97 & \textbf{75.5}$\pm$6.71 & 70.03$\pm$3.96 & 62.06$\pm$3.01 & 93.16$\pm$0.20 & 73.09$\pm$3.5 \\

9 & 86.02$\pm$0.48 & 70.23$\pm$0.04 & 76.77$\pm$0.0 & 75.58$\pm$2.86 & 76.54$\pm$2.96 & 81.55$\pm$4.12 & 79.06$\pm$1.61 & 81.44$\pm$2.0 & 79.53$\pm$6.38 & 80.81$\pm$4.32 & 81.39$\pm$5.24 & \textbf{84.09}$\pm$2.73 \\

10 & 36.00$\pm$0.0 & 55.6$\pm$0.0 & 61.29$\pm$0.0 & 48.78$\pm$2.27 & 47.44$\pm$1.44 & 47.37$\pm$2.29 & 49.54$\pm$2.61 & \textbf{68.71}$\pm$14.55 & 60.22$\pm$4.2 & 54.75$\pm$4.63 & 76.51$\pm$10.40 & 62.86$\pm$3.08 \\

11 & 64.67$\pm$0.16 & 74.21$\pm$0.04 & 80.55$\pm$0.0 & 76.25$\pm$0.46 & 76.24$\pm$0.81 & 76.38$\pm$1.09 & 80.82$\pm$0.71 & 85.24$\pm$2.83 & 82.93$\pm$7.68 & 66.78$\pm$3.34 & 89.21$\pm$5.62 & \textbf{89.15}$\pm$6.86 \\

12 & 67.27$\pm$0.09 & 70.41$\pm$0.0 & 79.92$\pm$0.0 & 75.31$\pm$3.75 & 76.33$\pm$3.09 & 79.98$\pm$3.32 & 84.15$\pm$3.13 & 89.93$\pm$4.29 & 92.87$\pm$3.31 & \textbf{93.83}$\pm$1.92 & 96.28$\pm$2.29 & 93.02$\pm$3.32 \\

13 & 89.23$\pm$0.43 & 67.72$\pm$0.0 & 70.88$\pm$0.0 & 73.19$\pm$2.15 & 69.12$\pm$1.61 & 71.37$\pm$3.54 & 72.63$\pm$3.68 & 74.88$\pm$5.14 & 86.21$\pm$2.65 & 82.34$\pm$2.49 & 86.78$\pm$6.67 & \textbf{89.61}$\pm$1.34 \\

14 & 100.0$\pm$0.0 & 98.79$\pm$0.0 & 100.0$\pm$0.0 & 99.84$\pm$0.32 & 100.0$\pm$0.0 & 99.11$\pm$0.47 & 99.92$\pm$0.16 & 99.68$\pm$0.16 & 98.92$\pm$1.8 & 96.31$\pm$3.67 & \textbf{100.0}$\pm$0.0 & 99.19$\pm$1.3 \\

15 & 90.06$\pm$0.45 & 95.56$\pm$0.0 & 96.41$\pm$0.0 & 97.8$\pm$0.51 & 97.59$\pm$0.47 & 98.14$\pm$0.31 & 98.22$\pm$0.59 & 98.48$\pm$0.24 & 77.63$\pm$2.91 & 75.85$\pm$2.69 & \textbf{100.0}$\pm$0.0 & 97.04$\pm$4.47 \\ \hline\hline

OA & 75.38$\pm$0.06 & 78.97$\pm$0.01 & 81.86$\pm$0.0 & 78.22$\pm$0.36 & 77.95$\pm$0.68 & 78.16$\pm$0.28 & 80.21$\pm$0.27 & 86.42$\pm$1.64 & 83.27$\pm$0.8 & 80.24$\pm$0.55 & 88.31$\pm$1.78 & \textbf{86.51}$\pm$0.71 \\

AA & 78.58$\pm$0.11 & 81.63$\pm$0.01 & 84.31$\pm$0.0 & 81.45$\pm$0.37 & 81.06$\pm$0.55 & 81.43$\pm$0.32 & 83.2$\pm$0.27 & 87.97$\pm$1.38 & 84.98$\pm$0.74 & 81.96$\pm$0.75 & 90.23$\pm$1.39 & \textbf{88.78}$\pm$0.68 \\

k(x100) & 73.49$\pm$0.07 & 77.3$\pm$0.01 & 80.43$\pm$0.0 & 76.55$\pm$0.39 & 76.23$\pm$0.71 & 76.52$\pm$0.3 & 78.66$\pm$0.29 & 85.27$\pm$1.77 & 81.89$\pm$0.86 & 78.62$\pm$0.59 & 87.33$\pm$1.92 & \textbf{85.4}$\pm$0.76 \\ \hline

\end{tabular}}
\label{table:compdisjointUH}
\end{table*}

\begin{figure*}[!t]
\resizebox{\textwidth}{!}{
\begin{tabular}{ccc}
\includegraphics{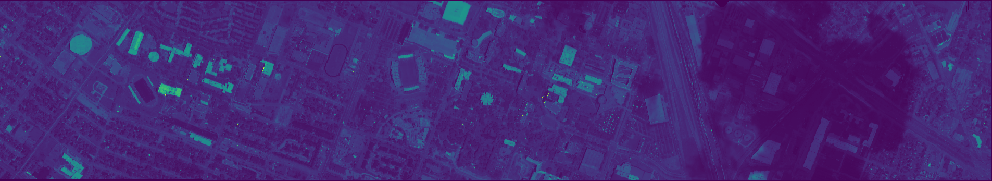} &
\includegraphics{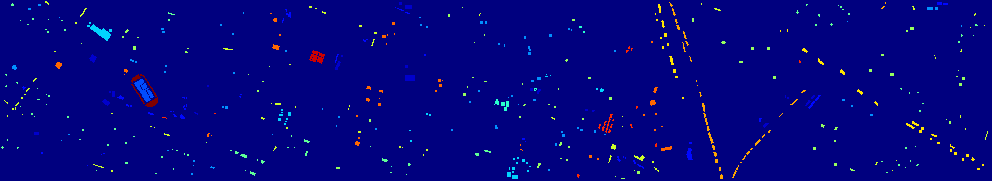} &
\includegraphics{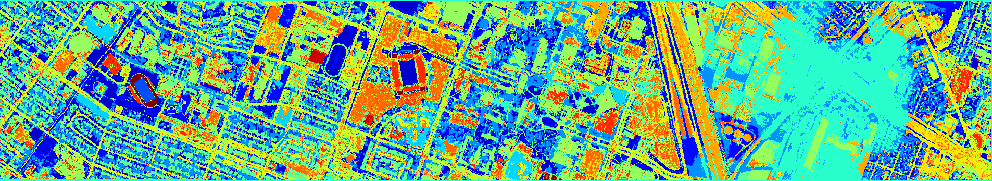} \\
[8pt]
\HUGESSS\textit{a}) 1PC &
\HUGESSS\textit{b}) GT & 
\HUGESSS\textit{c}) MLR \\
[8pt]
\includegraphics{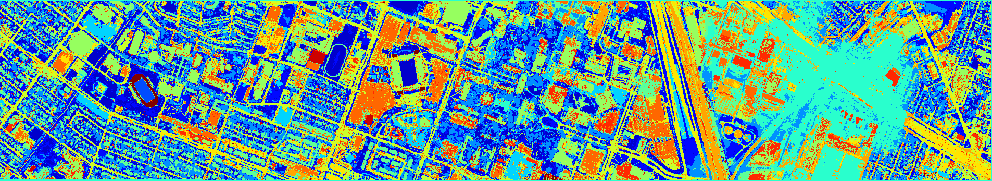} &
\includegraphics{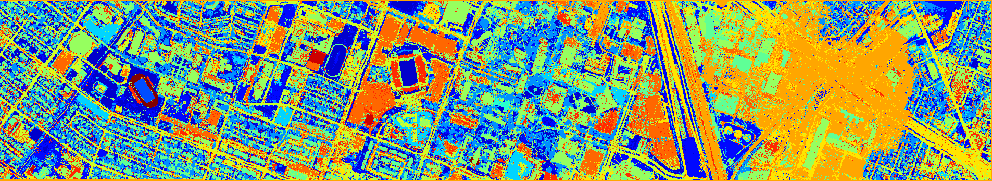} &
\includegraphics{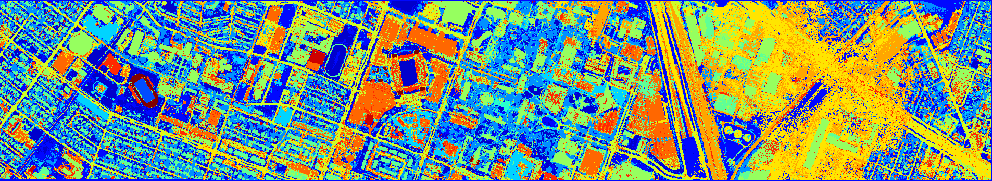} \\
[8pt]
\HUGESSS\textit{d}) SVM  &
\HUGESSS\textit{e}) MLP  &
\HUGESSS\textit{f}) RNN  \\
[8pt]
\includegraphics{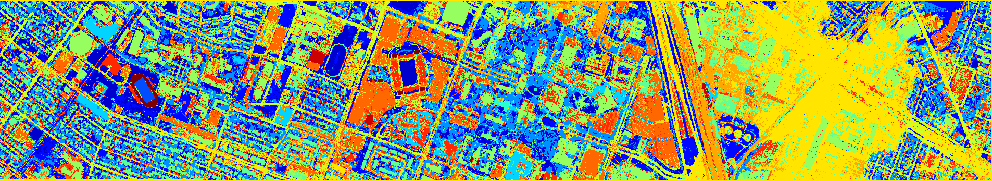} &
\includegraphics{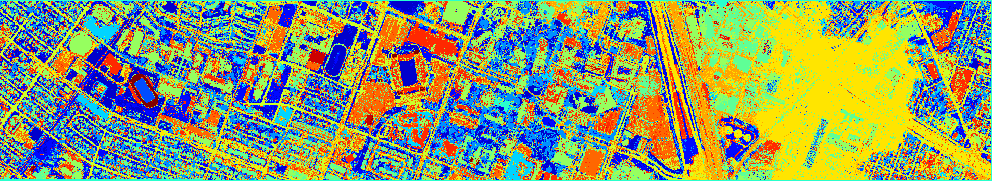} &
\includegraphics{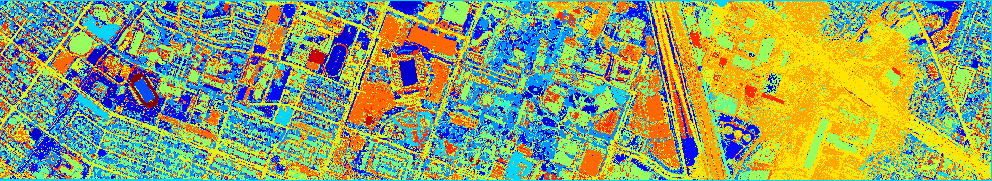} \\
[8pt]
\HUGESSS\textit{g}) LSTM  &
\HUGESSS\textit{h}) GRU  &
\HUGESSS\textit{i}) CNN1D  \\
[8pt]
\includegraphics{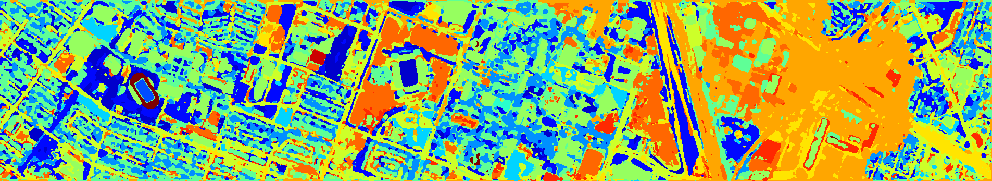} &
\includegraphics{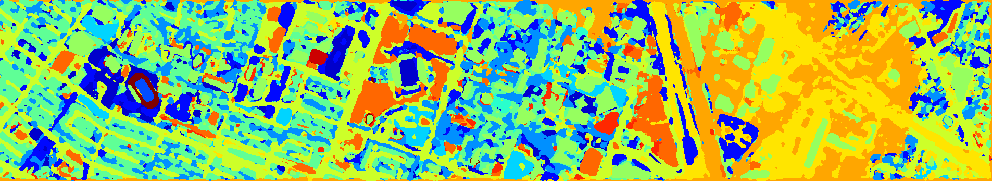} &
\includegraphics{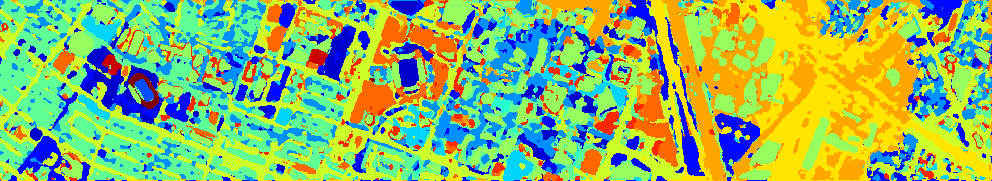} \\
[8pt]
\HUGESSS\textit{j}) CNN2D  &
\HUGESSS\textit{k}) CNN3D  &
\HUGESSS\textit{l}) MorphCNN  \\
\end{tabular}}
\caption{Classification Maps obtained by MLR \cite{li2010semisupervised}, SVM \cite{melgani2004classification}, MLP \cite{paoletti2019deep}, RNN \cite{hang2019cascaded}, LSTM \cite{hochreiter1997long}, GRU \cite{cho2014properties}, CNN-1D \cite{hong2021graph}, CNN-2D \cite{makantasis2015deep}, CNN-3D \cite{hamida2018deep} and MorphCNN \cite{roy2021morphological} on the disjoint train-test dataset for the UH scene.}
\label{fig:comparativeUH}
\end{figure*}

\begin{table*}[!t]
\let\center\empty
\let\endcenter\relax
\centering
\caption{Classification results obtained by RF \cite{ham2005investigation}, MLR \cite{li2010semisupervised}, SVM \cite{melgani2004classification}, MLP \cite{paoletti2019deep}, RNN \cite{hang2019cascaded}, LSTM \cite{hochreiter1997long}, GRU \cite{cho2014properties}, CNN-1D \cite{hong2021graph}, CNN-2D \cite{makantasis2015deep}, CNN-3D \cite{hamida2018deep}, HybridSN \cite{roy2019hybridsn}, and MorphCNN \cite{roy2021morphological} on the disjoint train-test dataset for the KSC scene.}
\resizebox{\linewidth}{!}{\color{black}
\begin{tabular}{c|cccccccccccc} \hline
\textbf{Class} & RF \cite{ham2005investigation} & MLR \cite{li2010semisupervised} & SVM \cite{melgani2004classification}& MLP \cite{paoletti2019deep}& RNN \cite{hang2019cascaded}& LSTM \cite{hochreiter1997long}& GRU \cite{cho2014properties}& CNN-1D \cite{hong2021graph}& CNN-2D \cite{makantasis2015deep}& CNN-3D \cite{hamida2018deep}& HybridSN \cite{roy2019hybridsn} & MorphCNN \cite{roy2021morphological} \\ \hline

1 & 99.69$\pm$0.0 & \textbf{100.0}$\pm$0.00 & 94.13$\pm$0.00 & 99.18$\pm$1.17 & 87.33$\pm$1.73 & 92.22$\pm$0.91 & 89.44$\pm$0.69 & 99.79$\pm$0.14 & 85.52$\pm$15.45 & 97.17$\pm$2.93 & \textbf{100.0}$\pm$0.0 & 97.63$\pm$2.26 \\

2 & 98.38$\pm$0.22 & 99.03$\pm$0.001 & 0.00$\pm$0.00 & 86.63$\pm$7.33 & 63.12$\pm$1.49 & 81.64$\pm$4.17 & 70.85$\pm$0.91 & 99.19$\pm$1.14 & 67.31$\pm$25.41 & 92.91$\pm$2.62 & \textbf{100.0}$\pm$0.0 & 86.79$\pm$9.64 \\

3 & 99.23$\pm$0.57 & 99.54$\pm$0.00 & 54.59$\pm$0.00  & 84.25$\pm$2.70 & 69.72$\pm$2.82 & 75.38$\pm$0.21 & 78.89$\pm$7.79 & 95.11$\pm$3.05 & 60.09$\pm$16.23 & 81.04$\pm$12.05 & \textbf{99.69}$\pm$0.21 & 98.31$\pm$2.37 \\

4 & 88.16$\pm$1.22 & 99.06$\pm$0.002 & 17.28$\pm$0.00  & 78.97$\pm$11.86 & 47.82$\pm$4.20 & 58.09$\pm$1.16 & 44.08$\pm$3.41 & 77.73$\pm$2.81 & 45.17$\pm$9.84 & 44.54$\pm$16.19 & \textbf{99.53}$\pm$0.66 & 88.94$\pm$15.64 \\

5 & 73.72$\pm$0.0 & \textbf{100.0}$\pm$0.00 & 0.00$\pm$0.00 & 13.38$\pm$18.92 & 68.37$\pm$5.63 & 74.21$\pm$5.20 & 65.21$\pm$3.28 & 80.53$\pm$4.85 & 67.40$\pm$12.84 & 85.15$\pm$6.98 & 98.78$\pm$0.34 & 48.66$\pm$34.29 \\

6 & 88.88$\pm$0.24 & \textbf{100.0}$\pm$0.00 & 0.00$\pm$0.00  & 78.12$\pm$8.79 & 56.24$\pm$1.97 & 65.12$\pm$6.03 & 59.82$\pm$2.30 & 91.97$\pm$1.74 & 65.47$\pm$29.63 & 62.74$\pm$15.45 & \textbf{100.0}$\pm$0.0 & 86.32$\pm$16.43 \\

7 & \textbf{100.0}$\pm$0.0 & 89.88$\pm$0.00 & 0.00$\pm$0.00  & 78.65$\pm$3.99 & 83.52$\pm$8.91 & 90.26$\pm$1.40 & 89.14$\pm$3.47 & 95.13$\pm$1.40 & 77.15$\pm$28.34 & 80.52$\pm$17.07 & 97.75$\pm$1.83 & 97.75$\pm$3.17 \\

8 & 85.51$\pm$0.22 & \textbf{100.0}$\pm$0.00 & 60.10$\pm$0.00  & 89.62$\pm$8.25 & 65.57$\pm$2.32 & 71.40$\pm$2.99 & 69.76$\pm$2.53 & 97.45$\pm$0.51 & 64.75$\pm$13.87 & 71.49$\pm$11.96 & 99.90$\pm$0.12 & 70.76$\pm$34.44 \\

9 & 96.68$\pm$0.42 & \textbf{100.0}$\pm$0.00 & 89.37$\pm$0.00 & 97.59$\pm$1.79 & 88.39$\pm$3.35 & 90.72$\pm$2.93 & 86.72$\pm$1.36 & 99.92$\pm$0.11 & 89.22$\pm$8.06 & 98.94$\pm$1.33 & \textbf{100.0}$\pm$0.0 & 91.93$\pm$7.16 \\

10 & 99.22$\pm$0.13 & \textbf{100.0}$\pm$0.00 & 98.83$\pm$0.00 & 96.50$\pm$3.33 & 92.42$\pm$3.12 & 88.92$\pm$3.37 & 88.53$\pm$1.58 & 99.90$\pm$0.13 & 73.08$\pm$23.37 & 90.67$\pm$8.13 & \textbf{100.0}$\pm$0.0 & \textbf{100.0}$\pm$0.00 \\

11 & \textbf{100.0}$\pm$0.0 & 98.03$\pm$0.001 & 94.94$\pm$0.00 & 98.50$\pm$0.86 & 83.89$\pm$2.29 & 90.26$\pm$1.08 & 84.83$\pm$3.64 & \textbf{100.0}$\pm$0.0 & 87.55$\pm$10.06 & 97.56$\pm$1.17 & 96.34$\pm$0.79 & \textbf{100.0}$\pm$0.0 \\

12 & 97.89$\pm$0.19 & \textbf{99.29}$\pm$0.001 & 89.25$\pm$0.00 & 98.52$\pm$0.79 & 81.31$\pm$4.48 & 87.46$\pm$2.05 & 83.57$\pm$4.87 & 98.36$\pm$1.06 & 82.48$\pm$19.17 & 99.30$\pm$0.99 & 99.06$\pm$0.99 & 97.89$\pm$2.06 \\

13 & \textbf{100.0}$\pm$0.0 & \textbf{100.0}$\pm$0.0 & \textbf{100.0}$\pm$0.0 & \textbf{100.0}$\pm$0.0 & 99.88$\pm$0.10 & \textbf{100.0}$\pm$0.0 & 99.92$\pm$0.05 & \textbf{100.0}$\pm$0.0 & 99.92$\pm$0.12 & \textbf{100.0}$\pm$0.0 & \textbf{100.0}$\pm$0.0 & \textbf{100.0}$\pm$0.0 \\ \hline\hline

OA & 96.17$\pm$0.07 & 99.45$\pm$0.001 & 72.84$\pm$0.00  & 91.76$\pm$0.56 & 81.47$\pm$1.17 & 86.10$\pm$0.40 & 82.76$\pm$0.72 & 97.18$\pm$0.18 & 79.98$\pm$13.38 & 89.71$\pm$1.30 & \textbf{99.48}$\pm$0.05 & 92.76$\pm$2.08 \\

AA & 94.41$\pm$0.08 & 98.83$\pm$0.001 & 53.73$\pm$0.00 & 84.61$\pm$0.62 & 75.96$\pm$1.71 & 81.97$\pm$0.34 & 77.75$\pm$0.70 & 95.00$\pm$0.34 & 74.24$\pm$16.27 & 84.77$\pm$2.60 & \textbf{99.31}$\pm$0.17 & 89.61$\pm$1.60 \\

k(x100) & 95.74$\pm$0.08 & 99.40$\pm$0.001 & 69.29$\pm$0.00 & 90.82$\pm$0.62 & 79.33$\pm$1.30 & 84.51$\pm$0.45 & 80.79$\pm$0.80 & 96.86$\pm$0.20 & 77.63$\pm$14.99 & 88.51$\pm$1.46 & \textbf{99.43}$\pm$0.06 & 91.94$\pm$2.31 \\ \hline

\end{tabular}}
\label{table:compdisjointKSC}
\end{table*}

\begin{figure*}[!t]
\resizebox{\textwidth}{!}{
\begin{tabular}{cccccc}
\includegraphics[clip=true, trim = 0 0 0 0, width=0.56\linewidth]{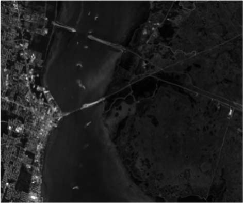} &
\includegraphics[]{gt/KSC.png} &
\includegraphics[]{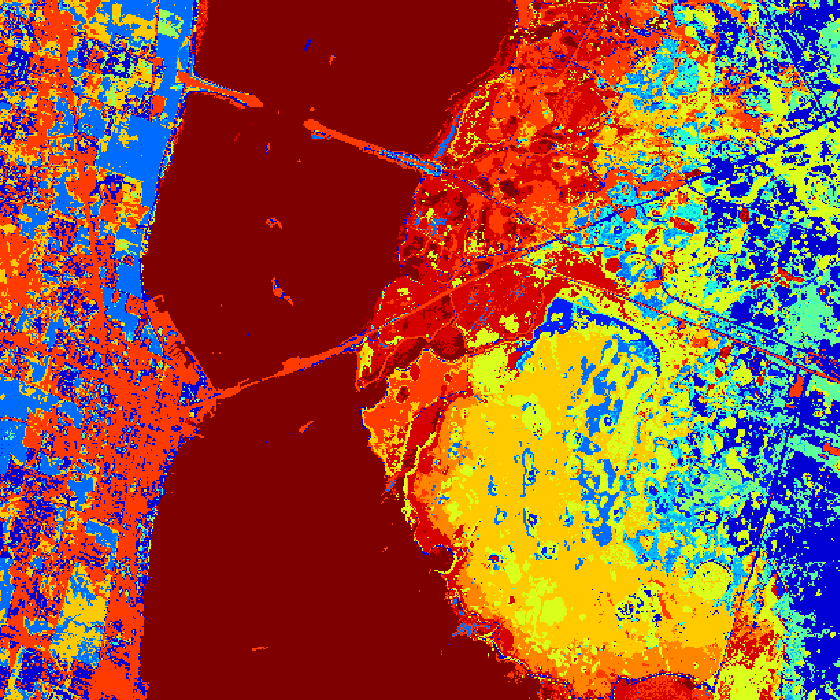} &
\includegraphics[]{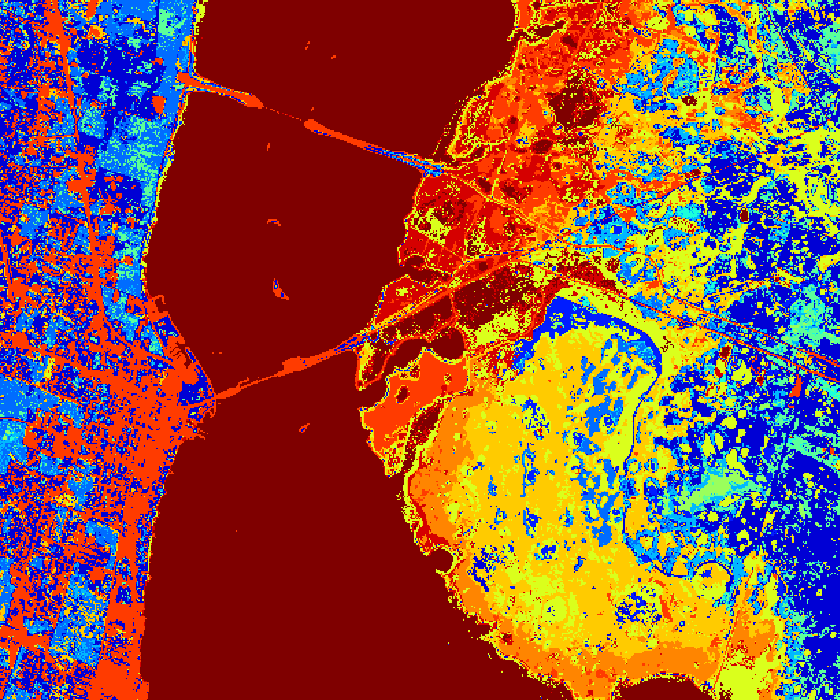} &
\includegraphics[]{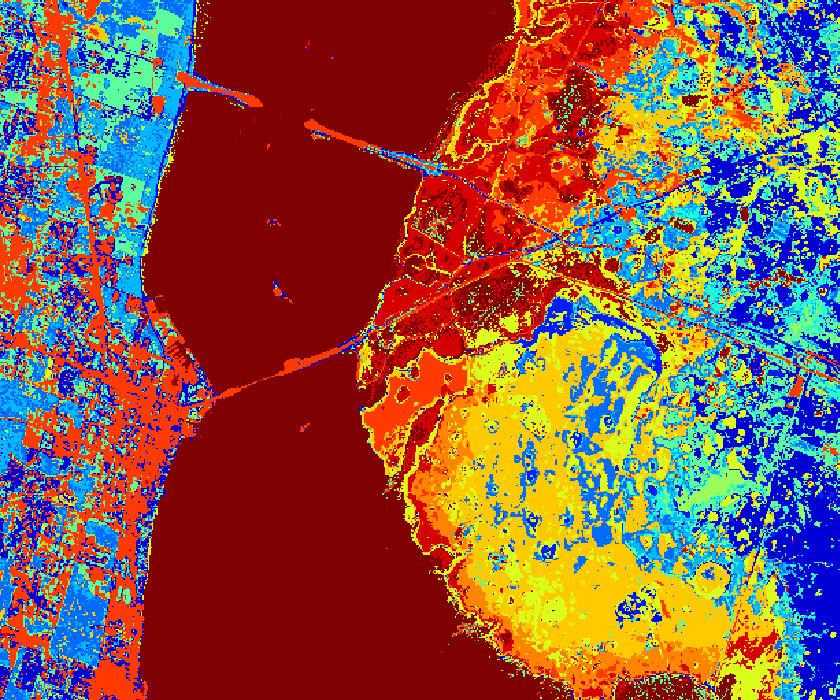} &
\includegraphics[]{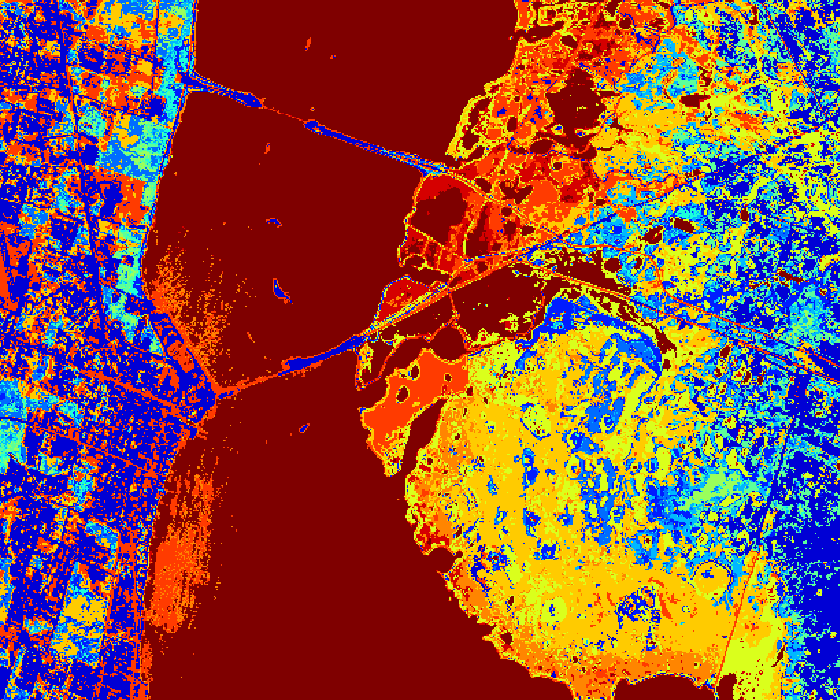} \\
[8pt]
\HUGESS\textit{a}) 1PC &
\HUGESS\textit{b}) GT &
\HUGESS\textit{c}) MLR  & 
\HUGESS\textit{d}) SVM  & 
\HUGESS\textit{e}) MLP  &
\HUGESS\textit{f}) RNN  \\
[8pt]
\includegraphics[]{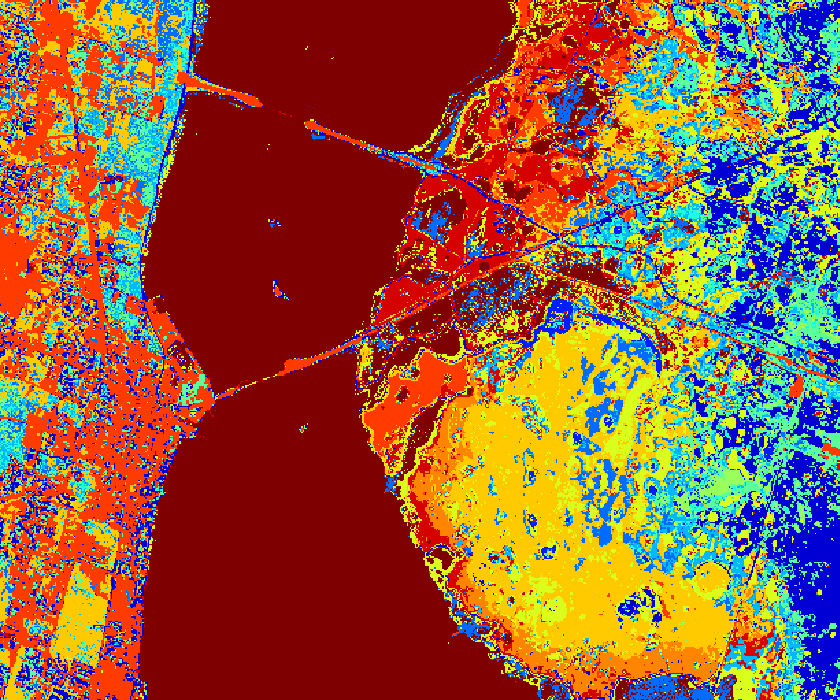} &
\includegraphics[]{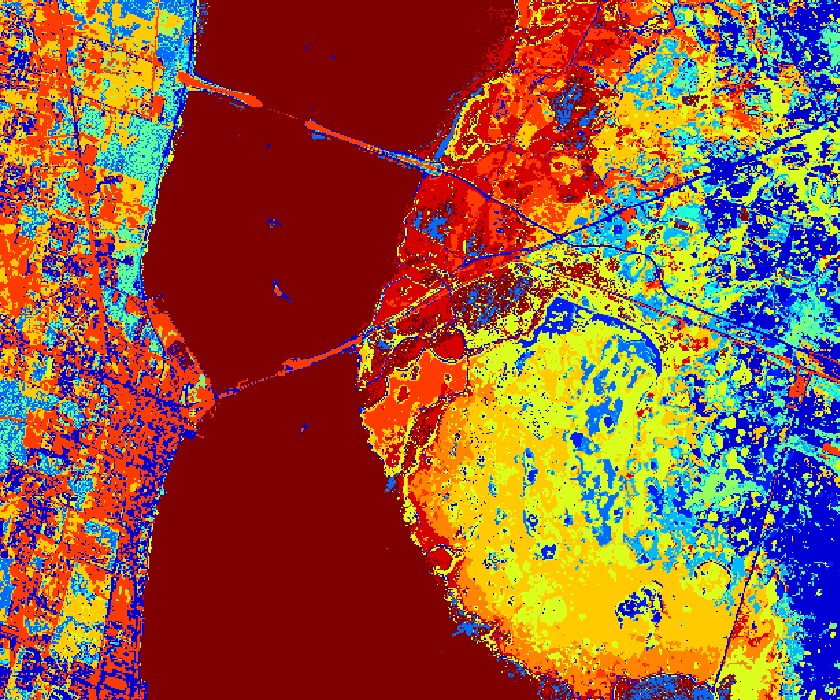} &
\includegraphics[]{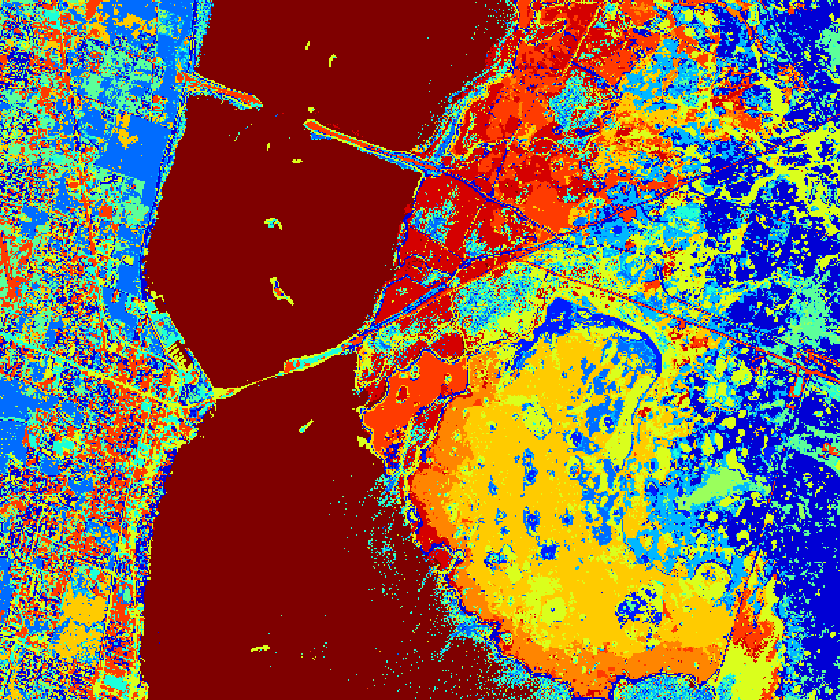} &
\includegraphics[]{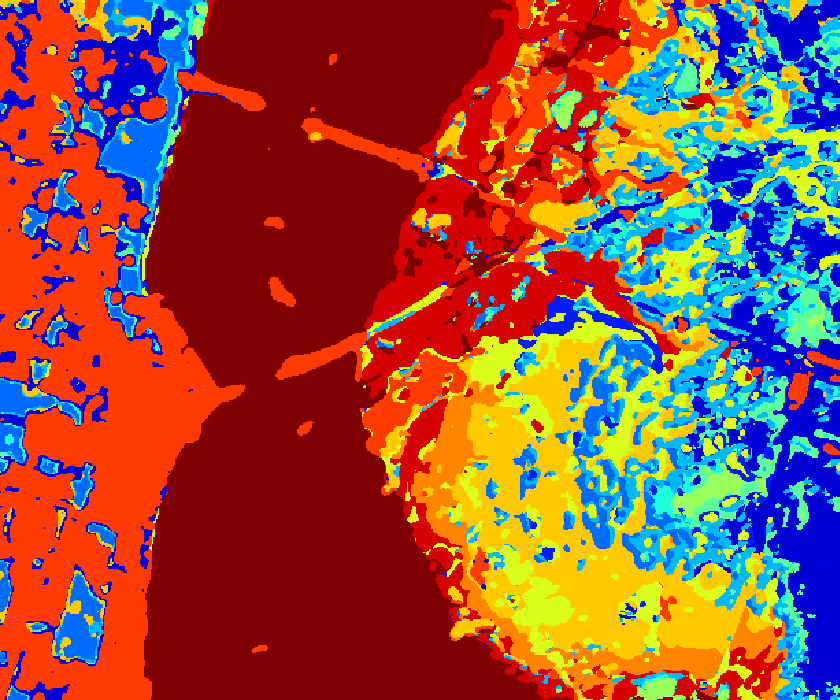} &
\includegraphics[]{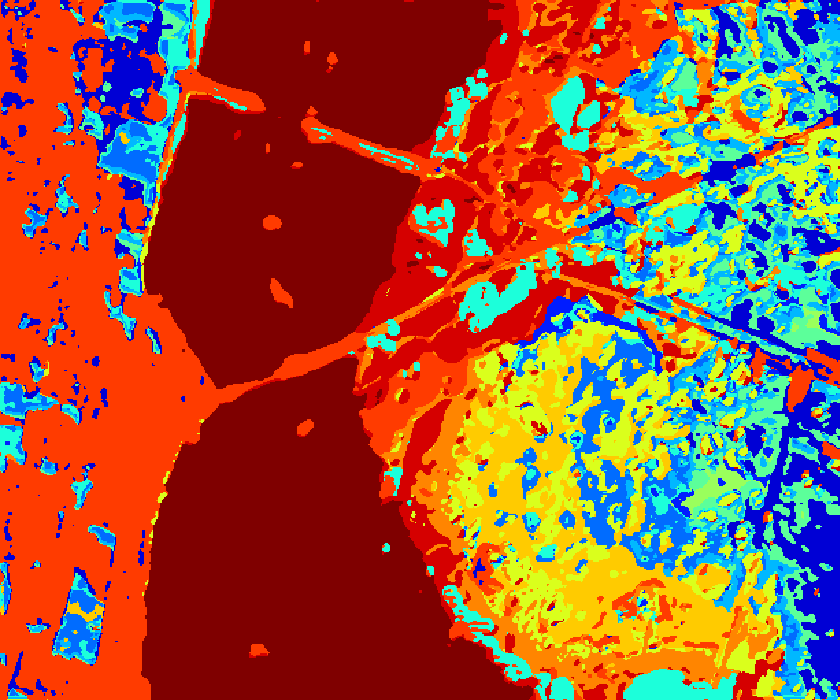} &
\includegraphics[]{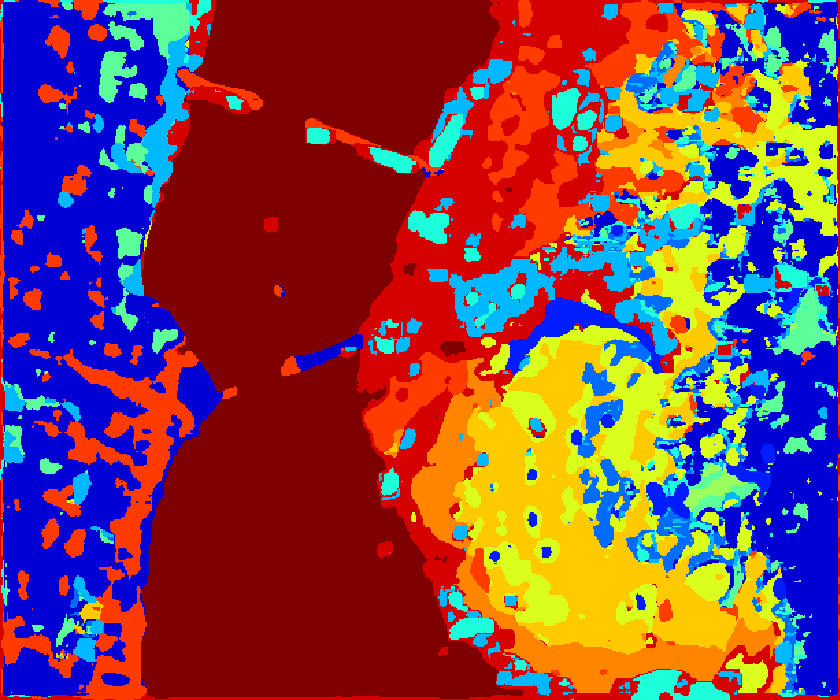} \\
[8pt]
\HUGESS\textit{g}) LSTM  &
\HUGESS\textit{h}) GRU  &
\HUGESS\textit{i}) CNN1D  &
\HUGESS\textit{j}) CNN2D  &
\HUGESS\textit{k}) CNN3D  & 
\HUGESS\textit{l}) MorphCNN  \\
\end{tabular}}
\caption{Classification Maps obtained by MLR \cite{li2010semisupervised}, SVM \cite{melgani2004classification}, MLP \cite{paoletti2019deep}, RNN \cite{hang2019cascaded}, LSTM \cite{hochreiter1997long}, GRU \cite{cho2014properties}, CNN-1D \cite{hong2021graph}, CNN-2D \cite{makantasis2015deep}, CNN-3D \cite{hamida2018deep} and MorphCNN \cite{roy2021morphological} on the disjoint train-test dataset for the KSC scene.}
\label{fig:comparativeDKSC}
\end{figure*}

Here we have enlisted the experimental results with detailed discussion on the obtained results. The obtained accuracies for disjoint training and test samples are shown in Tables \ref{table:compdisjointUP}, \ref{table:compdisjointIP}, \ref{table:compdisjointUH} and \ref{table:compdisjointKSC} and Figures \ref{fig:comparativeDUP}, \ref{fig:comparativeDIP} \ref{fig:comparativeUH}, and \ref{fig:comparativeDKSC}. All the results shown in the Tables and Figures are obtained using the 10-cross-validation process to compute the overall, average and kappa $(\kappa)$ accuracy for comparison purposes. For instance, let us assume the case of Pavia University results, for this particular case, the work \cite{roy2021morphological} has the highest average, overall and kappa $(\kappa)$ accuracies which are 95.51\%, 93.95\%, and 93.95\% respectively in comparison with the average, overall and kappa $(\kappa)$ accuracies for other comparative works; 92.55\%, 89.94\%, 89.9\% for \cite{makantasis2015deep}, 89.43\%, 86.25\%, 85.61\% for \cite{hamida2018deep}, 89.09\%, 89.5\%, 85.5\% for \cite{hong2021graph}, 82.05\%, 87.43\%, 76.89\% for \cite{paoletti2019deep}, 80.38\%, 83.63\%, 74.76\% for \cite{cho2014properties}, 80.38\%, 84.06\%, 74.32\% for \cite{hochreiter1997long}, 77.8\%, 86.12\%, 72.06\% for \cite{melgani2004classification}, 77.07\%, 83.83\%, 70.84\% for \cite{hang2019cascaded}, and 72.23\%, 82.12\%, 65.44\% for \cite{li2010semisupervised}. Similar observations can be made of the other experimental datasets.

\begin{table*}[!t]
\let\center\empty
\let\endcenter\relax
\centering
\caption{Classification results obtained by RF \cite{ham2005investigation}, MLR \cite{li2010semisupervised}, SVM \cite{melgani2004classification}, MLP \cite{paoletti2019deep}, RNN \cite{hang2019cascaded}, LSTM \cite{hochreiter1997long}, GRU \cite{cho2014properties}, CNN-1D \cite{hong2021graph}, CNN-2D \cite{makantasis2015deep}, CNN-3D \cite{hamida2018deep}, HybridSN \cite{roy2019hybridsn}, and MorphCNN \cite{roy2021morphological} on the disjoint train-test dataset for the University of Trento (UT) scene.}
\resizebox{\linewidth}{!}{\color{black}
\begin{tabular}{c|ccccccccccccccccccc} \hline
\textbf{Class} & RF \cite{ham2005investigation} & MLR \cite{li2010semisupervised} & SVM \cite{melgani2004classification}& MLP \cite{paoletti2019deep}& RNN \cite{hang2019cascaded}& LSTM \cite{hochreiter1997long}& GRU \cite{cho2014properties}& CNN-1D \cite{hong2021graph}& CNN-2D \cite{makantasis2015deep}& CNN-3D \cite{hamida2018deep}& HybridSN \cite{roy2019hybridsn} & MorphCNN \cite{roy2021morphological} \\ \hline 

1 & 97.27$\pm$0.33 & 92.57$\pm$0.00 & 95.03$\pm$0.00 & 70.68$\pm$3.69 & 93.46$\pm$3.67 & 95.71$\pm$0.32 & 95.78$\pm$0.97 & 98.06$\pm$0.72 & 96.35$\pm$0.22 & 92.64$\pm$7.45 & \textbf{99.15$\pm$0.16} & 97.83$\pm$0.37 
\\

2 & 89.50$\pm$0.11 & \textbf{90.92$\pm$0.00} & 88.66$\pm$0.00 & 76.22$\pm$2.90 & 83.09$\pm$1.50 & 84.73$\pm$0.87 & 84.40$\pm$1.51 & 89.16$\pm$1.99 & 85.87$\pm$1.87 & 75.70$\pm$7.23 & 81.65$\pm$2.93 & 90.41$\pm$1.83 
\\

3 & 75.04$\pm$0.33 & 90.10$\pm$0.00 & 91.71$\pm$0.00 & 83.95$\pm$2.08 & 69.51$\pm$0.87 & 59.26$\pm$14.31 & 65.15$\pm$4.84 & 70.67$\pm$8.99 & 66.84$\pm$2.00 & 60.07$\pm$3.57 & 74.86$\pm$10.58 & 83.86$\pm$10.69 
\\

4 & \textbf{99.96$\pm$0.01} & 92.77$\pm$0.00 & 85.18$\pm$0.00 & 37.55$\pm$2.74 & 98.44$\pm$0.97 & 97.72$\pm$1.41 & 99.65$\pm$0.18 & 99.86$\pm$0.11 & 98.67$\pm$0.61 & 98.88$\pm$0.80 & 99.64$\pm$0.28 & 99.06$\pm$0.57 
\\

5 & \textbf{99.97$\pm$0.00} & 99.14$\pm$0.00 & 97.76$\pm$0.00 & 99.98$\pm$.01 & 98.12$\pm$0.56 & 96.34$\pm$0.80 & 98.20$\pm$0.54 & 99.78$\pm$0.19 & 99.32$\pm$0.39 & 98.40$\pm$1.77 & 98.12$\pm$1.79 & 99.87$\pm$0.14 
\\

6 & 67.71$\pm$0.31 & 66.80$\pm$0.00 & 72.37$\pm$0.00 & 73.91$\pm$2.88 & 65.54$\pm$0.90 & 69.04$\pm$2.29 & 63.89$\pm$1.02 & 77.94$\pm$3.83 & 74.34$\pm$6.48 & 69.42$\pm$3.43 & 70.72$\pm$4.74 & \textbf{82.64$\pm$0.88} 

\\ \hline\hline

OA & 94.95$\pm$0.07 & 92.08$\pm$0.00 & 89.98$\pm$0.00 & 63.51$\pm$9.43 & 92.43$\pm$0.34 & 92.27$\pm$0.72 & 93.03$\pm$0.28 & 95.94$\pm$0.23 & 94.45$\pm$0.13 & 92.14$\pm$1.05 & 94.03$\pm$0.21 & \textbf{96.46$\pm$0.27} 
\\

AA & 88.24$\pm$0.07 & 88.72$\pm$0.00 & 88.45$\pm$0.00 & 63.22$\pm$5.90 & 84.69$\pm$0.51 & 83.80$\pm$1.91 & 84.51$\pm$0.85 & 89.25$\pm$1.34 & 86.90$\pm$0.53 & 82.52$\pm$4.27 & 87.36$\pm$1.83 & \textbf{92.28$\pm$1.60} 
\\ 

k(x100) & 93.23$\pm$0.09 & 89.43$\pm$0.00 & 86.74$\pm$0.00 & 48.58$\pm$13.70 & 89.86$\pm$0.48 & 89.67$\pm$0.95 & 90.66$\pm$0.37 & 94.55$\pm$0.31 & 92.56$\pm$0.18 & 89.45$\pm$1.43 & 92.00$\pm$0.27 & \textbf{95.26$\pm$0.36} 
\\ \hline

\end{tabular}}
\label{table:compdisjointUT}
\end{table*}


The comparative methods mostly misclassify the samples having similar spatial structures (i.e., Meadows and Bare Soil classes for Pavia University dataset) as shown in Table and Figure. Moreover, the overall accuracy for Grapes Untrained is lower than the other classes due to the reasons mentioned above. In a nutshell, one can say that higher accuracy can be achieved by increasing the number of labeled training samples. Thus a higher number of labeled training samples can produce better accuracies for all competing methods.

\begin{figure*}[!t]
\resizebox{\textwidth}{!}{
\begin{tabular}{ccc}
\includegraphics[width=10cm,height=3cm,keepaspectratio]{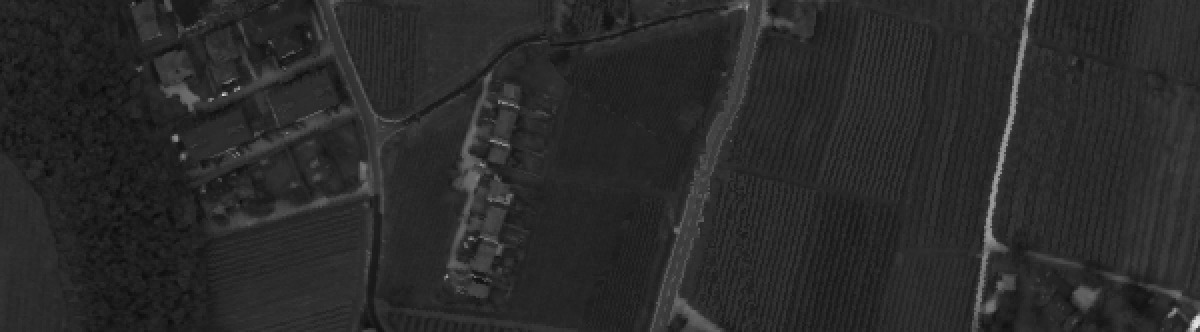} &
\includegraphics{gt/Trento_gt_.png} & \includegraphics{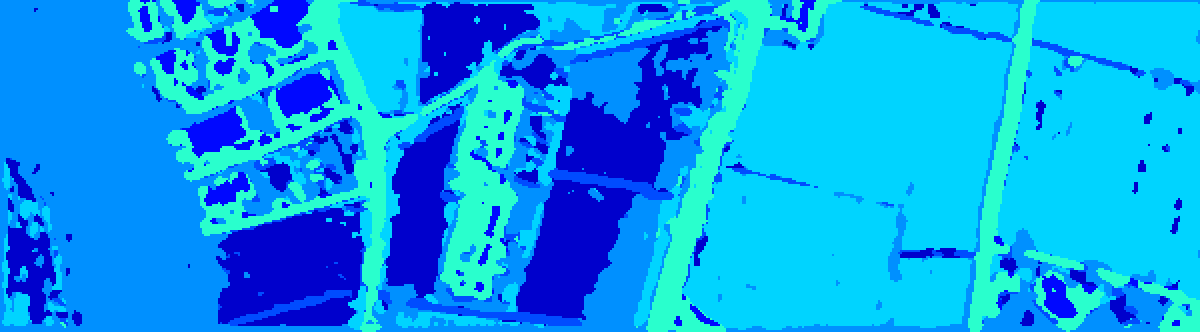}
 \\
[8pt]
\HUGESSS\textit{a}) 1PC &
\HUGESSS\textit{b}) GT & 
\HUGESSS\textit{c}) MLR \\
[8pt]
\includegraphics{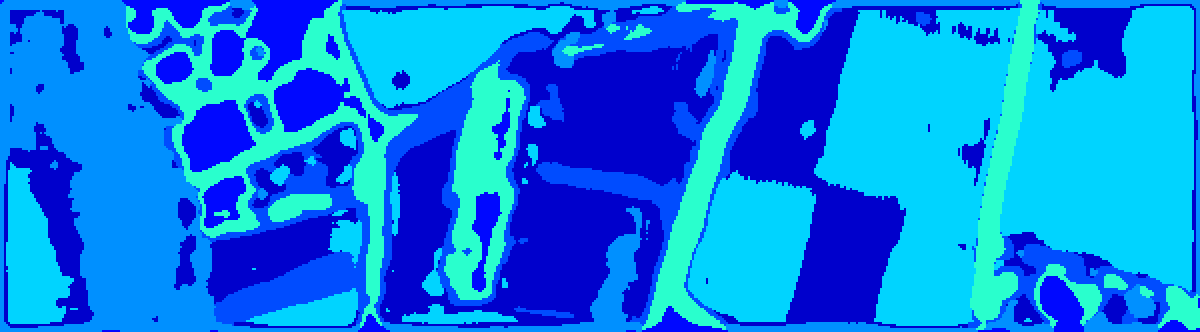} &
\includegraphics{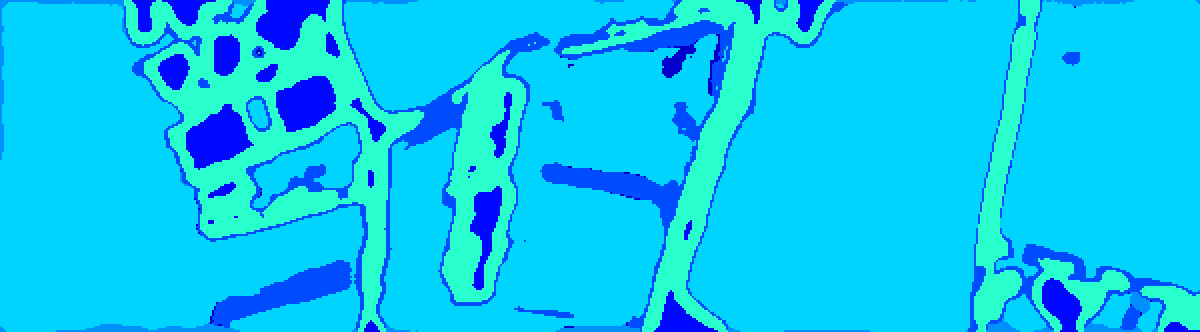}
&
\includegraphics{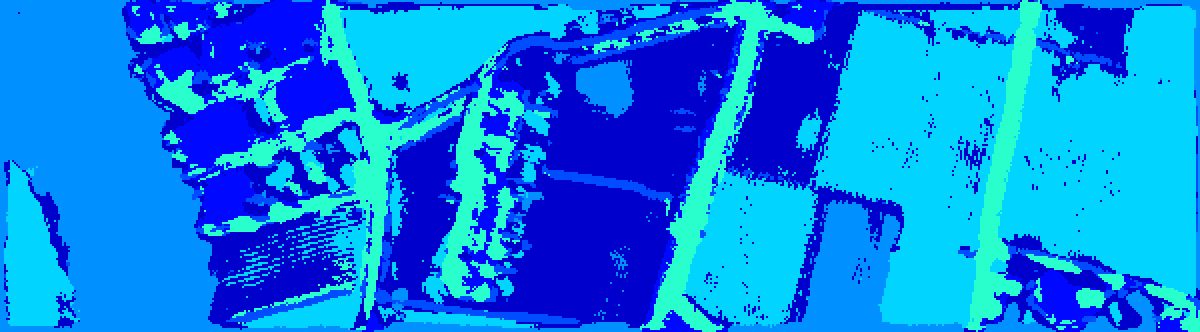}
 \\
[8pt]
\HUGESSS\textit{d}) SVM  &
\HUGESSS\textit{e}) MLP  &
\HUGESSS\textit{f}) RNN  \\
[8pt]
\includegraphics{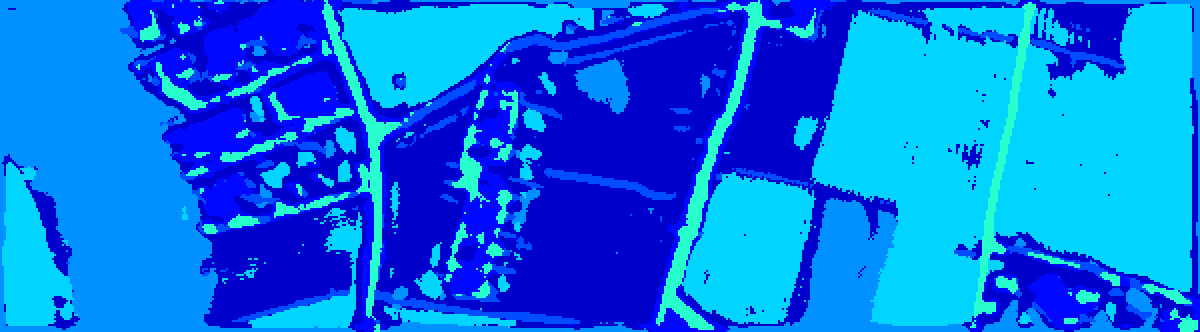} &
\includegraphics{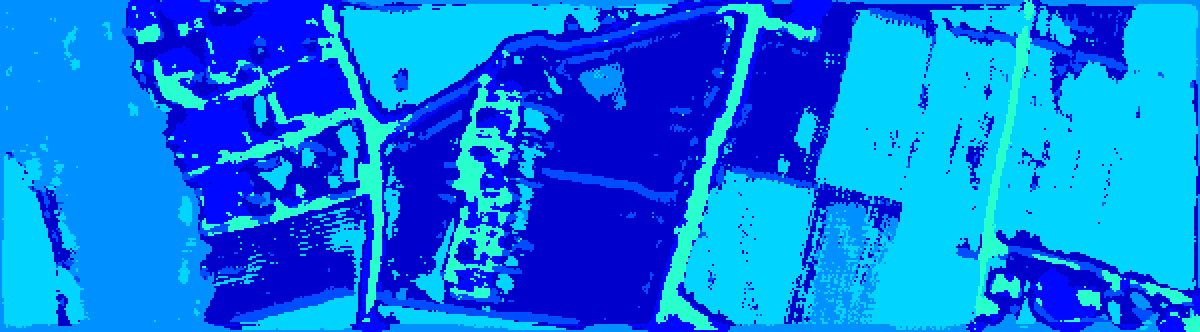} &
\includegraphics{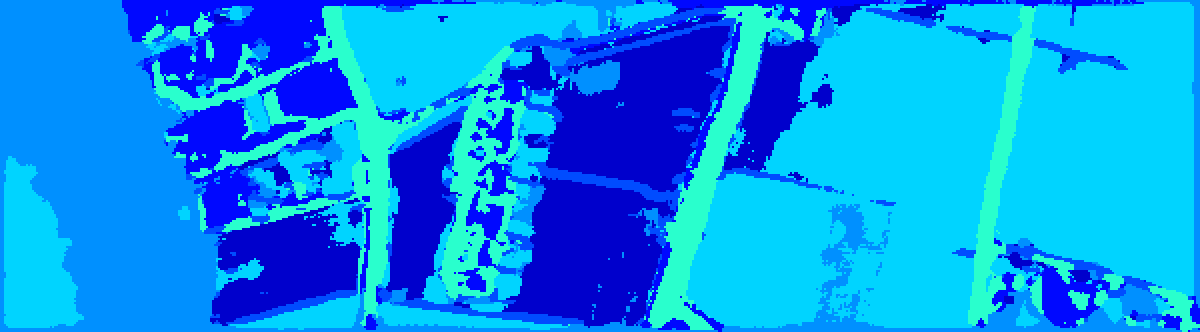} \\
[8pt]
\HUGESSS\textit{g}) LSTM  &
\HUGESSS\textit{h}) GRU  &
\HUGESSS\textit{i}) CNN1D  \\
[8pt]
\includegraphics{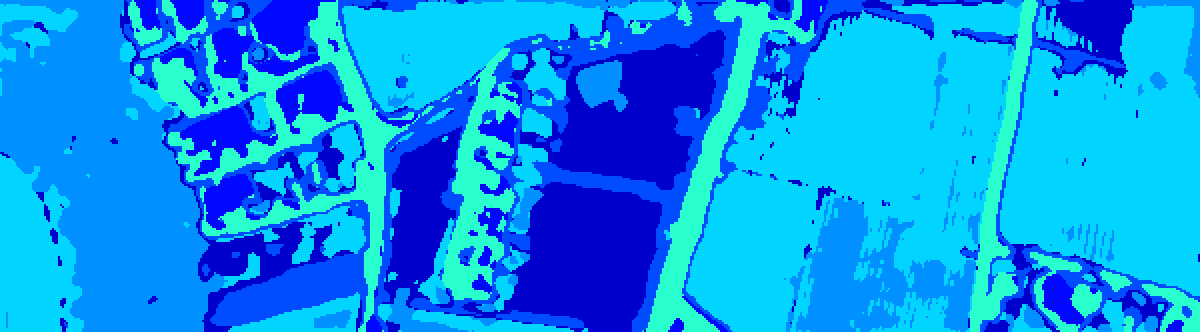}
 &
 \includegraphics{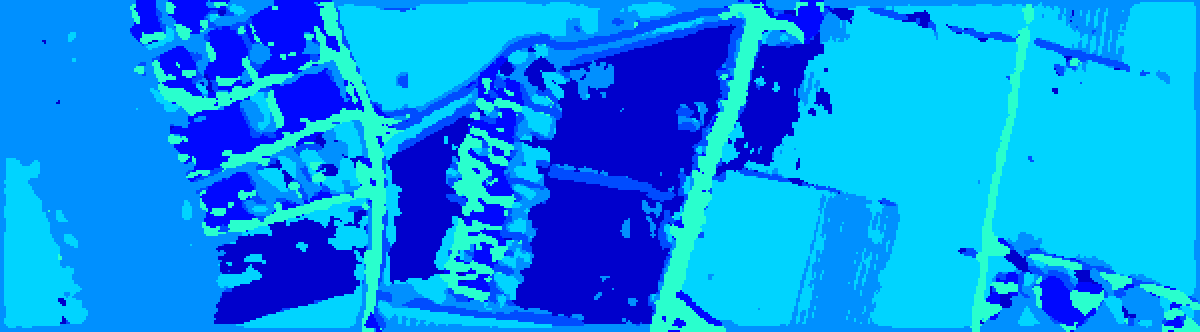}
 &
 \includegraphics{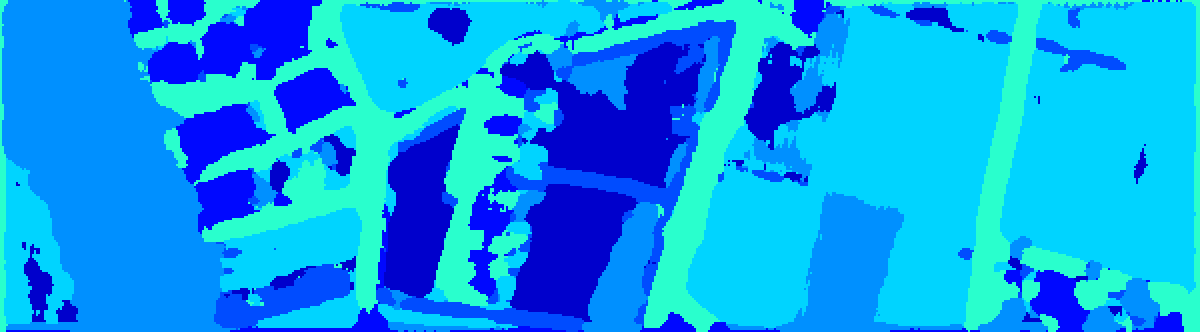}
 \\
[8pt]
\HUGESSS\textit{j}) CNN2D  &
\HUGESSS\textit{k}) CNN3D  &
\HUGESSS\textit{l}) MorphCNN  \\
\end{tabular}}
\caption{Classification Maps obtained by MLR \cite{li2010semisupervised}, SVM \cite{melgani2004classification}, MLP \cite{paoletti2019deep}, RNN \cite{hang2019cascaded}, LSTM \cite{hochreiter1997long}, GRU \cite{cho2014properties}, CNN-1D \cite{hong2021graph}, CNN-2D \cite{makantasis2015deep}, CNN-3D \cite{hamida2018deep} and MorphCNN \cite{roy2021morphological} on the disjoint train-test dataset for the UT scene.}
\label{fig:comparativeUT}
\end{figure*}

In general, the works \cite{roy2021morphological, makantasis2015deep} outperformed (i.e. stable results) than the other comparative methods especially in the case of less number of labeled training samples. The above leads to conclude that these works are not sensitive to the number of training samples. Moreover, as the number of training samples increases, the accuracies also increase for these methods however, other methods can work better with a higher number of training samples as compared to these methods. A similar trend has been observed with a higher number of training samples. Thus, one can conclude that the works \cite{roy2021morphological} and \cite{makantasis2015deep} can solve the limited availability of training samples issues to some extent while considering disjoint train/test samples. 

\begin{sidewaystable}
\let\center\empty
\let\endcenter\relax
\centering
\caption{Classification results obtained by CNN-2D \cite{makantasis2015deep} CNN-3D \cite{hamida2018deep}, G2C-Conv2D \cite{roy2021revisiting}, and G2C-Conv3D \cite{roy2021revisiting} on the disjoint train-test for IP, PU, Trento, UH, and KSC datasets. The higher accuracies are emphasised.}
\resizebox{\linewidth}{!}{\color{black}
\begin{tabular}{c||cccc||cccc||cccc||cccc||cccc}
\hline
\multirow{2}{*}{Class} & \multicolumn{4}{c||}{IP}    & \multicolumn{4}{c||}{PU}  & \multicolumn{4}{c||}{UH}  & \multicolumn{4}{c||}{Trento} & \multicolumn{4}{c}{KSC} \\ \cline{2-21} 

& Conv2D & Conv3D & G2C-2D & G2C-3D & Conv2D & Conv3D & G2C-2D & G2C-3D & Conv2D & Conv3D & G2C-2D & G2C-3D & Conv2D & Conv3D & G2C-2D & G2C-3D & Conv2D & Conv3D & G2C-2D & G2C-3D \\ \hline

1   &  46.66$\pm$0.09  & 56.00$\pm$8.64    &     53.33$\pm$18.57       &    \textbf{97.33$\pm$1.88} & 81.20$\pm$0.01  & 87.20$\pm$0.42     &     86.22$\pm$4.54       &    \textbf{91.26$\pm$3.40}  & 80.88$\pm$0.006  &   82.77$\pm$0.27     &     81.32$\pm$0.08       &      \textbf{82.90$\pm$0.20}      &  98.30$\pm$0.007  & 98.63$\pm$0.18     &     \textbf{98.75$\pm$0.58}       &      96.12$\pm$1.95   & 95.56$\pm$0.007 & 98.40$\pm$0.81  & 98.24$\pm$0.52  & \textbf{99.27$\pm$0.56} \\ 

2   &  48.04$\pm$0.03  & 64.64$\pm$4.75    &     \textbf{68.93$\pm$5.41}       &     57.33$\pm$4.91    &  89.52$\pm$0.01 &   89.90$\pm$0.35     &      \textbf{92.99$\pm$0.90}      &    83.84$\pm$0.49  & 81.64$\pm$0.005  &   82.64$\pm$0.85     &      81.79$\pm$1.44      &      \textbf{83.52$\pm$0.31}      &  85.74$\pm$0.029  & 84.78$\pm$2.61   &      90.29$\pm$2.68      &      \textbf{93.30$\pm$1.41}  & 79.06$\pm$0.03  & 95.16$\pm$1.18  & 90.82$\pm$2.04  & \textbf{98.87$\pm$0.60}   \\ 

3   &  24.33$\pm$0.009  & 44.96$\pm$8.99    &     35.97$\pm$6.92       &     \textbf{78.21$\pm$2.13}  &  53.53$\pm$0.006   &   59.98$\pm$2.20     &     51.33$\pm$2.60       &    \textbf{78.08$\pm$2.84} & 49.37$\pm$0.032   &   65.80$\pm$5.12     &      74.32$\pm$7.73      &      \textbf{91.48$\pm$1.12}      & 78.43$\pm$0.08  & \textbf{86.63$\pm$9.87}     &      63.81$\pm$7.90      &     83.51$\pm$12.16 & 74.92$\pm$0.016 &  90.36$\pm$5.99  & 80.58$\pm$3.37  & \textbf{97.09$\pm$0.21}    \\

4   &  28.28$\pm$0.13  &   41.75$\pm$4.15    &     18.18$\pm$2.18       &     \textbf{42.42$\pm$3.59}   & 96.92$\pm$0.009   &   \textbf{97.43$\pm$0.44}     &      96.97$\pm$1.05      &    94.05$\pm$1.16  & 86.26$\pm$0.006  &    \textbf{95.54$\pm$3.17}    &      91.60$\pm$0.29      &      90.49$\pm$2.39      &  93.59$\pm$0.006  &   98.31$\pm$0.30    &      97.11$\pm$0.76      &     \textbf{98.53$\pm$0.23} & 50.62$\pm$0.056  & 74.61$\pm$3.33  & 69.62$\pm$4.01  & \textbf{77.57$\pm$1.52}   \\

5   &  29.56$\pm$0.05  &   46.83$\pm$2.99    &     26.88$\pm$3.96       &     \textbf{52.79$\pm$3.88}       &  98.41$\pm$0.007  & 97.99$\pm$0.95     &      \textbf{99.28$\pm$0.38}      &    99.19$\pm$0.38  & 95.39$\pm$0.018  &    96.96$\pm$0.27    &      98.67$\pm$1.00     &      \textbf{99.87$\pm$0.08}      &  98.22$\pm$0.011   & \textbf{99.77$\pm$0.18}    &       99.57$\pm$0.22     &      98.70$\pm$0.60 & 71.04$\pm$0.03 & \textbf{88.32$\pm$3.31}  & 81.26$\pm$1.37  & 87.34$\pm$0.91   \\

6   &  72.31$\pm$0.10  &    92.56$\pm$3.01    &     94.35$\pm$2.59       &     \textbf{98.39$\pm$1.06}       & 47.65$\pm$0.08 & 62.30$\pm$2.37     &     60.68$\pm$4.23       &    \textbf{87.04$\pm$2.05}  & 81.11$\pm$0.02  &    92.07$\pm$1.83    &    82.51$\pm$4.67       &     \textbf{94.40$\pm$0.00}       &  65.09$\pm$0.036    & 75.74$\pm$3.30     &       58.87$\pm$2.95     &     73.78$\pm$2.48  & 79.82$\pm$0.03 & 90.25$\pm$1.10  & 78.63$\pm$4.43  & \textbf{93.84$\pm$1.10}   \\

7   &  0.00$\pm$0.00  &    0.00$\pm$0.00    &      0.00$\pm$0.00       &     0.00$\pm$0.00       &  61.36$\pm$0.02  &    76.55$\pm$1.11    &     65.30$\pm$3.66       &    \textbf{88.44$\pm$5.08}  & 79.57$\pm$0.02  &    84.48$\pm$2.21    &    82.64$\pm$1.12        &      \textbf{85.85$\pm$0.49}  &    &        &            &  &  88.01$\pm$0.010 & 98.50$\pm$1.40  & 95.88$\pm$1.05  & \textbf{99.62$\pm$0.52} \\

8   &  91.60$\pm$0.05  &    \textbf{98.93$\pm$0.67}    &     98.66$\pm$0.82       &     94.53$\pm$4.64       &  81.97$\pm$0.02  &  94.73$\pm$1.51    &     97.25$\pm$0.93       &    \textbf{97.34$\pm$0.92}  & 47.61$\pm$0.08  &    65.71$\pm$4.44    &   56.98$\pm$0.73         &      \textbf{71.06$\pm$0.98}   &    &        &            &   & 85.15$\pm$0.062 & 97.08$\pm$0.34  & 87.70$\pm$1.77  & \textbf{98.99$\pm$0.71} \\

9   &  36.66$\pm$0.04  &    33.33$\pm$12.47    &     50.00$\pm$8.16       &     \textbf{80.00$\pm$8.16}       & 88.38$\pm$0.04  & \textbf{96.98$\pm$0.81}    &     96.31$\pm$1.74       &    96.85$\pm$1.61  &  74.15$\pm$0.04    &    79.85$\pm$1.96    &     77.49$\pm$1.45       &      \textbf{82.37$\pm$1.60}  &     &        &            &  & 98.11$\pm$0.013 &  98.64$\pm$0.97  & 99.62$\pm$0.21  & \textbf{99.77$\pm$0.18}      \\

10   &  50.49$\pm$0.04  &    \textbf{62.42$\pm$5.93}    &     52.75$\pm$3.42       &     55.00$\pm$4.17     &   &        &            &    &  41.40$\pm$0.02   &    45.65$\pm$1.29    &        \textbf{52.67$\pm$3.07}    &     49.25$\pm$1.74  &     &        &            &  & 96.20$\pm$0.008  & 99.31$\pm$0.76  & 99.90$\pm$0.13  & \textbf{100.0$\pm$0.00}      \\

11   &  73.45$\pm$0.01  &    79.68$\pm$1.91    &     79.68$\pm$0.66       &     \textbf{81.72$\pm$0.19}    &   &        &            &    &  50.18$\pm$0.01   &    54.26$\pm$3.88    &        65.71$\pm$1.04    &     \textbf{70.55$\pm$1.47}   &    &        &            &  &  92.97$\pm$0.007 & 99.81$\pm$0.26  & 99.15$\pm$0.45  & \textbf{100.0$\pm$0.00}      \\

12   &  30.61$\pm$0.01  &    49.29$\pm$4.92    &     \textbf{56.38$\pm$5.87}       &     35.93$\pm$7.39   &    &        &            &   &  70.22$\pm$0.02    &    76.46$\pm$3.88    &        \textbf{88.98$\pm$4.50}    &     88.05$\pm$4.16  &     &        &            &    & 92.83$\pm$0.025 & 95.63$\pm$0.77  & \textbf{98.90$\pm$0.39}  & 98.13$\pm$0.66      \\

13   &  95.83$\pm$0.03  &    \textbf{97.08$\pm$1.17}    &     94.16$\pm$5.62       &     95.83$\pm$0.58       &    &    &            &     &  85.84$\pm$0.019  &    \textbf{91.92$\pm$0.28}    &        85.02$\pm$0.92    &     82.57$\pm$0.82   &    &        &            &    & \textbf{100.0$\pm$0.00} & \textbf{100.0$\pm$0.00}  & \textbf{100.0$\pm$0.00}  & \textbf{100.0$\pm$0.00}      \\

14   &  74.18$\pm$0.001  &    76.08$\pm$4.46    &     \textbf{95.59$\pm$0.98}       &     90.03$\pm$5.77       &    &    &            &     &  77.86$\pm$0.06 &   82.86$\pm$1.63 &         80.70$\pm$0.68    &     \textbf{99.05$\pm$0.76}  &     &   &      &            &            \\

15   &  19.52$\pm$0.01  &   54.54$\pm$16.05    &     21.21$\pm$8.12       &     \textbf{74.74$\pm$2.97}       &    &    &            &     &  48.34$\pm$0.06  &    70.68$\pm$3.24    &        59.83$\pm$6.72    &     \textbf{99.92$\pm$0.09}  &     &  &      &            &            \\

16   &  71.21$\pm$0.18  &    \textbf{78.78$\pm$2.83}    &     62.87$\pm$5.96       &     78.03$\pm$10.55       &   &     &            &    &     &        &      &      &            &   &     &            &            \\ \hline \hline

OA   &  57.02$\pm$0.12  &    69.25$\pm$1.32    &     68.33$\pm$0.60       &     \textbf{72.82$\pm$0.26}       & 81.23$\pm$0.36 & 85.96$\pm$0.27     &      86.55$\pm$0.44      &     \textbf{87.79$\pm$0.23} & 69.67$\pm$0.19  &   76.52$\pm$1.13     &     77.27$\pm$0.61       &     \textbf{82.26$\pm$0.56}   &  91.95$\pm$1.03  &    95.09$\pm$0.26    &     93.15$\pm$0.34       &      \textbf{95.21$\pm$0.33}   & 89.76$\pm$1.13 & 96.15$\pm$0.43  &  94.05$\pm$0.49  &  \textbf{97.65$\pm$0.02} \\

AA   &  49.55$\pm$0.01  &    61.05$\pm$2.92    &     56.81$\pm$1.73       &     \textbf{69.52$\pm$1.23}       &  77.66$\pm$0.002 & 84.78$\pm$0.18    &     82.92$\pm$0.38       &    \textbf{90.68$\pm$0.06} &  69.99$\pm$0.007  &    77.84$\pm$0.86    &        77.35$\pm$1.18    &     \textbf{84.76$\pm$0.52}  &  86.56$\pm$0.021   &   90.64$\pm$1.80     &      84.73$\pm$1.57      &     \textbf{90.66$\pm$2.24}   & 84.95$\pm$0.011 & 94.31$\pm$0.65  &  90.79$\pm$0.64  &  \textbf{96.19$\pm$0.01} \\

k(x100)   &  50.66$\pm$0.22  &    64.78$\pm$1.58    &     63.60$\pm$0.74       &     \textbf{68.87$\pm$0.37}   &  74.57$\pm$0.58  &   81.11$\pm$0.36     &     81.74$\pm$0.65       &    \textbf{83.95$\pm$0.30}  & 67.22$\pm$0.24  &   74.61$\pm$1.19     &    75.43$\pm$0.67        &     \textbf{80.87$\pm$0.60}   &  89.20$\pm$1.38  &   93.42$\pm$0.36     &     90.79$\pm$0.46       &     \textbf{93.62$\pm$0.45}    & 88.60$\pm$1.26 & 95.72$\pm$0.48  &  93.37$\pm$0.55  &  \textbf{97.39$\pm$0.02} \\ \hline
\end{tabular}
}
\label{table:compdisjointG2C}
\end{sidewaystable}

Moreover, one can conclude that the AE-based models do not perform well as compared to the other models, although the unsupervised methods do not require the samples to be labeled, if there are no constraints, these methods might learn nothing. Moreover, AE has a symmetric architecture that leads to the explosion of training parameters which increases the difficulty in training. The works \cite{rs11070864} and \cite{ABC} overcome the above-mentioned issues, however, the work \cite{sun2017encoding} does not adopt the greedy layer-wise approach thus producing the worst results, thus, there is room for improvement in such methods. 

In a nutshell, the classification results based on CNN are way better than AE-based methods while considering the limited availability of labeled training samples. Although the AEs can learn the internal structure of the unlabeled data, the final feature representation might not have task-driven characteristics which might be the reason for not performing well as compared to the supervised learning models. Moreover, AL and/or SL takes the benefits from the selection of the most important samples for training which enables the model to focus more attention on indistinguishable samples for HSIC. Whereas, FSL benefits from the exploration of the relationship between samples to find a discriminative decision boundary for HSIC. TL makes good use of similarity among different HSI’s to reduce the quantity required for training also reduces the number of trainable parameters while boosting the models' robustness. According to the raw data (i.e., processing the HSI without extracting/learning the features), DA generates more samples which bring a diversity of samples.

\subsection{Experiments with Convolutional Feature Extractors}

This section revisited several deep Hyperspectral feature extraction processes, i.e., a traditional convolutional process and a gradient centralized convolutional process. In this hierarchy, we have conducted several experiments using several state-of-the-art works published in recent years. This experiment is specifically designed to check the performance of the convolutional process rather than testing the model's performance. The baseline models apply convolutional feature extractors which include a 2D convolution neural network for HSI classification (Conv2D) introduce by Makantasis \textit{et al.} \cite{makantasis2015deep} and the 3D convolutional approach for remote sensing image classification (Conv3D) proposed by Hamida \textit{et al.} \cite{hamida2018deep} (a traditional 3D convolutional feature extractor), and recently Roy \textit{et al.} introduced generalized gradient centralized 2D convolution (G2C-Conv2D) \cite{roy2021revisiting}, and generalized gradient centralized 3D convolution (G2C-Conv3D) \cite{roy2021revisiting} to extract the fine-grained spectral-spatial feature representation. The generalized gradient centralized 3D convolution (G2C-Conv3D) operation is designed by using a weighted combination between the vanilla and gradient centralized 3D convolutions (GC-Conv3D) to extract both the intensity level semantic information and gradient level information from the HSIs.

All the aforementioned convolutional feature extractors have been evaluated on 5 different Hyperspectral datasets, namely, IP, PU, Trento, UH, and KSC datasets. The experimental results are illustrated in Table \ref{table:compdisjointG2C}. From all these results, one can easily conclude that the G2C-Conv3D convolutional process outperformed Conv2D and Conv3D followed by G2C-Conv2D. A similar trend has been observed for all datasets except the Trento dataset on which the 3D convolutional process slightly performed better as compared to the traditional Conv2D and G2C-Conv2D, respectively. The accuracy difference is not that high as compared to the G2C-Conv3D for other datasets. Most importantly, the G2C-Conv3D convolution operation is simple to implement and can easily be plugged into existing CNNs to boost both the robustness and classification performance.

\section{Conclusion and Future Directions}
\label{SecVIII}

The rich information contained in HSI data is a captivating factor that constitutes the utilization of HSI technology in real-world applications. Moreover, advances in machine learning methods strengthen the deployment potentials of such technologies. In this work, we surveyed recent developments of Hyperspectral Image Classification (HSIC) using state of the art Deep Neural Networks (for instance, Auto-encoder (AE), Deep Belief Network (DBN), Recurrent Neural Network (RNN), Convolutional Neural Network (CNN), Transfer Learning (TL), Few-shot Learning (FSL), Active/Self Learning (AL/SL), and Data Augmentation (DA)) in a variety of learning schemes (specifically, supervised, semi-supervised and unsupervised learning). In addition, we also analyzed the strategies to overcome the challenges of limited availability of training data like Data Augmentation, Few-shot Learning (FSL), Transfer Learning, and Active Learning, etc. According to the methodologies discussed above, we select some of the representative works to conduct the experiments on benchmark HSI datasets. 

Although the current HSIC techniques reflect a rapid, remarkable, and sophistication of the task, further developments are still required to improve the generalization capabilities. The main issue of deep neural network-based HSIC is the lack of labeled data. HSI data is infamous due to the limited availability of labeled data and deep neural networks demand a sufficiently large amount of labeled training data. Section \ref{SecVI} discussed some widely used strategies to combat the aforesaid issue but significant improvements are still needed to efficiently utilize limited available training data. One direction to solve this problem could be to explore the integration of various learning strategies discussed in section \ref{SecVI} to cash in the joint benefits. One more way is to exploit a few-shot or K-shot learning approaches that can accurately predict the class labels with only a few labeled samples. Moreover, there is a need to focus on the joint exploitation of spectral-spatial features of HSI to complement classification accuracies achieved from the aforementioned HSIC frameworks. Another future potential of HSIC is computationally efficient architectures. Therefore, the issue of the high computational complexity of deep neural networks is of paramount importance and it is crucial to implement parallel HSIC architectures to speed up the processing of deep neural networks to meet the computational stipulation of time-critical HSI applications. In this direction, high-performance computing platforms and specialized hardware modules like graphical processing units (GPUs) and field-programmable gate arrays (FPGAs) can be used to implement the parallel HSIC frameworks. Hence, to assimilate aforesaid aspects in the development of a new HSIC framework is to appropriately utilize the limited training samples while considering joint spectral-spatial features of HSI and maintaining the low computational burden.

\section*{Acknowledgment}
The authors thank Ganesan Narayanasamy who is leading IBM OpenPOWER/POWER enablement and ecosystem worldwide for his support to get the IBM AC922 system's access.

\scriptsize{
\bibliographystyle{ieeetr}
\bibliography{sample}}

\begin{IEEEbiography}[{\includegraphics[width=1in,height=1.25in,clip,keepaspectratio]{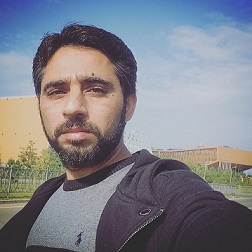}}]{Muhammad Ahmad} received his MS degree in Electronics Engineering from International Islamic University, Islamabad, Pakistan, a Ph.D. degree in Computer Science and Engineering, from Innopolis University, Innopolis, Russia, and another Ph.D. degree in Cyber-Physical Systems from the University of Messina, Messina, Italy.

Muhammad is currently working at the National University of Computer \& Emerging Sciences (FAST-NUCES). He has also served as an Assistant Professor, Lecturer, Instructor, Research Fellow, Research Associate, and Research Assistant for a number of international/national universities. He has also worked with Ericsson (Mobilink Project) as Radio Access Network (RAN) Supervisor. He authored and co-authored over 70 scientific contributions to international journals, conferences, and books. He is supervising/co-supervising several graduates (MS and Ph.D.). He served/serving as a lead/guest editor on several special issues in journals (SCI/E, JCR). He has delivered a number of invited and keynote talks and reviewed (reviewing) the technology-leading articles for journals. His research interest includes Hyperspectral Imaging, Remote Sensing, Machine Learning, Computer Vision, and Wearable Computing.
\end{IEEEbiography}

\vskip -2\baselineskip plus -1fil
\begin{IEEEbiography}[{\includegraphics[width=1in,height=1.25in,clip,keepaspectratio]{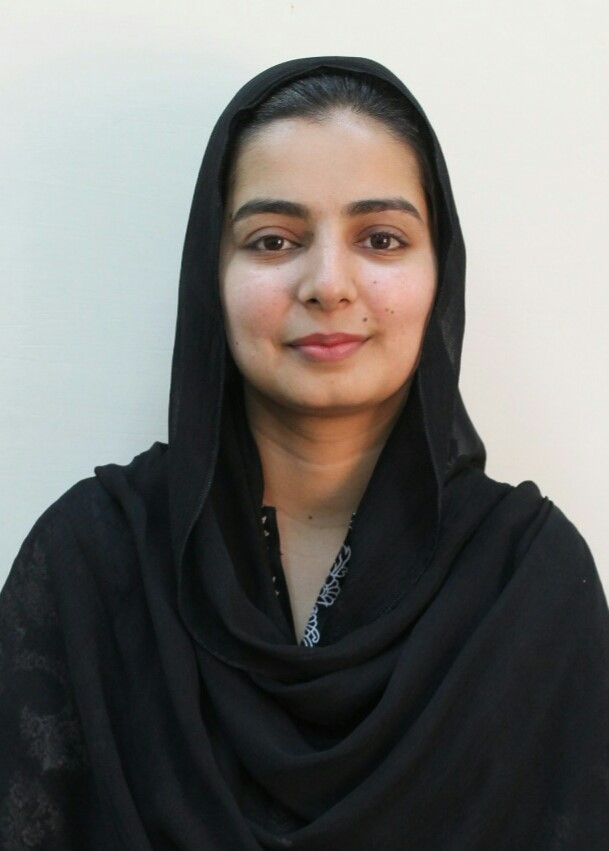}}]{Sidrah Shabir} received her bachelor’s degree in Computer Engineering from COMSATS University Islamabad and a master’s degree in Computer Engineering from Khwaja Fareed University of Engineering and Information Technology. Currently, she is working as Lab Engineer at the Department of Computer Engineering, Khwaja Fareed University of Engineering and Information Technology. Her research interests include Machine learning, Hyperspectral Imaging and Hardware Accelerator Design for Machine learning.
\end{IEEEbiography}

\vskip -2\baselineskip plus -1fil
\begin{IEEEbiography}[{\includegraphics[width=1in,height=1.25in,clip,keepaspectratio]{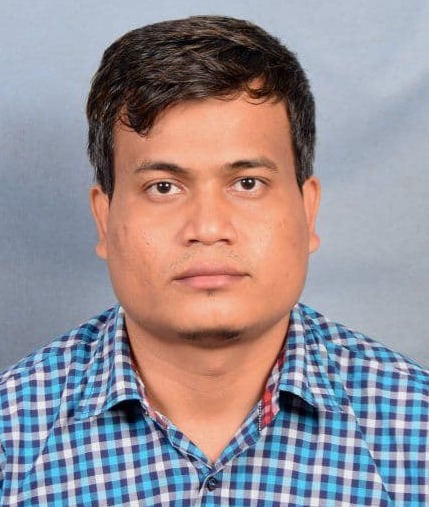}}]{Swalpa Kumar Roy}(S'15) received the bachelor’s and the master’s degree in Computer Science and Engineering from West Bengal University of Technology, Kolkata, India, in 2012, and Indian Institute of Engineering Science and Technology, Shibpur, Howrah, India, (IIEST Shibpur) in 2015 and also the Ph.D. degree in Computer Science and Engineering from University of Calcutta, Kolkata in 2021. 

From July 2015 to March 2016, he was a Project Linked Person with the Optical Character Recognition (OCR) Laboratory, Computer Vision and Pattern Recognition Unit, Indian Statistical Institute, Kolkata. He is currently working as an Assistant Professor with the Department of Computer Science and Engineering, Jalpaiguri Government Engineering College, West Bengal, India. Dr. Roy was nominated for the Indian National Academy of Engineering (INAE) engineering teachers mentoring fellowship program by INAE Fellows in 2021 and also a recipient of the Outstanding Paper Award in second Hyperspectral Sensing Meets Machine Learning and Pattern Analysis (HyperMLPA) at the Workshop on Hyperspectral Imaging and Signal Processing: Evolution in Remote Sensing (WHISPERS) in 2021. He has served as a reviewer for the IEEE Transactions on Geoscience and Remote Sensing and IEEE Geoscience and Remote Sensing Letters. His research interests include computer vision, deep learning and remote sensing.
\end{IEEEbiography}

\vskip -2\baselineskip plus -1fil
\begin{IEEEbiography}[{\includegraphics[width=1in,height=1.25in,clip,keepaspectratio]{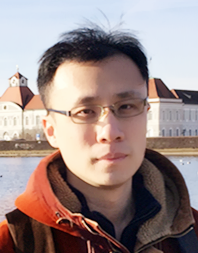}}]{Danfeng Hong}
(S'16--M'19--SM'21) received the M.Sc. degree (summa cum laude) in computer vision from the College of Information Engineering, Qingdao University, Qingdao, China, in 2015, the Dr. -Ing degree (summa cum laude) from the Signal Processing in Earth Observation (SiPEO), Technical University of Munich (TUM), Munich, Germany, in 2019. 

From 2015 to 2019, he was a Research Associate at the Remote Sensing Technology Institute (IMF), German Aerospace Center (DLR), Oberpfaffenhofen, Germany. Since 2019, He has been a Research Scientist and led a Spectral Vision Working Group at IMF, DLR. He is also an Adjunct Scientist at GIPSA-lab, Grenoble INP, CNRS, Univ. Grenoble Alpes, Grenoble, France, from 2020. He is currently with the Key Laboratory of Digital Earth Science, Aerospace Information Research Institute (AIR), Chinese Academy of Sciences (CAS). His research interests include signal/image processing and analysis, hyperspectral remote sensing, machine / deep learning, artificial intelligence, and their applications in Earth Vision.

Dr. Hong is an Editorial Board Member of Remote Sensing and a Topical Associate Editor of the IEEE Transactions on Geoscience and Remote Sensing (TGRS). He was a recipient of the Best Reviewer Award of the IEEE TGRS in 2021 and the Jose Bioucas Dias award for recognizing the outstanding paper at the Workshop on Hyperspectral Imaging and Signal Processing: Evolution in Remote Sensing (WHISPERS) in 2021. He is also a Leading Guest Editor of the International Journal of Applied Earth Observation and Geoinformation, the IEEE Journal of Selected Topics in Applied Earth Observations, and Remote Sensing.
\end{IEEEbiography}

\vskip -2\baselineskip plus -1fil

\begin{IEEEbiography}[{\includegraphics[width=1in,height=1.25in,clip,keepaspectratio]{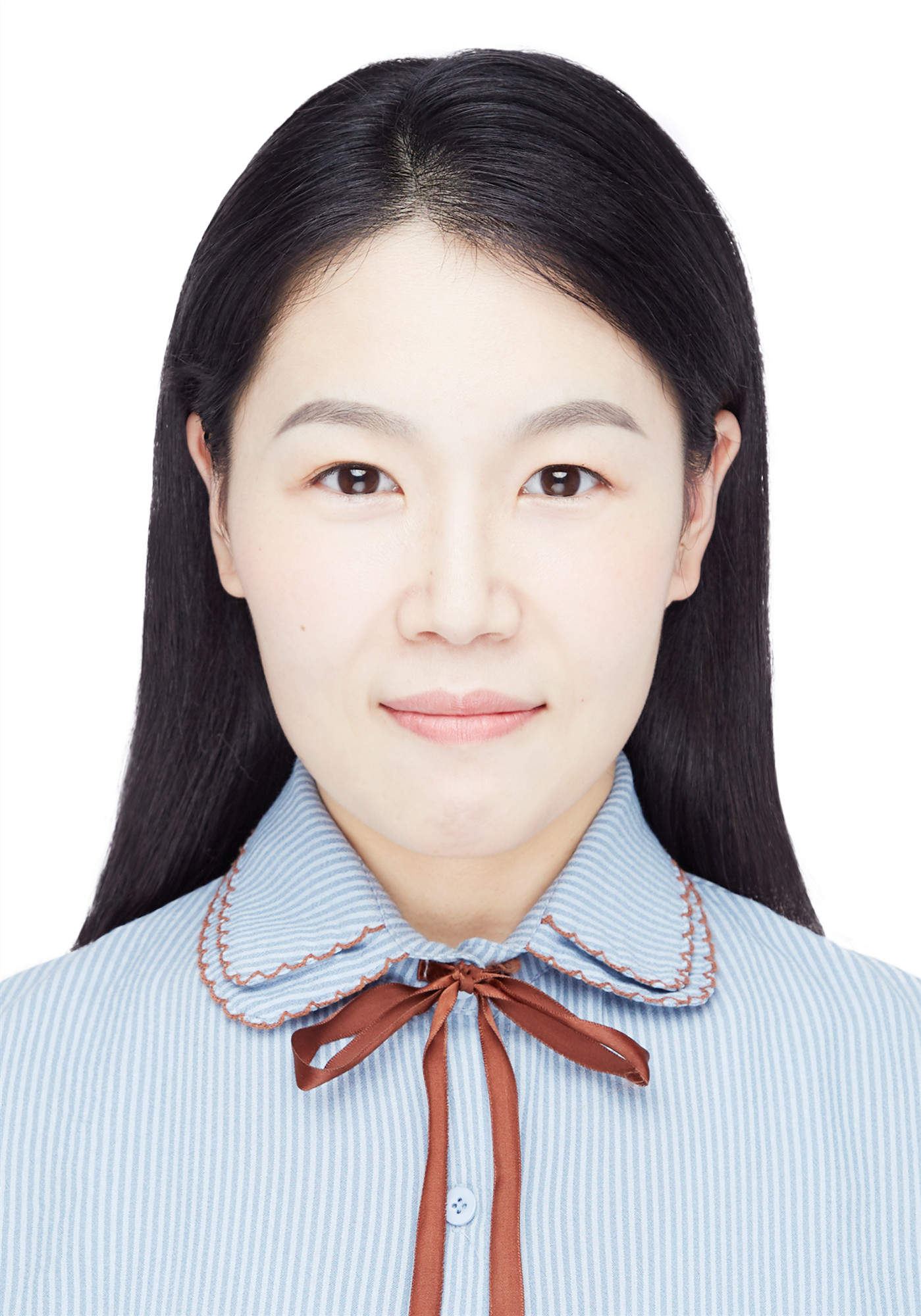}}]{Xin Wu} (S'19--M'20) received the M.Sc. degree in Computer Science and Technology from the College of Information Engineering, Qingdao University, Qingdao, China, in 2014, the Ph.D. degree from the School of Information and Electronics, Beijing Institute of Technology (BIT), Beijing, China, in 2020.

In 2018, she was a visiting student at the Photogrammetry and Image Analysis department of the Remote Sensing Technology Institute (IMF), German Aerospace Center (DLR), Oberpfaffenhofen, Germany. She is currently a Postdoctoral Researcher in the School of Information and Electronics, BIT, Beijing, China. Her research interests include signal/image processing, fractional Fourier transform, deep learning and their applications in biometrics and geospatial object detection.

She was a recipient of the Jose Bioucas Dias award for recognizing the outstanding paper at the Workshop on Hyperspectral Imaging and Signal Processing: Evolution in Remote Sensing (WHISPERS) in 2021.
\end{IEEEbiography}

\vskip -2\baselineskip plus -1fil

\begin{IEEEbiography}[{\includegraphics[width=1in,height=1.25in,clip,keepaspectratio]{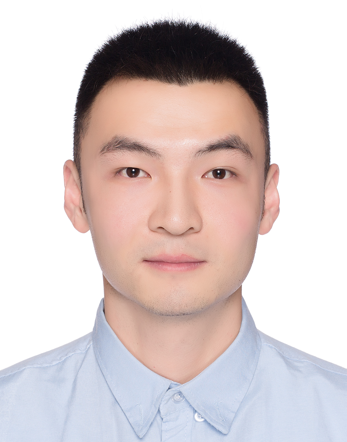}}]{Jing Yao} received the B.Sc. degree from Northwest University, Xi’an, China, in 2014, and the Ph.D. degree in the School of Mathematics and Statistics, Xi’an Jiaotong University, Xi’an, China, in 2021. 

He is currently an Assistant Professor with the Key Laboratory of Digital Earth Science, Aerospace Information Research Institute, Chinese Academy of Sciences, Beijing, China. From 2019 to 2020, he was a visiting student at Signal Processing in Earth Observation (SiPEO), Technical University of Munich (TUM), Munich, Germany, and at the Remote Sensing Technology Institute (IMF), German Aerospace Center (DLR), Oberpfaffenhofen, Germany.

His research interests include low-rank modeling, hyperspectral image analysis and deep learning-based image processing methods.
\end{IEEEbiography}

\vskip -2\baselineskip plus -1fil
\begin{IEEEbiography}[{\includegraphics[width=1in,height=1.25in,clip,keepaspectratio]{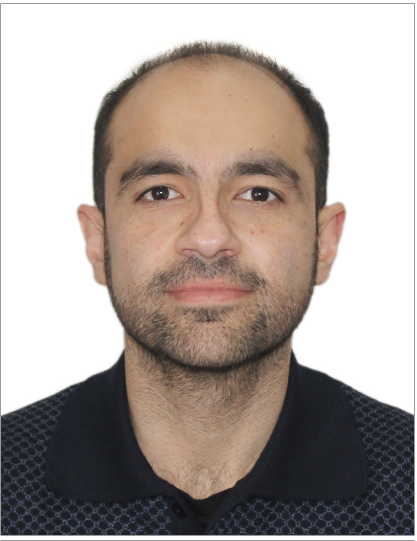}}]{Adil Mehmood Khan} received his B.S. degree in Information Technology from National University of Sciences and Technology (NUST), Pakistan in 2005. He completed his M.Sc. and Ph.D. degrees in Computer Engineering from Kyung Hee University, South Korea in 2011. He is currently a Professor at the Institute of Artificial Intelligence and Data Science, Innopolis University, Russia. His research interests are machine learning and deep learning.
\end{IEEEbiography}

\vskip -2\baselineskip plus -1fil
\begin{IEEEbiography}[{\includegraphics[width=1in,height=1.25in,clip,keepaspectratio]{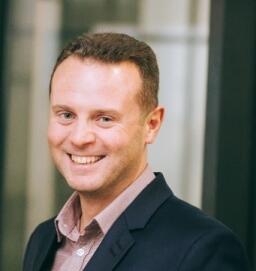}}]{Manuel Mazzara} is a professor of Computer Science at Innopolis University (Russia) with a research background in software engineering, service-oriented architectures and programming, concurrency theory, formal methods, software verification and Artificial Intelligence.

Manuel received a PhD in computing science from the University of Bologna and cooperated with European and US industry, plus governmental and inter-governmental organizations such as the United Nations, always at the edge between science and software production.

The work conducted by Manuel and his team in recent years focuses on the development of theories, methods, tools and programs covering the two major aspects of Software Engineering and Artificial Intelligence: the process side, describing how we develop software, and the product side, describing the results of this process.

\end{IEEEbiography}

\vskip -2\baselineskip plus -1fil
\begin{IEEEbiography}[{\includegraphics[width=1in,height=1.25in,clip,keepaspectratio]{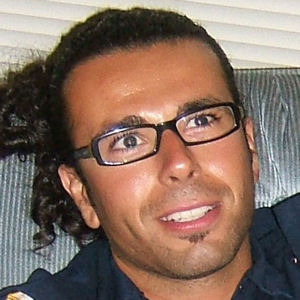}}]{Salvatore Distefano}is an Associate Professor at the University of Messina (Italy). He authored and co-authored more than 250 scientific papers and contributions to international journals, conferences, and books. He visited as a scholar and professor different universities and research centers 
such as collaborating with top scientists such as UMass Dartmouth, UCLA, Duke, Innopolis, and kazan Federal University.

He took part in several national and international projects, such as Reservoir, Vision (EU FP7), SMSCOM (EU FP7 ERC Advanced Grant), Beacon, IoT-Open.EU (EU H2020). He is a member of international conference committees and he is on the editorial boards of IEEE Transactions on Dependable and Secure Computing, Journal of Cloud Computing, International Journal of Big Data. His main research interests include non-Markovian modeling; Quality of Service/Experience; Parallel and Distributed Computing, Grid, Cloud, Autonomic, Volunteer, Crowd, Edge, Fog Computing; Internet of Things; Cyber-Physical Social Systems; Smart Cities; Intelligent Transportation Systems; Big Data, Stream Processing; Software-Defined and virtualized ecosystems; Hyper Spectral Imaging; Machine Learning. During his research activity, he contributed to the development of several tools such as WebSPN, ArgoPerformance, GS3 and Stack4Things. He is also one of the co-founders of the SmartMe.io start-up, a spin-off of the University of Messina established in 2017.
\end{IEEEbiography}

\vskip -2\baselineskip plus -1fil

\begin{IEEEbiography}[{\includegraphics[width=1in,height=1.25in,clip,keepaspectratio]{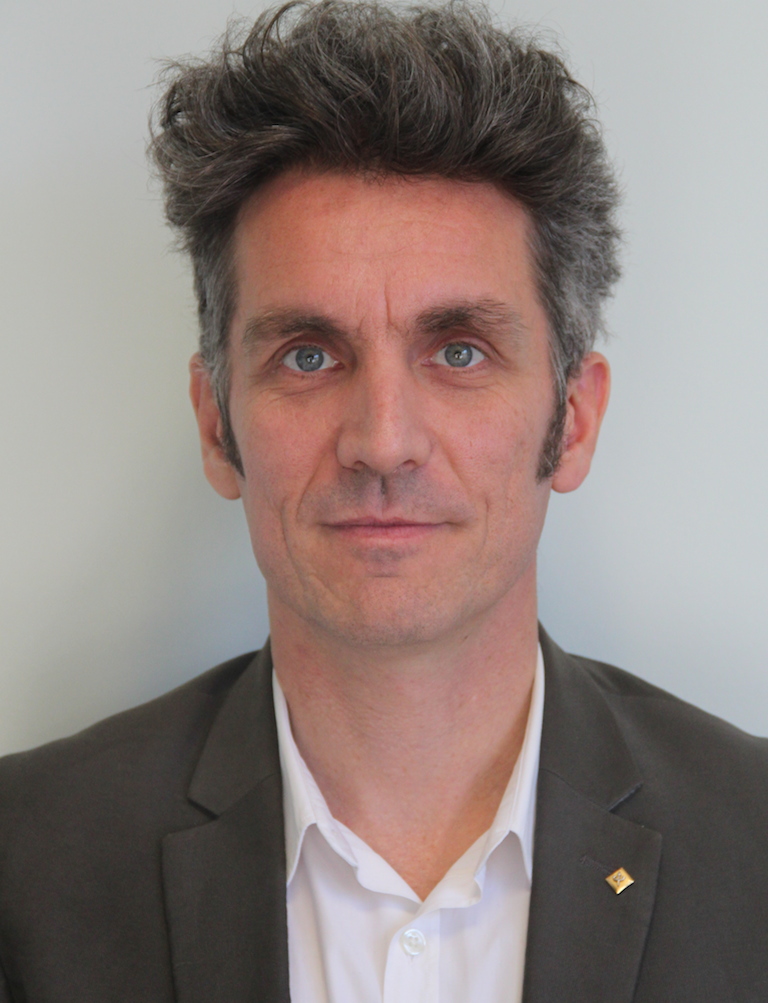}}]{Jocelyn Chanussot}
(M'04--SM'04--F'12) received the M.Sc. degree in electrical engineering from the Grenoble Institute of Technology (Grenoble INP), Grenoble, France, in 1995, and the Ph.D. degree from the Université de Savoie, Annecy, France, in 1998. Since 1999, he has been with Grenoble INP, where he is currently a Professor of signal and image processing. His research interests include image analysis, hyperspectral remote sensing, data fusion, machine learning and artificial intelligence. He has been a visiting scholar at Stanford University (USA), KTH (Sweden) and NUS (Singapore). Since 2013, he is an Adjunct Professor of the University of Iceland. In 2015-2017, he was a visiting professor at the University of California, Los Angeles (UCLA). He holds the AXA Chair in remote sensing and is an Adjunct Professor at the Chinese Academy of Sciences, Aerospace Information Research Institute, Beijing.

Dr. Chanussot is the founding President of IEEE Geoscience and Remote Sensing French chapter (2007-2010) which received the 2010 IEEE GRS-S Chapter Excellence Award. He has received multiple outstanding paper awards. He was the Vice-President of the IEEE Geoscience and Remote Sensing Society, in charge of meetings and symposia (2017-2019). He was the General Chair of the first IEEE GRSS Workshop on Hyperspectral Image and Signal Processing, Evolution in Remote sensing (WHISPERS). He was the Chair (2009-2011) and  Cochair of the GRS Data Fusion Technical Committee (2005-2008). He was a member of the Machine Learning for Signal Processing Technical Committee of the IEEE Signal Processing Society (2006-2008) and the Program Chair of the IEEE International Workshop on Machine Learning for Signal Processing (2009). He is an Associate Editor for the IEEE Transactions on Geoscience and Remote Sensing, the IEEE Transactions on Image Processing and the Proceedings of the IEEE. He was the Editor-in-Chief of the IEEE Journal of Selected Topics in Applied Earth Observations and Remote Sensing (2011-2015). In 2014 he served as a Guest Editor for the IEEE Signal Processing Magazine. He is a Fellow of the IEEE, a member of the Institut Universitaire de France (2012-2017) and a Highly Cited Researcher (Clarivate Analytics/Thomson Reuters).
\end{IEEEbiography}

\end{document}